\documentclass[preprint2]{aastex62}
\usepackage{natbib}
\newcommand{\halpha}{H$\alpha$}
\newcommand{\NII}{[\hbox{{\rm N}\kern 0.1em{\sc ii}}] }
\newcommand{\zfire}{{\sc ZFIRE}}
\newcommand{\vrot}{$V_{rot}$}
\newcommand{\vtwo}{$V_{2.2}$}

\newcommand{\sigmag}{$\sigma_g$}

\newcommand{\Mstar}{M$_{\star}$}

\newcommand{\kms}{km~s$^{-1}$}

\newcommand{\zsimtwo}{$z\sim2$}

\newcommand{\sof}{S$_{0.5}$}

\newcommand{\voversig}{$V_{2.2}/\sigma_g$}
\newcommand{\sersic}{S{\'e}rsic}
\newcommand{\disknum}{25}
\newcommand{\irrnum}{19}
\newcommand{\totnum}{44}
\newcommand{\jdisk}{$j_{disk}$}

\revised{\today}

\shorttitle{\zfire\ 2D Kinematics at $z\sim2$}
\shortauthors{Alcorn et al.}

\begin{document}

\title{ZFIRE: 3D Modeling of Rotation, Dispersion, and Angular Momentum of Star-Forming Galaxies at $\MakeLowercase{z}\sim2$}

\correspondingauthor{Leo Y. Alcorn}
\email{lyalcorn@tamu.edu}

\author[0000-0002-2250-8687]{Leo Y. Alcorn}
\affiliation{Department of Physics and Astronomy, Texas A\&M University, College Station, TX, 77843-4242 USA}
\affiliation{George P.\ and Cynthia Woods Mitchell Institute for Fundamental Physics and Astronomy, Texas A\&M University, College  Station, TX, 77843-4242}
\affiliation{LSSTC Data Science Fellow}

\author[0000-0001-9208-2143]{Kim-Vy Tran}
\affiliation{Department of Physics and Astronomy, Texas A\&M University, College Station, TX, 77843-4242 USA}
\affiliation{George P.\ and Cynthia Woods Mitchell Institute for Fundamental Physics and Astronomy, Texas A\&M University, College  Station, TX, 77843-4242}
\affiliation{School of Physics, University of New South Wales, Sydney, NSW 2052, Australia}

\author[0000-0002-3254-9044]{Karl Glazebrook}
\affiliation{Swinburne University of Technology, Hawthorn, VIC 3122, Australia}

\author[0000-0001-5937-4590]{Caroline M. Straatman}
\affiliation{Max Planck Institut f{\"u}r Astronomie, K{\"o}nigstuhl 17, 69117 Heidelberg, Germany}

\author[0000-0002-4653-8637]{Michael Cowley}
\affiliation{Australian Astronomical Observatory, PO Box 915, North Ryde, NSW 1670, Australia}
\affiliation{Department of Physics and Astronomy, Faculty of Science and Engineering, Macquarie University, Sydney, NSW 2109, Australia}

\author[0000-0001-6003-0541]{Ben Forrest}
\affiliation{Department of Physics and Astronomy, Texas A\&M University, College Station, TX, 77843-4242 USA}
\affiliation{George P.\ and Cynthia Woods Mitchell Institute for Fundamental Physics and Astronomy, Texas A\&M University, College  Station, TX, 77843-4242}

\author{Glenn G. Kacprzak}
\affiliation{Swinburne University of Technology, Hawthorn, VIC 3122, Australia}

\author{Lisa J. Kewley}
\affiliation{Research School of Astronomy and Astrophysics, The Australian National University, Cotter Road, Weston Creek, ACT 2611, Australia}

\author{Ivo Labb{\'e}}
\affiliation{Swinburne University of Technology, Hawthorn, VIC 3122, Australia}
\affiliation{Leiden Observatory, Leiden University, P.O. Box 9513, NL 2300 RA Leiden, The Netherlands}

\author[0000-0003-2804-0648]{Themiya Nanayakkara}
\affiliation{Swinburne University of Technology, Hawthorn, VIC 3122, Australia}
\affiliation{Leiden Observatory, Leiden University, P.O. Box 9513, NL 2300 RA Leiden, The Netherlands}

\author[0000-0001-5185-9876]{Lee R. Spitler}
\affiliation{Australian Astronomical Observatory, PO Box 915, North Ryde, NSW 1670, Australia}
\affiliation{Department of Physics and Astronomy, Faculty of Science and Engineering, Macquarie University, Sydney, NSW 2109, Australia}

\author[0000-0003-2008-1752]{Adam Tomczak}
\affiliation{Department of Physics, University of California, Davis, CA, 95616, USA}

\author[0000-0002-9211-3277]{Tiantian Yuan}
\affiliation{Swinburne University of Technology, Hawthorn, VIC 3122, Australia}

\begin{abstract}

We perform a kinematic and morphological analysis of \totnum\ star-forming galaxies at \zsimtwo\ in the COSMOS legacy field using near-infrared spectroscopy from Keck/MOSFIRE and F160W imaging from CANDELS/3D-HST as part of the ZFIRE survey. 
Our sample consists of cluster and field galaxies from $2.0 < z < 2.5$ with K band multi-object slit spectroscopic measurements of their \halpha\ emission lines. 
\halpha\ rotational velocities and gas velocity dispersions are measured using the Heidelberg Emission Line Algorithm (HELA), which compares directly to simulated 3D data-cubes. 
Using a suite of simulated emission lines, we determine that HELA reliably recovers input \sof\ and angular momentum at small offsets, but \voversig\ values are offset and highly scattered.
We examine the role of regular and irregular morphology in the stellar mass kinematic scaling relations, deriving the  kinematic measurement \sof, and finding $\log(S_{0.5}) = (0.38\pm0.07)\log(M/M_{\odot}-10) + (2.04\pm0.03)$ with no significant offset between morphological populations and similar levels of scatter ($\sim0.16$ dex).
Additionally, we identify a correlation between \Mstar and \voversig\ for the total sample, showing an increasing level of rotation dominance with increasing \Mstar, and a high level of scatter for both regular and irregular galaxies.
We estimate the specific angular momenta (\jdisk) of these galaxies and find a slope of $0.36\pm0.12$, shallower than predicted without mass-dependent disk growth, but this result is possibly due to measurement uncertainty at \Mstar $<$ 9.5.
However, through a K-S test we find irregular galaxies to have marginally higher \jdisk\ values than regular galaxies, and high scatter at low masses in both populations.

\end{abstract}


\section{Introduction} \label{sec:intro}

The $\Lambda$CDM model predicts galaxies build their angular momentum through tidal interactions until the dark matter halo virializes \citep{White1978a,Fall1980, Mo1997}.
Dark matter-dominated gravitational potentials accrete primordial gas, which collapses into galaxy disks. 
The angular momentum of the baryonic disk of a galaxy has been shown to correlate with the angular momentum of the dark matter halo in the overall population of star-forming galaxies (SFGs), and is therefore a fundamental indicator of the total (baryonic and dark matter) growth of galaxies \citep{Emsellem2007a, Romanowsky2012, Obreschkow2014, Cortese2016}. 

As the baryonic matter collapses to form a disk, angular momentum will be subject to change due to gas accretion or merging events \citep{Vitvitska2002, Lagos2017, Penoyre2017}. 
In the case of cold gas accretion, as matter accretes onto the gravitational potential, a torque on the galaxy can be exerted and the angular momentum increases with time \citep{White1984a, Keres2004, Sales2012, Stewart2013, Danovich2015}. 
In the case of minor or major mergers, the angular momentum can increase or decrease based on the geometry of the merger itself \citep{Vitvitska2002, Puech2007, Naab2014, Rodriguez-Gomez2017}.
However in a number of cases, both observed and simulated, galaxies with clear signs of disrupted morphology show coherent rotation \citep{Hung2015, Turner2017a, Rodriguez-Gomez2017}.
This could be caused by a merger that is at the correct orientation to increase the angular momentum of the system.
If major mergers are a significant part of galaxy evolution, then we should see a large scatter in angular momentum relations.

The mass - angular momentum plane can be mapped to the Fundamental Plane for spiral galaxies \citep{Obreschkow2014}, and the projection of this plane forms the Tully-Fisher Relation \citep[TFR,][]{Tully1977}. 
However, high gas masses drive fundamental differences between local and high-redshift galaxies, most notably by increasing the star-formation rate (SFR), the increasing thickness of disks, the formation of large star-forming clumps, and the increased contribution of the gas velocity dispersion (\sigmag) to the total kinematics of SFGs \citep{Tacconi2010, Daddi2010, Obreschkow2016}.
The increase in \sigmag\ could also be affected by cold-mode accretion or merging events, which could cause disk instabilities or loss of angular momentum \citep{Hung2015}.
\citet{Kassin2007} accounted for the increased scatter of the TFR by including \sigmag\ in the kinematic quantity \sof. 
The scatter of the \sof-\Mstar\ relation is smaller than the scatter of the stellar - mass TFR at all redshifts.
\voversig\ is also used in multi-object slit spectroscopic surveys to quantify the rotation support against random motions \citep{Price2015, Simons2017}. 
However, significant scatter still remains in the TFR, \sof, and \voversig\ spaces explored by recent high-redshift surveys.
Median values of these datasets demonstrate the decrease of \sigmag\ and increase of \vrot\ with time and stellar mass, possibly indicating kinematic downsizing and the formation of disky SFGs \citep{Kassin2007, Simons2016a, Simons2017}.

In this work, we investigate the relationship between irregular morphology and kinematics.
Due to the availablility high-resolution photometry by the Hubble Space Telescope (HST), we can examine the morphologies of galaxies at \zsimtwo, in conjunction with the kinematic signatures provided by Keck/MOSFIRE \citep{McLean2012}.
This will provide morphological signatures of recent merging events and irregular structure for our sample, which will allow us to determine if these morphologies are correlated with any kinematic effects such as increased \sigmag, or an increased scatter in kinematic scaling relations in possible merging events.

These processes have been explored extensively and with great spatial precision in IFU surveys \citep{Epinat2009a, Law2009, Schreiber2009, Swinbank2012, Wisnioski2015} \citep[for a thorough review of these surveys, see][]{Glazebrook2013}.
However, since IFU data requires light from a source to be separated into different spaxels rather than integrated into a single slit, low-mass (log(\Mstar/$M_{\odot}$) $<$ 10.5) and faint galaxies are not well-represented by these data \citep{Wisnioski2015, Burkert2016}.
Additionally, these surveys also tend to exclude morphologically complex galaxies and galaxies with misaligned kinematic and morphological position angles ($PA$), as well as galaxies with \voversig $<2$.

In contrast, surveys utilizing slit spectroscopy are more sensitive to low-mass and faint galaxies.
Multi-object slit surveys demonstrate that the low-mass population is sensitive to the processes which affect angular momentum \citep{Simons2016a}. These processes include star-formation feedback, disk instabilities caused by rapid accretion of surrounding gas, or mergers. This population is often more dispersion-supported and irregularly shaped than the higher mass population at \zsimtwo.
These low-mass objects can provide evidence for which processes shape galaxy evolution at the peak of cosmic star-formation history.
In addition,slit surveys can measure larger data sets, over a variety of properties such as mass, luminosity, and environment.
Here, we attempt to bridge the gap between IFU and slit surveys. 
To investigate the effects of slit against IFU spectroscopy, we simulate IFU data cubes, and project them through a slit to create a slit observation of an emission line. 

Our data consist of objects from the COSMOS field \citep{Capak2007} measured by the \zfire\ survey \citep{Nanayakkara2016}, including a $z=2.095$ confirmed over-dense region in the COSMOS field \citep{Spitler2012,Yuan2014}. 
\zfire\footnote[1]{zfire.swinburne.edu.au} targets galaxy clusters at $z\sim2$ to explore galaxy evolution as a function of environment.  
\zfire\ combines deep multi-wavelength imaging with spectroscopy obtained from MOSFIRE to measure galaxy properties including sizes, stellar masses, star formation rates, gas-phase metallicities, and the interstellar medium \citep{Kacprzak2015,Kewley2015,Tran2015,Kacprzak2016, Alcorn2016, Nanayakkara2016, Tran2016, Straatman2017, Nanayakkara2017}.

In this work, we assume a flat $\Lambda$CDM cosmology with $\Omega_{M}$=0.3, $\Omega_{\Lambda}$ =0.7, and H$_{0}$=70. 
At the cluster redshift, $z = 2.095$, one arcsecond corresponds to an angular scale of 8.33 kpc.

\section{Data}

\begin{figure*}[h]
   \centering
 \begin{tabular}{l}
 \includegraphics[width=\textwidth]{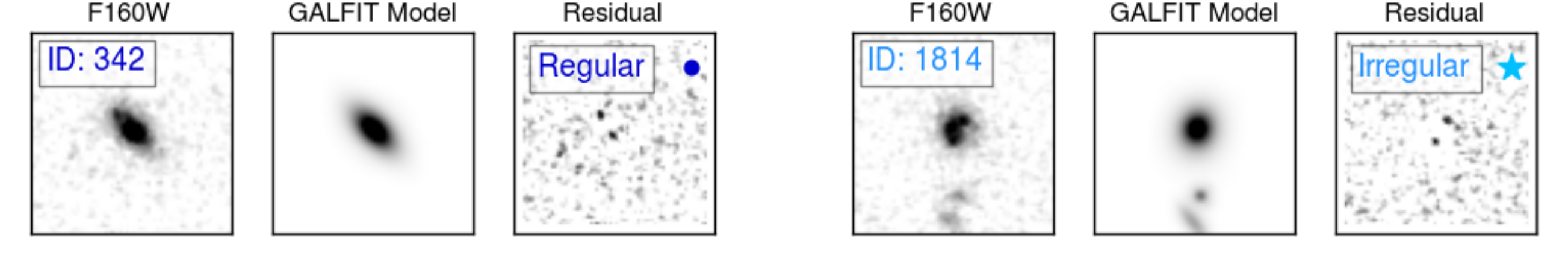}\\
 \includegraphics[width=\textwidth]{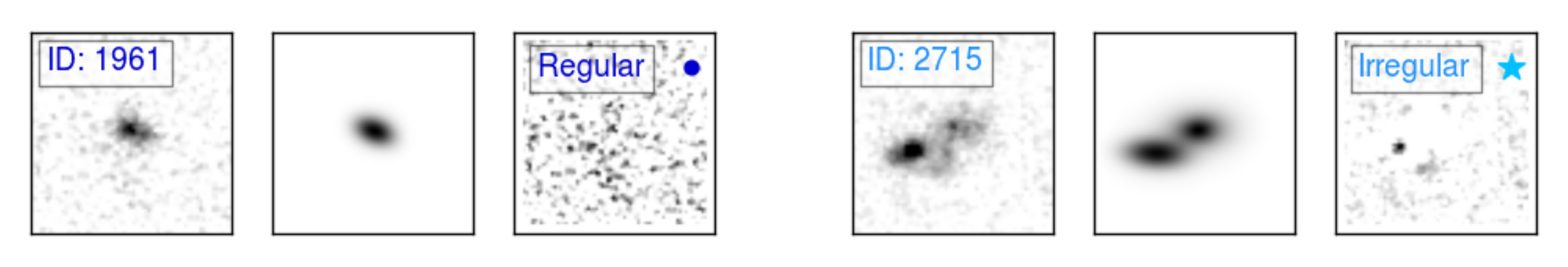}\\
 \includegraphics[width=\textwidth]{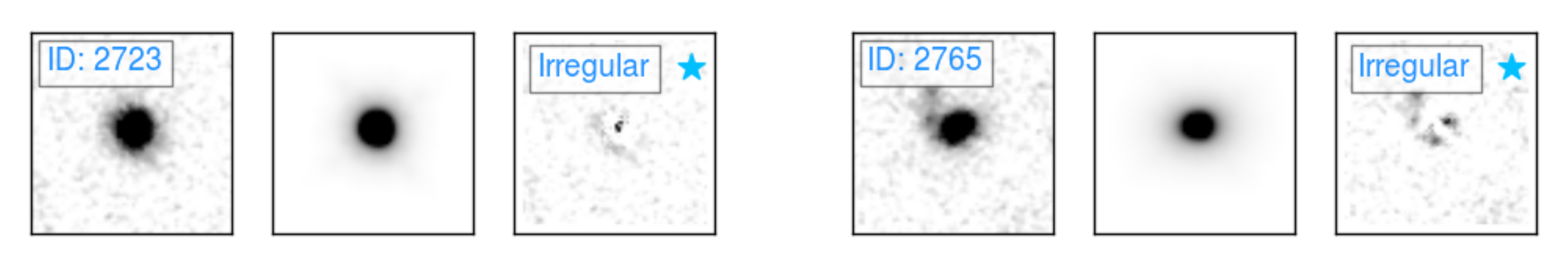}\\
 \includegraphics[width=\textwidth]{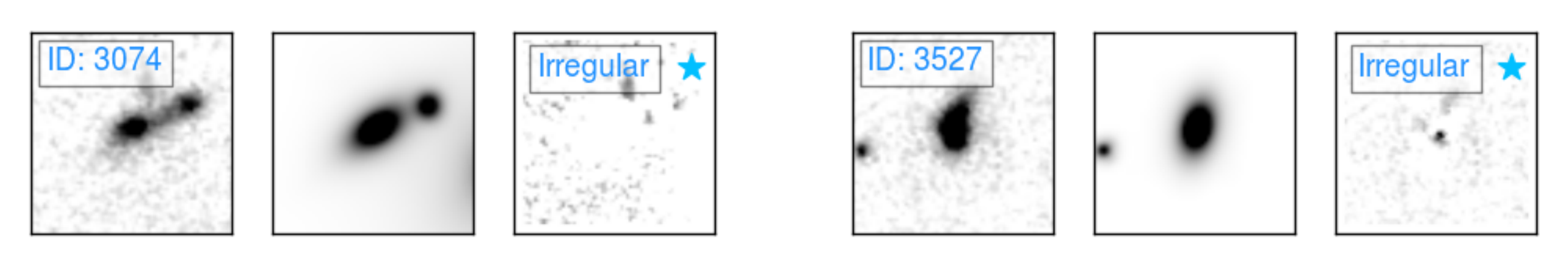}\\
   \includegraphics[width=\textwidth]{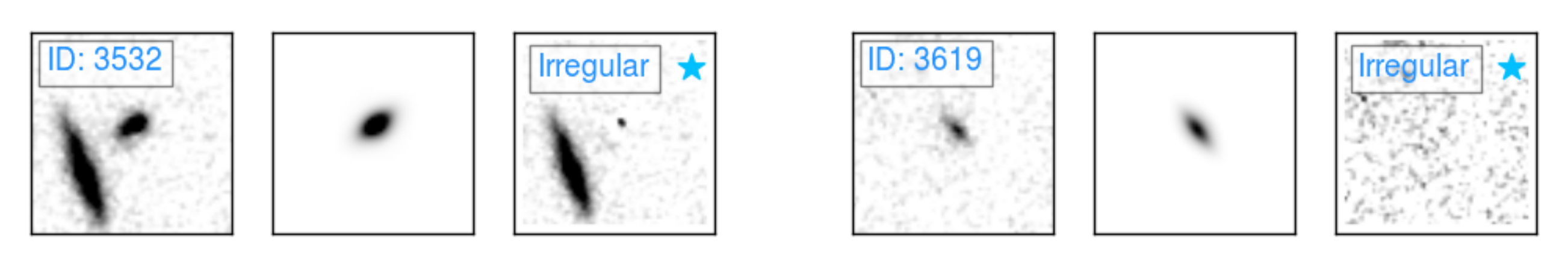}\\
       \end{tabular}
     \caption{Imaging of our sample. Two galaxies are shown per row. From left for each galaxy: The F160W imaging from CANDELS/3D-HST. Center: Best-fit GALFIT model, and if the galaxy is considered ``compact'', it is noted. Right: Residual of the fit from the data. The residual is used to determine whether an object is regular or irregularly-shaped, and its classification is noted in this panel. Regular galaxies are in dark blue, and are plotted as dark blue circles in the text. Irregular galaxies are in light blue, and are plotted as light blue stars in the text. Compact galaxies of either classification are unfilled circles or stars.}
\end{figure*}
\begin{figure*}[h]
   \centering
  \addtocounter{figure}{-1}
 \begin{tabular}{l}

 \includegraphics[width=\textwidth]{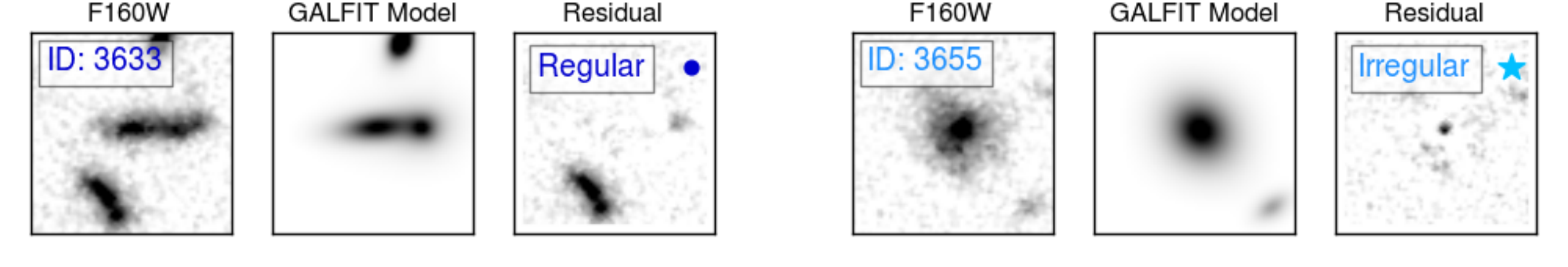}\\
 \includegraphics[width=\textwidth]{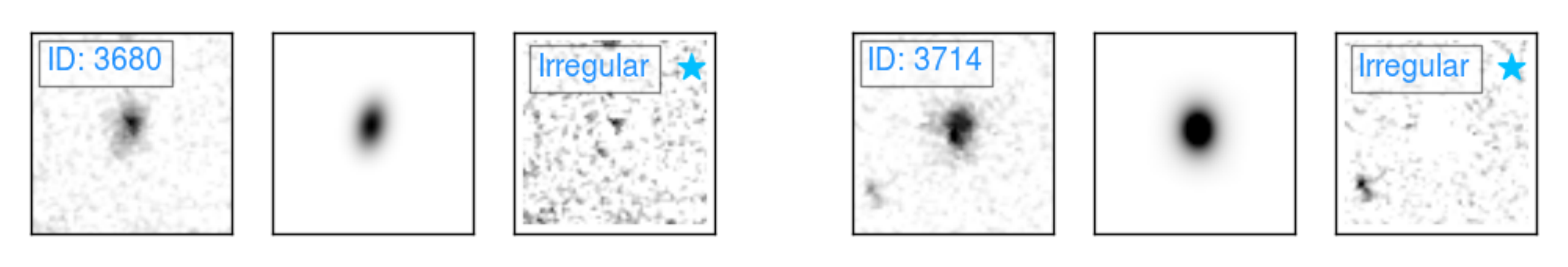}\\
 \includegraphics[width=\textwidth]{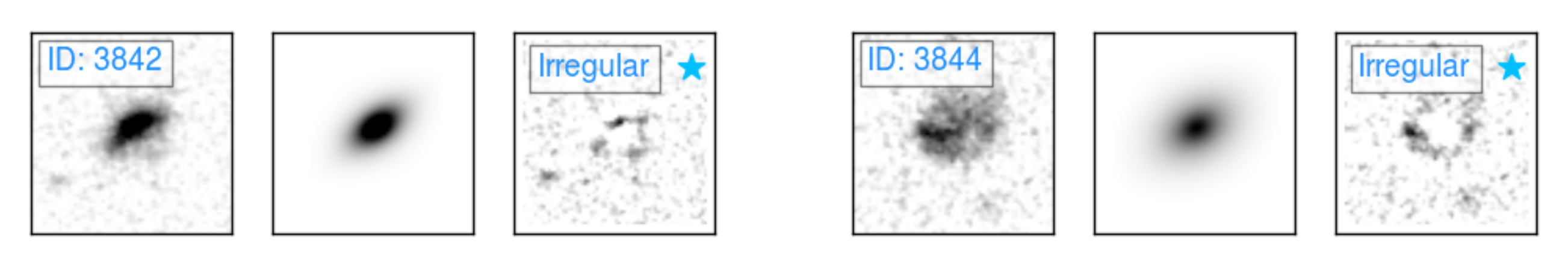}\\
 \includegraphics[width=\textwidth]{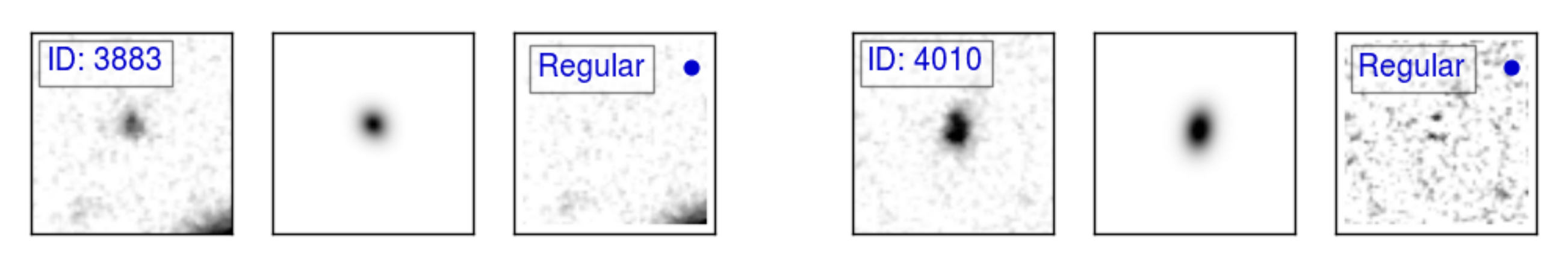}\\
  \includegraphics[width=\textwidth]{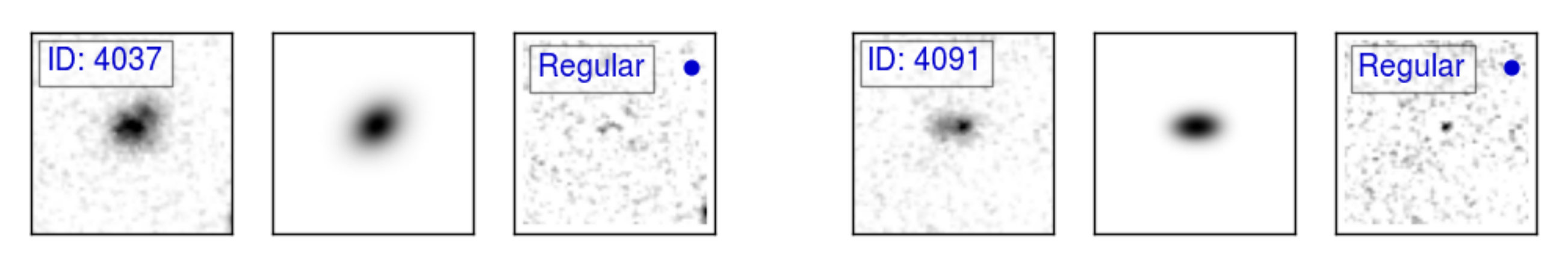}\\
 \includegraphics[width=\textwidth]{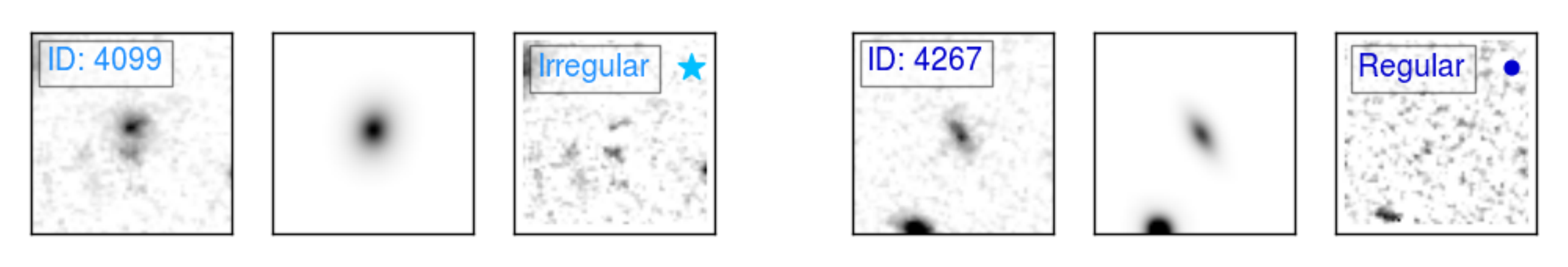}\\
 \includegraphics[width=\textwidth]{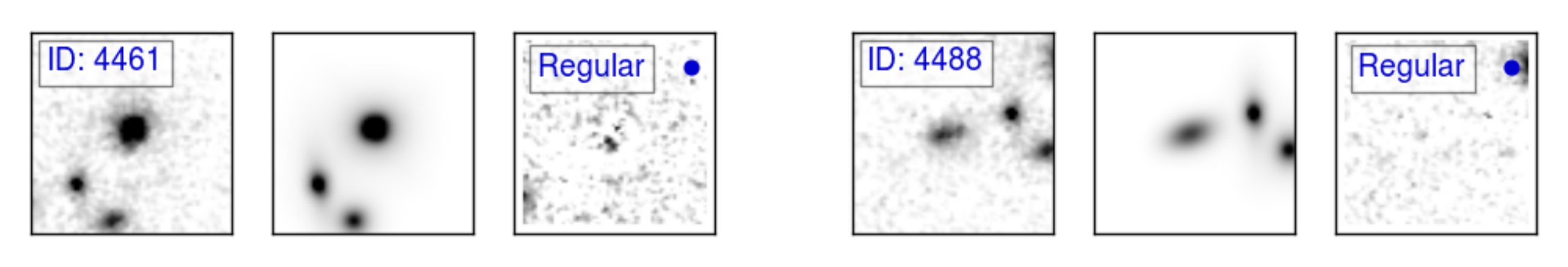}\\
       \end{tabular}
     \caption{Continued}
\end{figure*}
\begin{figure*}[h]
   \centering
  \addtocounter{figure}{-1}
 \begin{tabular}{l}

 \includegraphics[width=\textwidth]{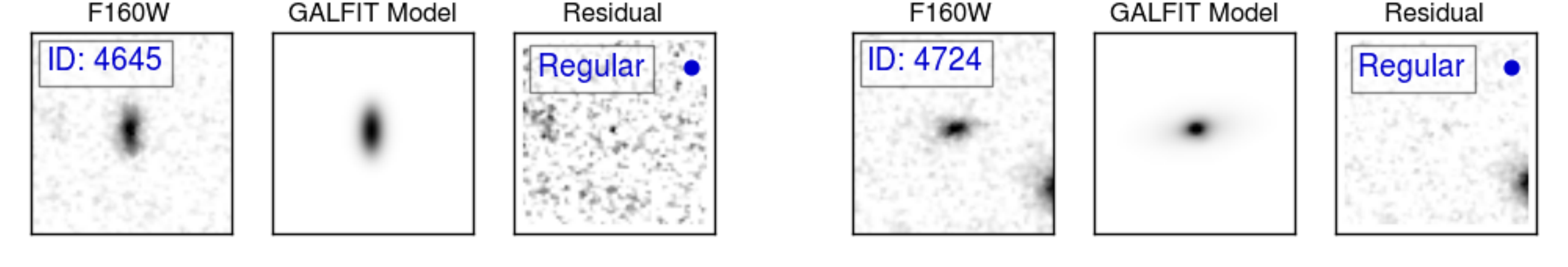}\\
  \includegraphics[width=\textwidth]{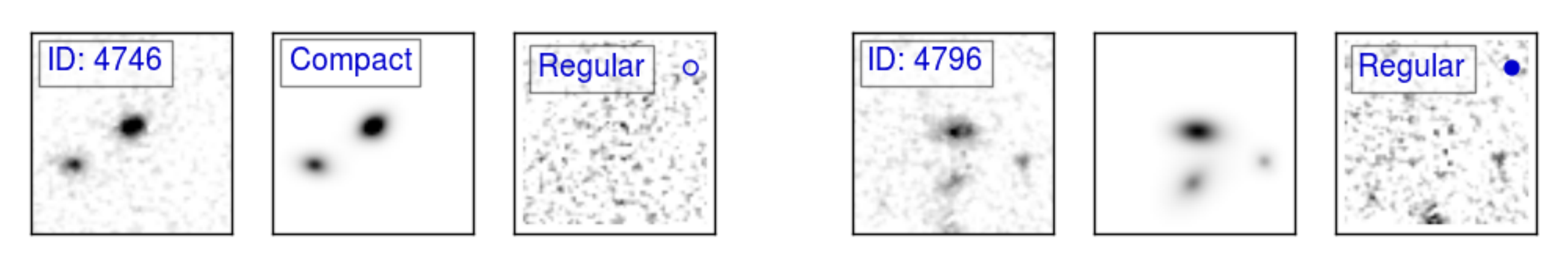}\\
 \includegraphics[width=\textwidth]{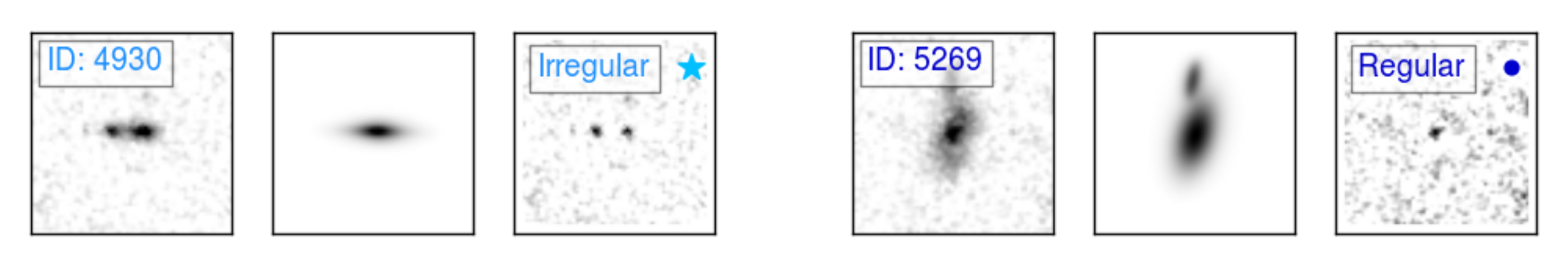}\\
 \includegraphics[width=\textwidth]{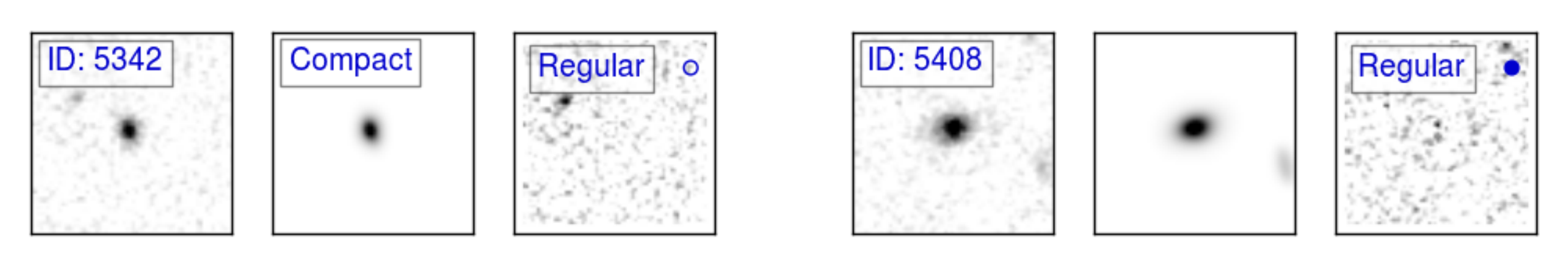}\\
 \includegraphics[width=\textwidth]{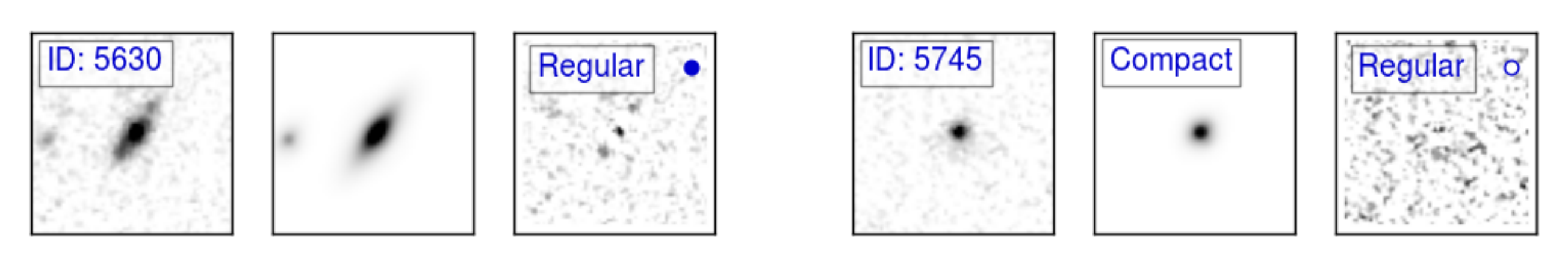}\\
 \includegraphics[width=\textwidth]{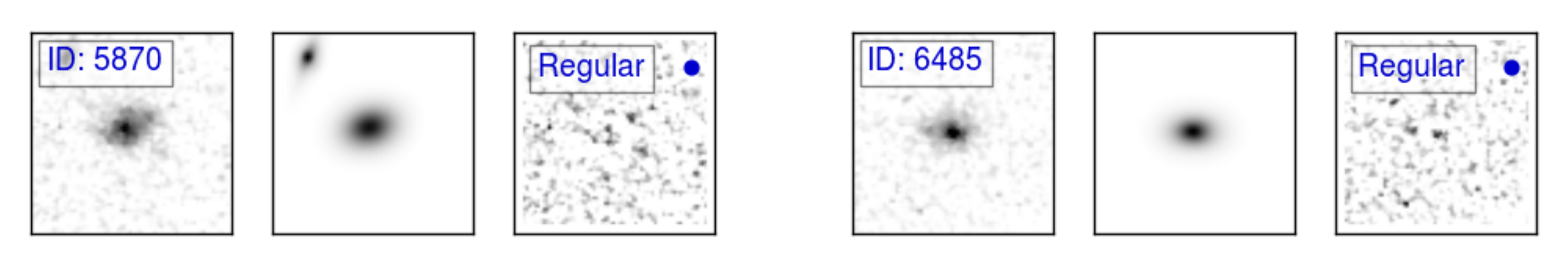}\\
  \includegraphics[width=\textwidth]{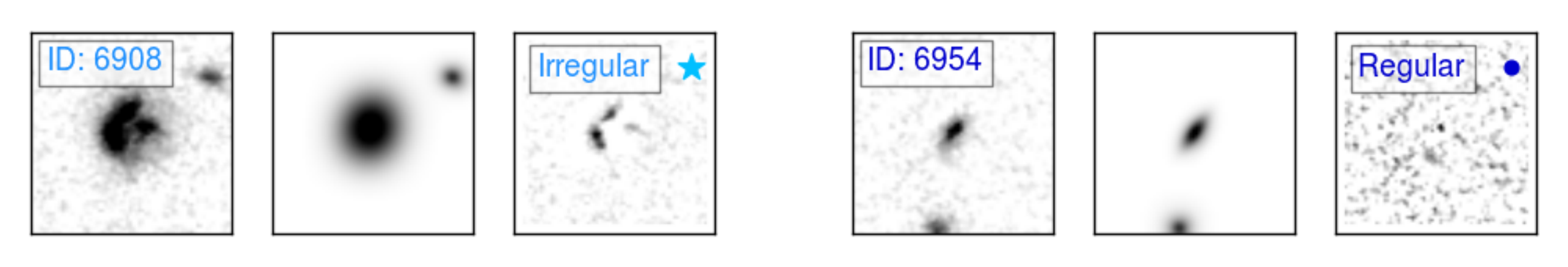}\\
       \end{tabular}
     \caption{Continued}
\end{figure*}
\begin{figure*}[h]
   \centering
  \addtocounter{figure}{-1}
 \begin{tabular}{l}

 \includegraphics[width=\textwidth]{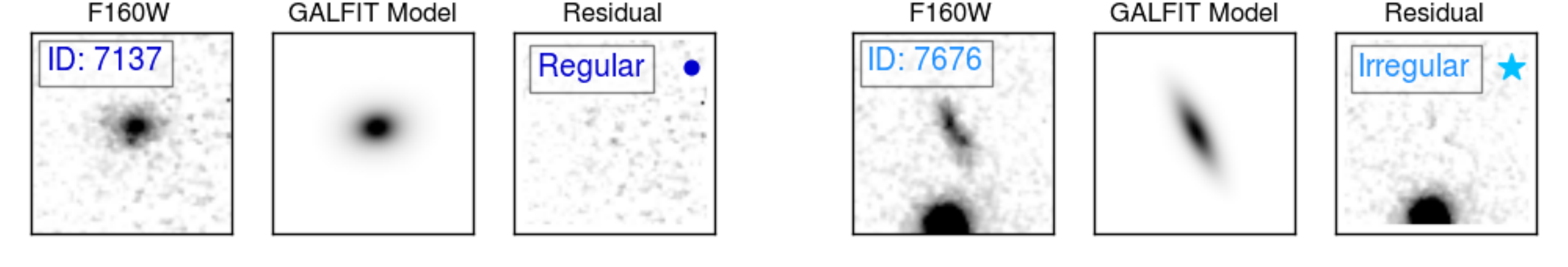}\\
  \includegraphics[width=\textwidth]{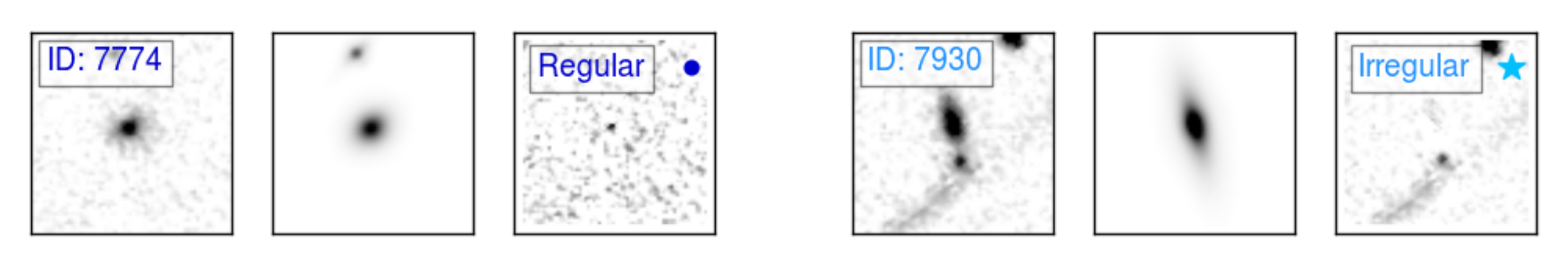}\\
 \includegraphics[width=\textwidth]{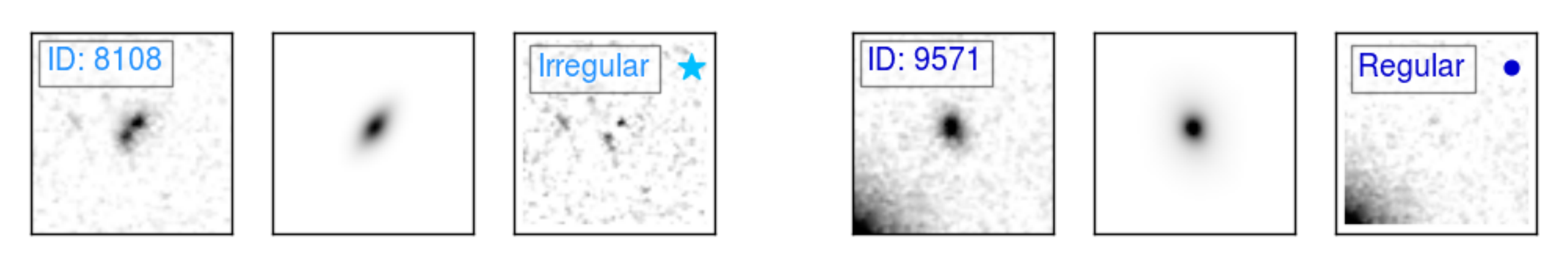}\\
       \end{tabular}
     \caption{Continued}
\end{figure*}

\subsection{Sample Selection} \label{sec:sample}

Our sample is drawn from the ZFIRE survey \citep{Nanayakkara2016}, a spectroscopic follow-up of ZFOURGE photometry \citep{Straatman2016}.
To summarize, we identify star-forming galaxies (SFGs) within a photometric redshift range of $1.7<z<2.5$ in ZFOURGE NIR imaging of COSMOS fields. 
ZFOURGE combines broad-band imaging in $K_s$ and the medium-band J$_{1}$, J$_{2}$, J$_{3}$, H$_{s}$, and H$_{l}$ filters to select objects using $K_s$-band images with a 5$\sigma$ limit of 25.3 AB magnitudes.
Rest-frame UVJ colors are used to identify SFGs, which will have prominent emission lines.
Objects with radio, infrared, ultraviolet, or x$-$ray indications of AGN activity \citep[identified via][]{Cowley2016} are rejected from this analysis.

The COSMOS protocluster was initially identified in \citet{Spitler2012} using photometric redshifts from ZFOURGE and subsequently confirmed with spectroscopic redshifts from MOSFIRE \citep{Yuan2014}.
This over-density consists of four merging groups, and is projected to evolve into a Virgo-like cluster at $z=0$.
Cluster members are identified to redshifts within $2.08<z<2.12$.

ZFOURGE uses FAST \citep{Kriek2009a} to fit \citet{Bruzual2003} stellar population synthesis models to the galaxy spectral energy distributions to estimate observed galaxy properties. 
After spectroscopic redshifts were obtained on MOSFIRE, objects were run in FAST using the spectroscopically confirmed redshifts rather than the photometric redshifts, providing our stellar masses and attenuation values ($A_V$).
We assume a \citet{Chabrier2003} initial mass function with constant solar metallicity and an exponentially declining star formation rate, and a \citet{Calzetti1999} dust law. 

\subsection{HST Imaging} \label{sec:imaging}

Our morphological measurements are from the Cosmic Assembly Near-Infrared Deep Extragalactic Survey \citep[][CANDELS]{Grogin2011, Koekemoer2011a} imaging processed by the 3D-HST team (v4.1 data release) \citet{Skelton2014}.
Our PSF is also constructed by the 3D-HST team.
We use GALFIT software \citep{Peng2010} to measure galaxy sizes from the F160W imaging.
At \zsimtwo, F160W corresponds to rest-frame $g$-band.
Our morphological fitting is summarized in \citet{Alcorn2016} but we briefly repeat here.

We generate a custom pipeline to fit the 161 COSMOS galaxies in ZFIRE with F160W imaging using initial measurements of size, axis ratio ($q$), position angle (PA), and magnitude from SExtractor.
Objects within 2$\arcsec$ of a target galaxy are simultaneously fit with the central object. 
Residual images are visually inspected to determine the best possible fits for each galaxy. 
Galaxies with poor residuals are re-fit using a modified set of initial parameters. 
Galaxies were restricted to \sersic\ indices ($n$) between $0.2-8.0$.
If objects iterated to the boundaries of our \sersic\ constraints, they were refit with a fixed \sersic\ index ($n=1.0$ for objects which went to $n=0.2$, and $n=4.0$ for objects which went to $n=8.0$)
Our results are consistent within 2$\sigma$ to \citet{VanDerWel2014} (see Table 1).

\begin{deluxetable*}{lllrrrr}
\tabletypesize{\scriptsize} 
\tablecolumns{7} 
\tablewidth{0pt} 
\tablecaption{Morphological measurements from F160W imaging.} 
\tablehead{ 
\colhead{ID} &\colhead{Cluster/Field} & \colhead {Regular/Irregular} & \colhead{R$_e$ (arcseconds)} & \colhead{Sersic Index} & \colhead{Axis Ratio} & \colhead{PA}} 
\startdata 
1814 & Field & Irregular & 0.29$\pm$0.01 & 1.0$\pm$0.0 & 0.8$\pm$0.0 & -11.6$\pm$3.7 \\
1961 & Field & Regular & 0.28$\pm$0.01 & 0.4$\pm$0.1 & 0.6$\pm$0.0 & 68.6$\pm$2.3 \\
2715 & Cluster & Irregular & 0.46$\pm$0.01 & 0.9$\pm$0.1 & 0.6$\pm$0.0 & -87.4$\pm$1.3 \\
2723 & Cluster & Irregular & 0.13$\pm$0.11 & 2.6$\pm$5.4 & 0.9$\pm$0.9 & 20.7$\pm$32.7 \\
2765 & Field & Irregular & 0.34$\pm$0.01 & 4.0$\pm$0.0 & 0.7$\pm$0.0 & -87.8$\pm$2.0 \\
3074 & Field & Irregular & 0.46$\pm$0.01 & 1.0$\pm$0.0 & 0.5$\pm$0.0 & -55.7$\pm$0.8 \\
342 & Field & Regular & 0.38$\pm$0.01 & 0.8$\pm$0.0 & 0.5$\pm$0.0 & 44.7$\pm$0.6 \\
3527 & Field & Irregular & 0.38$\pm$0.01 & 0.9$\pm$0.0 & 0.5$\pm$0.0 & -12.8$\pm$0.5 \\
3532 & Cluster & Irregular & 0.20$\pm$0.01 & 0.9$\pm$0.1 & 0.4$\pm$0.0 & -54.4$\pm$0.9 \\
3619 & Field & Irregular & 0.25$\pm$0.01 & 0.7$\pm$0.2 & 0.2$\pm$0.0 & 37.1$\pm$1.3 \\
3633 & Cluster & Regular & 0.59$\pm$0.01 & 0.8$\pm$0.1 & 0.3$\pm$0.0 & -85.1$\pm$0.6 \\
3655 & Field & Irregular & 0.54$\pm$0.01 & 0.7$\pm$0.0 & 0.9$\pm$0.0 & 44.5$\pm$2.6 \\
3680 & Field & Irregular & 0.34$\pm$0.01 & 0.6$\pm$0.1 & 0.5$\pm$0.0 & -11.6$\pm$1.6 \\
3714 & Field & Irregular & 0.32$\pm$0.01 & 0.9$\pm$0.0 & 0.7$\pm$0.0 & 1.3$\pm$0.2 \\
3842 & Cluster & Irregular & 0.43$\pm$0.01 & 0.9$\pm$0.0 & 0.5$\pm$0.0 & -54.9$\pm$0.6 \\
3844 & Field & Irregular & 0.66$\pm$0.02 & 1.0$\pm$0.0 & 0.7$\pm$0.0 & -60.8$\pm$1.8 \\
3883 & Field & Regular & 0.19$\pm$0.01 & 0.9$\pm$0.2 & 0.8$\pm$0.1 & 29.3$\pm$9.3 \\
4010 & Field & Regular & 0.29$\pm$0.01 & 0.6$\pm$0.1 & 0.6$\pm$0.0 & -8.7$\pm$1.0 \\
4037 & Field & Regular & 0.38$\pm$0.01 & 0.6$\pm$0.0 & 0.7$\pm$0.0 & -52.9$\pm$1.7 \\
4091 & Cluster & Regular & 0.33$\pm$0.01 & 0.3$\pm$0.1 & 0.5$\pm$0.0 & -88.5$\pm$0.5 \\
4099 & Field & Irregular & 0.38$\pm$0.01 & 1.2$\pm$0.1 & 0.8$\pm$0.0 & -11.1$\pm$3.8 \\
4267 & Field & Regular & 0.30$\pm$0.01 & 1.0$\pm$0.0 & 0.3$\pm$0.0 & 29.4$\pm$1.3 \\
4461 & Field & Regular & 0.30$\pm$0.01 & 4.0$\pm$0.0 & 0.9$\pm$0.1 & -83.6$\pm$1.5 \\
4488 & Field & Regular & 0.35$\pm$0.01 & 0.6$\pm$0.1 & 0.5$\pm$0.0 & -71.3$\pm$1.2 \\
4645 & Cluster & Regular & 0.33$\pm$0.01 & 0.4$\pm$0.1 & 0.3$\pm$0.0 & -0.6$\pm$0.9 \\
4724 & Field & Regular & 0.68$\pm$0.22 & 8.0$\pm$2.0 & 0.3$\pm$0.0 & -82.4$\pm$1.5 \\
4746 & Field & Regular & 0.14$\pm$0.01 & 0.9$\pm$0.1 & 0.5$\pm$0.0 & -59.2$\pm$2.7 \\
4796 & Field & Regular & 0.29$\pm$0.01 & 0.8$\pm$0.1 & 0.4$\pm$0.0 & 85.8$\pm$1.7 \\
4930 & Cluster & Irregular & 0.39$\pm$0.01 & 1.0$\pm$0.0 & 0.1$\pm$0.0 & 88.6$\pm$0.5 \\
5269 & Cluster & Regular & 0.54$\pm$0.01 & 0.5$\pm$0.0 & 0.5$\pm$0.0 & -15.9$\pm$0.8 \\
5342 & Field & Regular & 0.14$\pm$0.01 & 1.0$\pm$0.2 & 0.4$\pm$0.0 & 10.3$\pm$2.3 \\
5408 & Cluster & Regular & 0.24$\pm$0.01 & 1.0$\pm$0.1 & 0.6$\pm$0.0 & -76.9$\pm$2.1 \\
5630 & Field & Regular & 0.38$\pm$0.01 & 1.4$\pm$0.1 & 0.3$\pm$0.0 & -34.8$\pm$0.5 \\
5745 & Cluster & Regular & 0.10$\pm$0.01 & 2.7$\pm$0.6 & 0.8$\pm$0.1 & -37.1$\pm$12.1 \\
5870 & Cluster & Regular & 0.38$\pm$0.01 & 0.7$\pm$0.0 & 0.7$\pm$0.0 & -75.8$\pm$2.1 \\
6485 & Field & Regular & 0.33$\pm$0.01 & 1.1$\pm$0.1 & 0.6$\pm$0.0 & 89.5$\pm$0.5 \\
6908 & Field & Irregular & 0.51$\pm$0.01 & 0.5$\pm$0.0 & 0.9$\pm$0.0 & -15.2$\pm$2.1 \\
6954 & Field & Regular & 0.24$\pm$0.01 & 0.6$\pm$0.1 & 0.3$\pm$0.0 & -34.9$\pm$1.0 \\
7137 & Field & Regular & 0.36$\pm$0.01 & 1.1$\pm$0.1 & 0.7$\pm$0.0 & -83.6$\pm$1.7 \\
7676 & Field & Irregular & 0.54$\pm$0.01 & 0.7$\pm$0.1 & 0.2$\pm$0.0 & 26.4$\pm$0.5 \\
7774 & Field & Regular & 0.24$\pm$0.01 & 1.2$\pm$0.2 & 0.8$\pm$0.1 & -52.1$\pm$10.4 \\
7930 & Cluster & Irregular & 0.53$\pm$0.03 & 2.5$\pm$0.2 & 0.2$\pm$0.0 & 13.1$\pm$0.5 \\
8108 & Field & Irregular & 0.29$\pm$0.01 & 1.0$\pm$0.0 & 0.4$\pm$0.0 & -34.8$\pm$1.2 \\
9571 & Cluster & Regular & 0.48$\pm$0.03 & 4.0$\pm$0.0 & 0.6$\pm$0.0 & 11.0$\pm$2.9 \\
\enddata 
\vspace{-0.8cm} 
\end{deluxetable*}

\disknum\ objects in our final sample are considered to be regular galaxies by evaluation of GALFIT residuals. 
Examples of our sample showing regular and irregular galaxies by our criteria are shown in Figure 1.
To determine the presence of irregular morphology or tidal features, we examine residual images.
Using segmentation maps from SExtractor, we isolate the individual galaxies and measure the residual, the sky flux, and the flux of the original object.
If residual levels are at more than 2 times the level of the sky, and more than 25\% of the flux of the original object remains, we determine the presence of significant artifacts.
If residual images show significant artifacts, which indicate that a \sersic\ profile is a poor or unreliable fit to the object, they are flagged as irregulars, although this population could include both irregulars and merging objects.
Conversely, regulars show no significant residuals (residual levels are less than 2 times sky levels and less than 25\% the flux levels of the object) when fit with a \sersic\ profile.
These values were determined empirically, although small changes do not significantly change our results.

In both cases, the presence of close companions was neglected in the absence of strong residuals, as we cannot spectroscopically confirm the redshifts of nearby objects.
This method is possibly biased toward classifying smaller galaxies ($<0.3\arcsec$) as regular galaxies, because residual values are only measured in areas identified as being associated with the original object.
Additionally, objects that are photometrically irregular may be kinematically regular, such as clumpy disks, and may not be distinct from regular galaxies apart from their photometry.
When comparing our populations through a two-population KS test, we find a similar distribution of stellar masses from $9.0\leq$ log(\Mstar) $\leq11.0$ and \sersic\ index from $0.2<n<8.0$. See Figure 2.

We include a category of  ``compactness'' in our final sample, where objects with an effective radius $r_e$ smaller than the HST F160W PSF FWHM ($r_e < 0.19 \arcsec$, or 1.58 kpc at $z=2.095$) \citep{Skelton2014} are compact.
These objects are marked as unfilled points in our figures and are morphologically unresolved.
From \citet{VanDerWel2014} the median size of late-type galaxies at \zsimtwo\ in our $M_{\ast}$ range is 2-4 kpc, thus we are confident that our adopted compactness threshold of 1.58 kpc is appropriate.
This is in contrast to objects that are kinematically unresolved, where their diameter is less than the seeing limit (See Table 2). 21 galaxies in this sample are kinematically unresolved.
The velocity of these unresolved sources is often underestimated \citep{Newman2012}, but we include compact objects with reliable velocity measurements (Section 3.1).

\begin{figure*}[t]
\figurenum{2}
    \centering
        \includegraphics[width=0.8\textwidth]{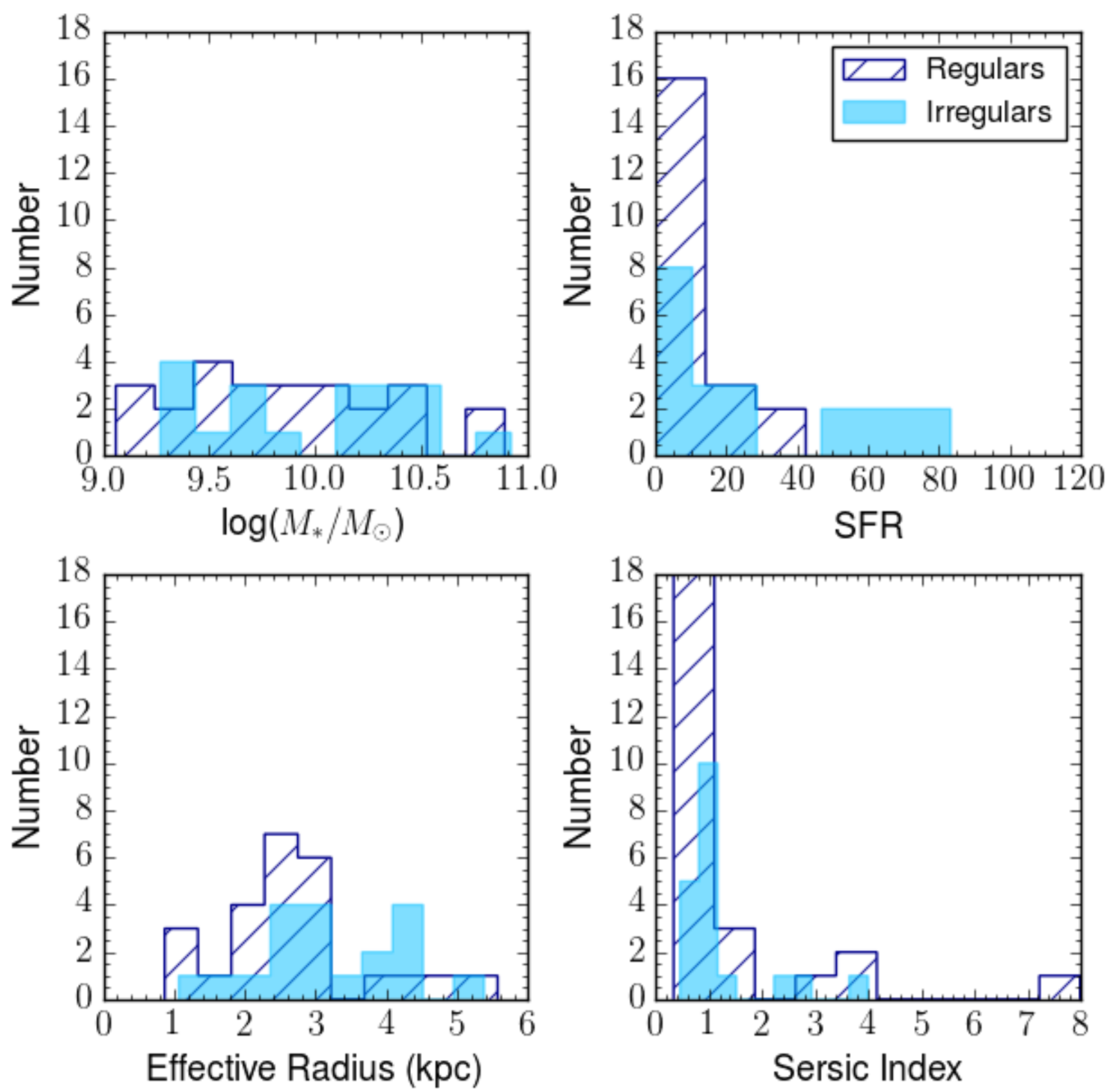}
    \caption{Histograms of our galaxy populations. Light blue solid bins are irregular galaxies, and dark blue hatched bins are regular galaxies. By applying a two-population KS test, we find similar properties in both populations, although irregulars are marginally more likely to have higher star-formation rates.}\label{fig:histograms}
\end{figure*}

\subsection{MOSFIRE NIR Spectroscopy} \label{sec:spectra}

Observations were taken in December 2013 and February 2014 in the K-band filter covering 1.93-2.45 $\mu$m, the wavelength range we would expect to see \halpha\ and \NII at the cluster redshift.
Seeing varied from  $\sim0.4\arcsec$ to  $\sim1.3\arcsec$ over the course of our observations.

The spectra are flat-fielded, wavelength calibrated, and sky subtracted using the MOSFIRE data reduction pipeline (DRP)\footnote[2]{http://keck-datareductionpipelines.github.io/MosfireDRP/}. 
A custom ZFIRE pipeline corrected for telluric absorption and performed a spectrophotometric flux calibration using a type A0V standard star.
We flux calibrate our objects to the continuum of the standard star, and use ZFOURGE photometry as an anchor to correct offsets between photometric and spectroscopic magnitudes. 
The final result of the DRP are flux-calibrated 2D spectra and 2D 1$\sigma$ images used for error analysis. 
For more information on ZFIRE spectroscopic data reduction and spectrophotometric calibrations, see \citet{Nanayakkara2016}.
1D spectra and catalogs are available to the public on the \zfire\ website.

From spectroscopic observations, we reject objects with only one identified emission line, without morphological measurements, or with AGN signatures \citep{Cowley2016}, leaving 92 SFGs with K band spectroscopy.

\subsection{PSF Fitting} \label{sec:psffitting}
The assumed PSF for an observation plays a role in the recovery of accurate velocities, as the mischaracterization of the shape of the PSF can result in an underestimation of the velocity.
In most cases, a Gaussian PSF with a FWHM given by seeing conditions is convolved with the emission-line fit, but in recent work it has been shown that on the MOSFIRE instrument, a Moffat profile is a better fit to the PSF \citep{Straatman2017}.
Therefore we fit and apply Moffat PSFs to all objects in our sample.

To determine our PSF, we create a 2D Moffat-profile simulated star.
We collapse this star into a flat spectral profile and sum along the wavelength component to estimate the spatial 1D profile of the star, and subtract the profile on either side of the peak at the positions of our dithering pattern (1.25$\arcsec$) to correctly account for any effect of the dither pattern on the wings of the PSF.
Then for each observed mask, we sum along the wavelength plane to determine the spatial profile of our flux monitor star.
We leave the Moffat parameters $\alpha$ and $\beta$ free and fit the Moffat profile given as 
\begin{equation} \label{eq:inclination}
PSF(r)=\frac{\beta-1}{\pi \alpha^2}\left[1+\left(\frac{r}{\alpha}\right)^2\right]^{-\beta},
\end{equation}

to our observed flux monitor stars, and use the best-fit values for the Moffat parameters to apply to our Moffat convolution kernel when we fit our emission lines.
If the wings of the best-fit Moffat profile appear to over-fit the observed star, we fix $\beta=2.5$ and refit to find $\alpha$.
The best fit Moffat parameters used to generate our emission line models can be seen in Table 2.

\begin{deluxetable*}{llrrrr}
\tabletypesize{\scriptsize} 
\tablecolumns{7} 
\tablewidth{0pt} 
\tablecaption{Mask properties and best-fit Moffat parameters.} 
\tablehead{ 
\colhead{Date} & \colhead{Mask} &\colhead{Average Seeing ($\arcsec$)} & \colhead {$\alpha$} & \colhead{$\beta$} & \colhead{Slit $PA$ ($^o$)\tablenotemark{a}}} 
\startdata 
Dec 2013 & Shallowmask1 & 0.7 & 0.601 & 2.487 & 134 \\
Dec 2013 & Shallowmask2 & 0.68 & 0.581 & 2.5 & -47.3 \\
Dec 2013 & Shallowmask3 & 0.7 & 0.674 & 2.778 & 14.8 \\
Dec 2013 & Shallowmask4 & 0.67 & 0.516 & 2.574 & -63 \\
Feb 2014 & DeepKband1 & 1.27 & 1.031 & 2.5 & 2 \\
Feb 2014 & DeepKband2 & 0.7 & 0.656 & 2.599 & -62 \\
Feb 2014 & KbandLargeArea3 &1.1 & 1.021 & 2.5 & 59 \\
Feb 2014 & KbandLargeArea4 & 0.66 & 0.489 & 2.525 & 2
\enddata 
\tablenotetext{a}{$PA$ is defined as east of north.}
\end{deluxetable*}

\section{Methods}
\subsection{Spectroscopic Fitting Method}
Our fitting procedure for our sample and our simulated observations are based around HELA (Heidelberg Emission-Line Algorithm), which was developed by C.M. Straatman \citep{Straatman2018}.
Information on the models generated by HELA is located in the Appendix.

We emphasize that there are many ways to refer to the velocity of a galaxy.
In this text, we refer to velocity in three main ways.
$V_{rot}(r)$ is the rotational velocity at a given radius of a galaxy, referred to as simply the rotational velocity in this text.
This is in contrast to $V_t$, which is the asymptotic velocity (at the flat part of the rotation curve). 
Additionally we use \vtwo, which is the velocity at 2.2$r_s$, where the rotation curve of an ideal disk peaks \citep{Freeman1970a}, and is used widely in literature as a common reference point for velocity \citep{Miller2011}.

To determine best-fit parameters for our emission line, our procedure is thus:
\begin{enumerate}
\item Identify the position of the \halpha\ emission line. Subtract continuum values if present (see Section 3.3).
\item Mask wavelengths which are strongly contaminated by sky emission in the observed spectra, or which are bad pixels.
\item Determine fitting bounds: -600 \kms $< V_t<$ 600 \kms, 10 \kms $< \sigma_g <$ 150 \kms, 0.1$\arcsec$ $< r_s <$ 1$\arcsec$, and 0.03$\arcsec$ $<r_t<$ $r_s$ (we also perform fitting where $r_t$ is fixed to $r_t=0.33r_s$ or $r_t=0.4r_s$). Position of the intensity peak cannot shift more than three pixels from given coordinates. These values and the intensity are all free parameters.
\item Run the simulated emission line through HELA (see Appendix) to derive best-fit parameters. We use a Markov-Chain Monte-Carlo analysis (MCMC) initializing 30 walkers over 1000 steps. Our walkers are initialized as a clump, values randomly distributed around the given wavelength and spatial position, and initial guess for $V_t$, \sigmag, $r_s=r_e/1.678$ (where $r_e$ is the effective radius measured from GALFIT), and $r_t=0.3r_s$, or $r_t$ fixed. We use the Python package emcee for our MCMC algorithm\footnote[3]{http://dan.iel.fm/emcee/current/} \citep{Foreman-Mackey2012a}.
\item Discard the first 200 iterations out of a total of 1000 - where the MCMC algorithm tends to be far from convergence. Our best-fit model is taken to be the median of the posterior likelihood output of all our free parameters after convergence, and errors are the 16th and 84th percentiles of the walkers. The value for \vtwo\ is determined by fitting the velocity curve function (Equation \ref{eq:vcurve}) to each walker and step, and then measuring the median value. 
\item In the case of multiple peaks in the posterior likelihood, we isolate one peak and fit a Gaussian to the largest peak to determine the best-fit values. Errors on the fit are determined from the $\sigma$ value on this Gaussian fit.
\end{enumerate}

We reject four compact galaxies with errors greater than $0.8V_{2.2}$ where \vtwo$>35$ km s$^{-1}$, which are considered unreliable. Six morphologically resolved galaxies with similar kinematics were kept in the sample and are shown as upper limits on the TFR (Figure 3).

\subsection{Fitting ZFIRE Data} \label{sec:datafitting}

Our fitting algorithm is applied to the 2D telluric and spectrophotometrically corrected emission lines.
Faint continua are seen in a small number of objects, so we subtract a flat continuum when one is detected.
Continuum subtraction is performed in the same method as \citet{Straatman2017}.
Summarized, for each row of pixels in a stamp 300 $\AA$ wide, we determine a median flux with outlier pixels $>2.5\sigma$ above the median rejected, and any sky or \halpha\ \NII\ emission masked.
This procedure is repeated a total of three times, then the median values are subtracted from each row.

The measured axis ratio from GALFIT is used to determine the inclination for use in our fitting procedure:
\begin{equation} \label{eq:inclination}
\sin i=\sqrt{\frac{1-q^2}{1-q_0^2}},
\end{equation}

where $q_0=0.19$ \citep{Miller2011}.
40 objects with galaxy PA-slit offset $\Delta\alpha>45^o$ or $\Delta\alpha<-45^o$, where PA is determined from GALFIT modeling, are rejected from the final sample, although objects with large PA uncertainties (mostly objects with low inclination or high $q$) that could overlap within this range are not rejected.
We also reject objects with significant sky emission (3 objects where more than 50\% of the line is masked, Appendix B.1) or where SNR $< 5$ (5 objects).

\section{Results}

\begin{figure*}[t] \label{fig:kinematics}
\centering
\figurenum{3}
\includegraphics[width=0.9\textwidth]{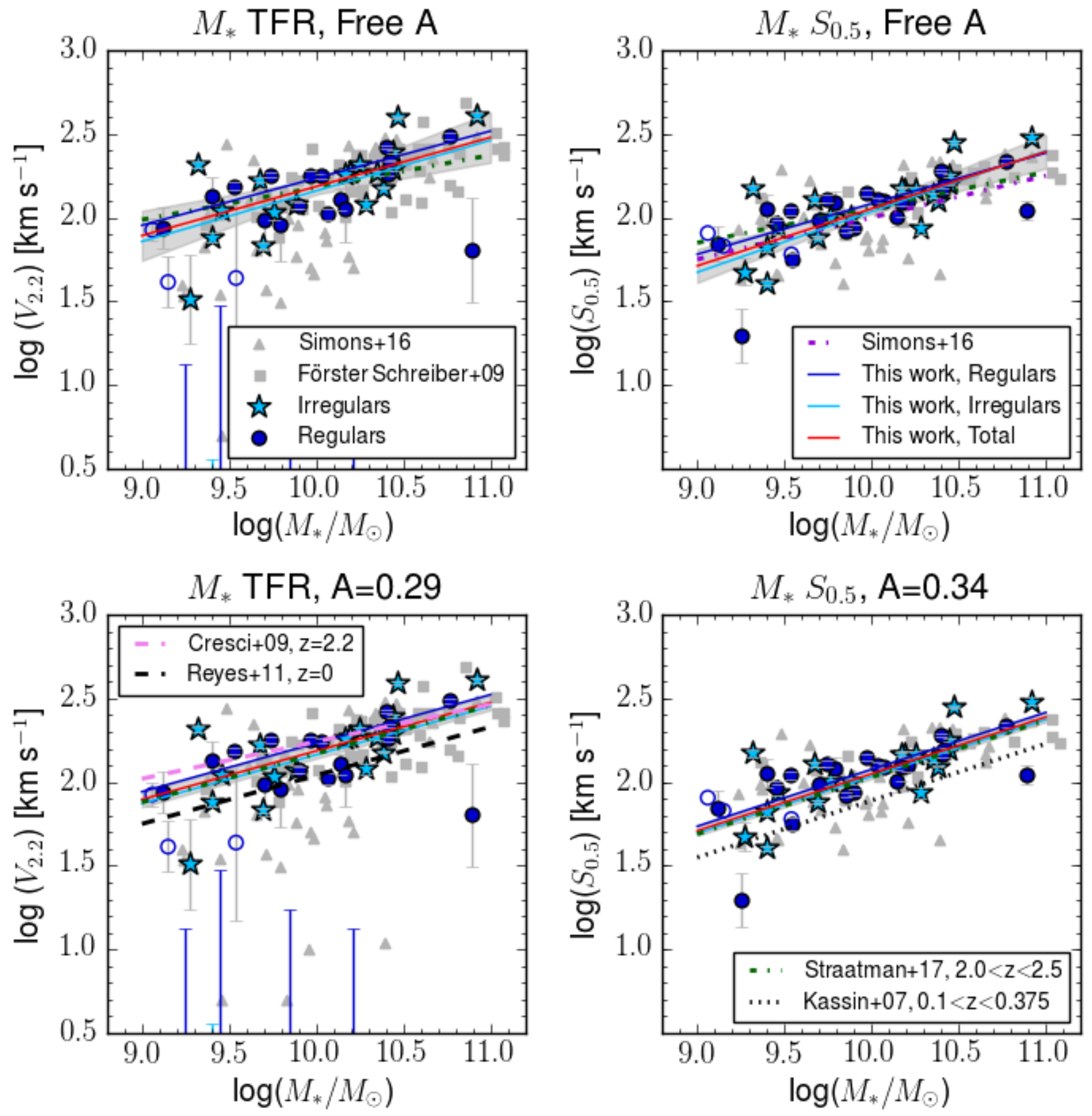}
	\caption{Kinematic scaling relations of the ZFIRE sample. Irregular galaxies are light blue stars, and the linear fit to irregular galaxies is the light blue line. Regular galaxies are dark blue circles, and the fit is the dark blue line. Compact galaxies of either population are unfilled circles or stars. Galaxies with unreliable velocity measurements are shown as upper limits. The best-fit linear relation to the total sample is the solid red line, and the grey shaded regions show the uncertainty in the best-fit line. The best-fit lines from \citet{Straatman2017} are the green dashed line. Upper Left: The stellar-mass TFR. We compare to the SIGMA sample (grey triangles) \citep{Simons2016a} and the SINS data points (grey squares) \citep{Schreiber2009}. Lower Left: As upper left, with slope fixed to $A=0.29$ for consistency with the $z=0$ TFR (black dashed) \citep{Reyes2011} and the SINS IFU survey (pink dashed) \citep{Cresci2009}. Upper Right: The stellar-mass \sof\ relation from \citet{Kassin2007}, which includes the contribution of \sigmag\ to the total kinematics of the system, and a comparison to \citet{Simons2016a}. Lower Right: Slope is fixed to $A=0.34$. We compare to their relation at $0.1<z<1.2$ and find an offset of 0.16$\pm$0.04 dex higher \sof\ at a given stellar mass.}
\end{figure*}

Our final sample consists of \totnum\ galaxies within $-45^o<\Delta\alpha<45^o$ and with less than half the emission line masked and SNR $> 5$. 
14 of these objects are associated with an over-density at $z=2.095$, and 30 are field objects. 
Due to the small number of cluster objects in our sample, as well as the lack of 1D environmental distinctions in this sample \citep{Alcorn2016}, we do not include any environmental analysis in this work.
We identify \disknum\ regular-type galaxies in our sample, and \irrnum\ galaxies which could include both merging galaxies and irregular galaxies - anything that is not well-described by a S{\'e}rsic profile.
\citet{Wisnioski2015} determines a disk fraction of 58\% at $z\sim2$, similar to our estimated disk (regular) fraction (56.8\%) determined from measuring the residual values after subtracting a \sersic\ fit.

\subsection{Measured Kinematic Scaling Relations} \label{sec:scalingrelations}

We derive a best fit linear relation using the Levenberg - Marquardt algorithm for the TFR of the form
\begin{equation} \label{eq:tfr}
	\log(V_{2.2}) = A\log(M_{\ast}/M_{\odot}-10) + B,
\end{equation}
weighted by the errors on \vtwo\ (Figure 3, left). 
We reject objects greater than 3$\sigma$ from the fit, and iterate the fit until the process converges.
Ranges on the fitting parameters are determined by bootstrapping the sample 1000 times.
In the case where $A$ and $B$ are both free parameters of the linear fit, we derive $A=0.29\pm0.1$ and $B=2.19\pm0.04$ for the total sample.
The irregular and regular populations are offset by 0.08 dex.
Scatter in all populations is high, at $0.5\pm0.02$ dex for the total sample, $0.6\pm0.02$ for regulars, and $0.39\pm0.03$ for irregulars.
Given this high level of scatter, we do not think our offsets are significant.
There are a number of low-mass objects that are significantly offset from the relation - these are the compact galaxies that could have underestimated velocities \citep{Newman2012}.

To compare our values for the TFR to literature values, in particular to determine a possible offset to local relations and IFU observations, we hold $A=0.29$, determined by \citet{Reyes2011} for the local TFR.
We derive an offset of $\Delta M/M_{\odot}$=-0.34$\pm$0.22 from local relations.

In both free and fixed slope cases, we do not find any statistically significant difference between irregulars and regulars.
Our results for the TFR do not change if we remove compact objects from our fitting.

In addition, given the values of both \vtwo\ and \sigmag, we derive a best-fit relation for \sof, defined in \citet{Kassin2007} as $S_{0.5}=\sqrt{0.5V_{2.2}^2 + \sigma_g^2}$. 
This equation is derived from a combined velocity scale $S_K$ \citep{Weiner2006a}, $S_K^2=KV_{rot}^2 + \sigma^2$, where $K$ is a constant $\leqslant1$.
Where rotation curves have been measured, $K=0.3-0.5$, consistent with the prediction for an isothermal potential and a flat rotation curve.
This suggests that $S_K$ is a good tracer for the gravitational potential, and for consistency with the literature we use $K=0.5$.

When we derive our equation of the form $\log(S_{0.5}) = A\log(M/M_{\odot}-10) + B$ to the data, we find best fit parameters of $0.38\pm0.07$ and $2.04\pm0.03$ (Figure 3, Right). 
When we fix $A=0.34$ (seen in $0.1<z<1.2$ from \citet{Kassin2007}) we measure $B=2.05\pm0.03$.
Scatter in all populations decreases significantly when we include the contribution of \sigmag\ to the total kinematics (from 0.5 dex for the TFR to 0.15 dex for \sof).
\citet{Kassin2007} derives a scatter of 0.16 dex in \sof\ for  $0.1<z<1.2$, similar to \cite{Price2015} who find a scatter of 0.17 dex at $1.4 < z < 2.6$.
\citet{Straatman2017} finds consistent values with these at $2.0 < z < 2.5$ (0.15 dex), using 22 galaxies drawn from the same ZFIRE sample as this paper, 20 of which are in common with our sample.
Our offset implies a zero-point evolution of $\Delta M/M_{\odot}$=-0.47$\pm$0.14.

When we hold $r_t=1/3 r_s$ and $r_t=0.4 r_s$, we find our results for both the \Mstar-TFR and \sof\ do not significantly change.
Our simulated MOSFIRE observations (Appendix B), show that we tend to overestimate our values for \sof\ to a median offset of $\sim$10\% (Figure 12, top two rows).
However, this offset is stable for SNR$>$10 and less than half the emission line masked (see Appendix B.1), indicating our \sof\ values are reliable.

\begin{figure}[h] \label{fig:voversig}
\figurenum{4}
	\centering
	\plotone{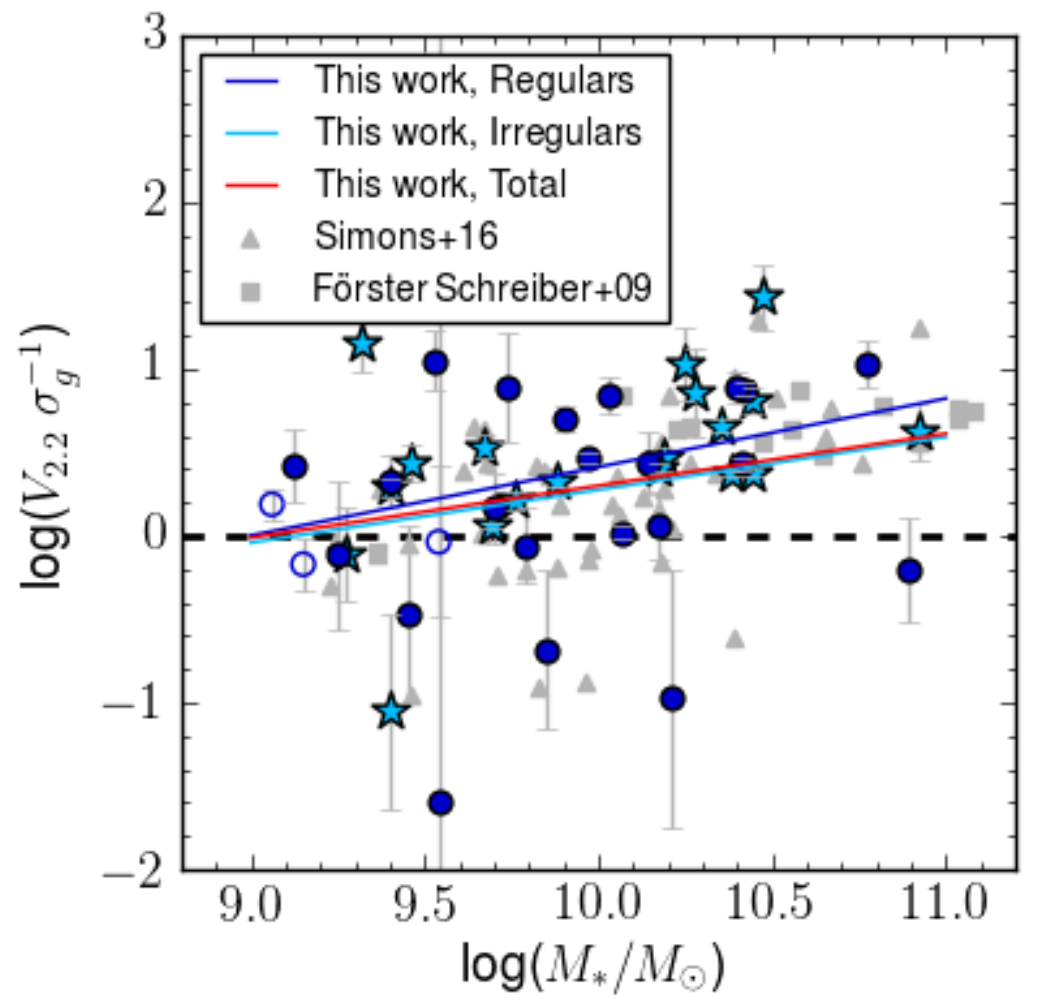}
	\caption{\voversig\ of galaxies in the ZFIRE sample, showing the ratio of rotational support (measured at \vtwo) and \sigmag, pressure support. We find consistent values between regulars and irregulars, and a clear relation between the rotational support and stellar mass. Colors and markers are as described in Figure 3. The black dashed line shows equal rotation and pressure support.}
\end{figure}

The \voversig\ parameter derived from \vtwo\ and \sigmag\ is an instructive measurement for determining the amount of rotational dominance in integrated kinematics. 
Higher \voversig\ indicates a well-ordered rotating disk with minimal random motion within the disk, whereas lower \voversig\ signals a stronger presences of random motion.
In \citet{Schreiber2009, Wisnioski2015, Turner2017a} galaxies are considered rotation-dominated at \voversig$>1$ and pressure-dominated at \voversig$<1$. 
Within our sample we observe both pressure-dominated and rotation-dominated galaxies.

We see a highly scattered trend between \Mstar\ and \voversig, where objects with smaller \Mstar\ are more likely to have log(\voversig)$<0$ (Figure 4).
We can see a clear trend in all populations of increasing rotation support at increasing stellar mass.
In Figure 8 we can see this is not due to a decrease in pressure support at high mass, as \sigmag\ values are unrelated to the stellar mass of a galaxy.
Scatter is large for all populations, $0.67\pm0.04$ dex for irregulars and $0.53\pm0.01$ dex for regulars.
The median values of \voversig\ for regular and irregular galaxies were 1.55 and 2.75, respectively, but given high levels of scatter in both populations, it is unclear if this difference is significant.
The median value of \voversig\ for the total sample was 2.48.
Again, our results are not significantly affected by holding $r_t$ to a fixed position relative to $r_s$.

Our MOSFIRE simulations (Appendix B) show difficulty in recovering \voversig\ using slit spectroscopy.
In the bottom two panels of Figure 12, we see that we tend to overestimate \voversig\ values by 25\% of the input, with scatter of around 20\%. 
This leads us to believe our values could be unreliable and could be related to the heavy scatter in our measured values for \voversig.

\startlongtable
\begin{deluxetable*}{lllcccccc} 
\tabletypesize{\footnotesize} 
\tablecolumns{9} 
\tablewidth{0pt} 
\tablecaption{Kinematic measurements of ZFIRE galaxies using HELA} 
\tablehead{ 
\colhead{ID} & \colhead{Date} & \colhead{Mask} & \colhead{$z_{spec}$} & \colhead{$M_{\ast}$} & \colhead{SFR\tablenotemark{a}} & \colhead{$V_{2.2}$} & \colhead{$\sigma_g$} & \colhead{$j_{disk}$}} 
\startdata 
1814 & feb2014 & KbandLargeArea4 & 2.17 & 9.76 & 14.6 & 108.44$\pm$13.19 & 66.19$\pm$3.55 & 321.87$\pm$39.63 \\
1961 & feb2014 & KbandLargeArea3 & 2.31 & 9.79 & N/A & 90.85$\pm$46.57 & 103.72$\pm$8.78 & 241.22$\pm$123.8 \\
2715 & dec2013 & mask2 & 2.08 & 9.88 & 13.7 & 119.38$\pm$5.98 & 55.18$\pm$4.51 & 555.5$\pm$30.6 \\
2723 & dec2013 & mask2 & 2.09 & 10.92 & N/A & 406.46$\pm$16.23 & 96.47$\pm$37.72 & 717.42$\pm$616.43 \\
2765 & dec2013 & mask1 & 2.23 & 10.44 & 83.3 & 193.38$\pm$4.42 & 80.17$\pm$2.42 & 1227.22$\pm$46.26 \\
3074 & dec2013 & mask1 & 2.23 & 10.19 & N/A & 186.93$\pm$9.12 & 63.69$\pm$9.9 & 879.78$\pm$45.46 \\
342 & feb2014 & KbandLargeArea4 & 2.15 & 10.42 & 31.3 & 218.5$\pm$3.04 & 28.66$\pm$2.65 & 823.63$\pm$15.56 \\
3527 & feb2014 & KbandLargeArea4 & 2.19 & 10.38 & 56.1 & 151.4$\pm$1.39 & 64.26$\pm$1.65 & 579.59$\pm$7.11 \\
3532 & dec2013 & mask1 & 2.1 & 9.4 & 9.9 & 3.57$\pm$4.8 & 40.39$\pm$1.31 & 7.27$\pm$9.77 \\
3619 & feb2014 & KbandLargeArea3 & 2.29 & 9.27 & 3.3 & 32.43$\pm$20.07 & 41.56$\pm$6.78 & 81.71$\pm$50.66 \\
3633 & dec2013 & mask1 & 2.1 & 10.4 & 42.4 & 315.97$\pm$8.34 & 33.83$\pm$11.64 & 1887.98$\pm$68.03 \\
-- & feb2014 & DeepKband2 & 2.1 & 10.4 & 42.4 & 211.17$\pm$2.87 & 34.27$\pm$1.88 & 1261.76$\pm$35.41 \\
3655 & feb2014 & KbandLargeArea3 & 2.13 & 10.35 & 17.7 & 185.23$\pm$6.35 & 40.64$\pm$3.43 & 1008.2$\pm$37.63 \\
3680 & dec2013 & mask3 & 2.18 & 9.32 & 5.0 & 209.49$\pm$9.65 & 14.38$\pm$6.1 & 689.72$\pm$36.88 \\
3714 & dec2013 & mask3 & 2.18 & 10.17 & 66.3 & 184.03$\pm$8.43 & 72.01$\pm$3.47 & 590.34$\pm$27.06 \\
3842 & dec2013 & mask1 & 2.1 & 10.25 & 8.8 & 206.85$\pm$7.2 & 19.19$\pm$9.42 & 904.32$\pm$33.96 \\
3844 & feb2014 & DeepKband2 & 2.44 & 10.44 & N/A & 248.01$\pm$6.05 & 38.03$\pm$6.42 & 1655.84$\pm$55.51 \\
3883 & dec2013 & mask3 & 2.3 & 9.12 & 2.9 & 87.01$\pm$24.22 & 32.9$\pm$13.64 & 169.18$\pm$47.63 \\
4010 & feb2014 & KbandLargeArea4 & 2.22 & 10.07 & N/A & 105.24$\pm$7.93 & 100.06$\pm$4.04 & 295.17$\pm$22.75 \\
4037 & dec2013 & mask2 & 2.17 & 10.77 & N/A & 307.45$\pm$11.23 & 28.65$\pm$9.45 & 1156.14$\pm$45.06 \\
4091 & dec2013 & mask1 & 2.1 & 9.4 & 3.6 & 133.45$\pm$36.72 & 62.65$\pm$13.94 & 425.12$\pm$117.36 \\
4099 & dec2013 & mask3 & 2.44 & 10.28 & N/A & 119.92$\pm$8.5 & 16.67$\pm$10.01 & 472.09$\pm$37.65 \\
4267 & feb2014 & KbandLargeArea3 & 2.41 & 10.14 & N/A & 128.02$\pm$19.22 & 46.88$\pm$19.41 & 388.38$\pm$59.6 \\
4461 & feb2014 & DeepKband2 & 2.3 & 10.89 & 10.2 & 63.69$\pm$45.49 & 101.6$\pm$7.37 & 351.76$\pm$251.73 \\
4488 & dec2013 & mask2 & 2.31 & 10.21 & 7.8 & 13.41$\pm$23.53 & 126.4$\pm$10.4 & 46.35$\pm$81.35 \\
4645 & feb2014 & DeepKband1 & 2.1 & 9.53 & 5.5 & 154.04$\pm$7.65 & 13.71$\pm$5.65 & 488.73$\pm$25.11 \\
4724 & dec2013 & mask2 & 2.3 & 9.54 & 3.1 & 1.42$\pm$18.89 & 56.5$\pm$4.74 & 42.13$\pm$561.56 \\
4746 & dec2013 & mask4 & 2.18 & 9.54 & 6.1 & 28.37$\pm$45.49 & 56.59$\pm$6.91 & 40.63$\pm$65.17 \\
-- & feb2014 & DeepKband2 & 2.18 & 9.54 & 6.1 & 58.66$\pm$47.28 & 43.72$\pm$5.9 & 84.02$\pm$67.76 \\
4796 & feb2014 & DeepKband2 & 2.17 & 9.45 & 6.6 & 30.02$\pm$36.87 & 89.7$\pm$7.13 & 88.28$\pm$108.46 \\
4930 & feb2014 & DeepKband2 & 2.1 & 9.46 & 7.2 & 110.08$\pm$12.99 & 40.05$\pm$8.31 & 438.98$\pm$52.75 \\
5269 & dec2013 & mask3 & 2.11 & 10.03 & 13.7 & 176.48$\pm$6.07 & 25.23$\pm$6.27 & 928.94$\pm$34.88 \\
5342 & dec2013 & mask3 & 2.16 & 9.06 & 2.5 & 84.3$\pm$13.09 & 54.48$\pm$8.77 & 119.25$\pm$18.9 \\
5408 & dec2013 & mask4 & 2.1 & 9.74 & 20.9 & 180.32$\pm$7.71 & 23.24$\pm$17.28 & 442.28$\pm$21.42 \\
5630 & feb2014 & KbandLargeArea4 & 2.24 & 9.97 & 23.6 & 179.28$\pm$9.08 & 61.15$\pm$3.8 & 733.62$\pm$40.15 \\
5745 & feb2014 & DeepKband2 & 2.09 & 9.15 & 8.6 & 41.15$\pm$14.31 & 60.86$\pm$2.79 & 60.21$\pm$21.35 \\
5870 & dec2013 & mask4 & 2.1 & 9.9 & 7.8 & 118.68$\pm$5.07 & 23.47$\pm$3.97 & 444.18$\pm$21.64 \\
6485 & dec2013 & mask2 & 2.16 & 10.41 & 17.1 & 182.66$\pm$6.08 & 67.17$\pm$6.05 & 619.32$\pm$29.24 \\
6908 & feb2014 & DeepKband2 & 2.06 & 10.47 & 59.9 & 395.94$\pm$8.19 & 14.55$\pm$6.5 & 1985.46$\pm$43.6 \\
6954 & feb2014 & DeepKband1 & 2.13 & 9.25 & 6.7 & 13.28$\pm$12.42 & 17.29$\pm$6.82 & 32.11$\pm$30.04 \\
7137 & dec2013 & mask2 & 2.16 & 9.85 & 9.3 & 17.32$\pm$19.15 & 83.12$\pm$2.37 & 64.45$\pm$71.27 \\
7676 & dec2013 & mask3 & 2.16 & 9.4 & 4.4 & 76.67$\pm$5.14 & 39.34$\pm$4.75 & 416.2$\pm$29.36 \\
7774 & feb2014 & DeepKband1 & 2.2 & 10.17 & 10.9 & 111.81$\pm$50.62 & 95.29$\pm$10.76 & 278.55$\pm$126.33 \\
7930 & dec2013 & mask3 & 2.1 & 9.69 & 8.2 & 68.3$\pm$2.32 & 58.92$\pm$1.79 & 492.14$\pm$28.89 \\
8108 & dec2013 & mask2 & 2.16 & 9.67 & 6.1 & 167.71$\pm$5.73 & 48.6$\pm$7.88 & 502.34$\pm$23.26 \\
9571 & dec2013 & mask3 & 2.09 & 9.7 & 7.8 & 97.75$\pm$42.2 & 66.87$\pm$12.8 & 876.68$\pm$383.16 \\
\enddata 
\tablenotetext{a}{SFR is determined from the \halpha flux and corrected for dust assuming a \citet{Calzetti1999} dust law.}
\end{deluxetable*}

We notice a slight difference between the regular and irregular populations in recovered \sigmag, where regulars are more likely to have high values of \sigmag\ than irregulars (Figure 5).
A logistic regression analysis was inconclusive.

\begin{figure}[h] \label{fig:msig}
\figurenum{5}
	\centering
	\plotone{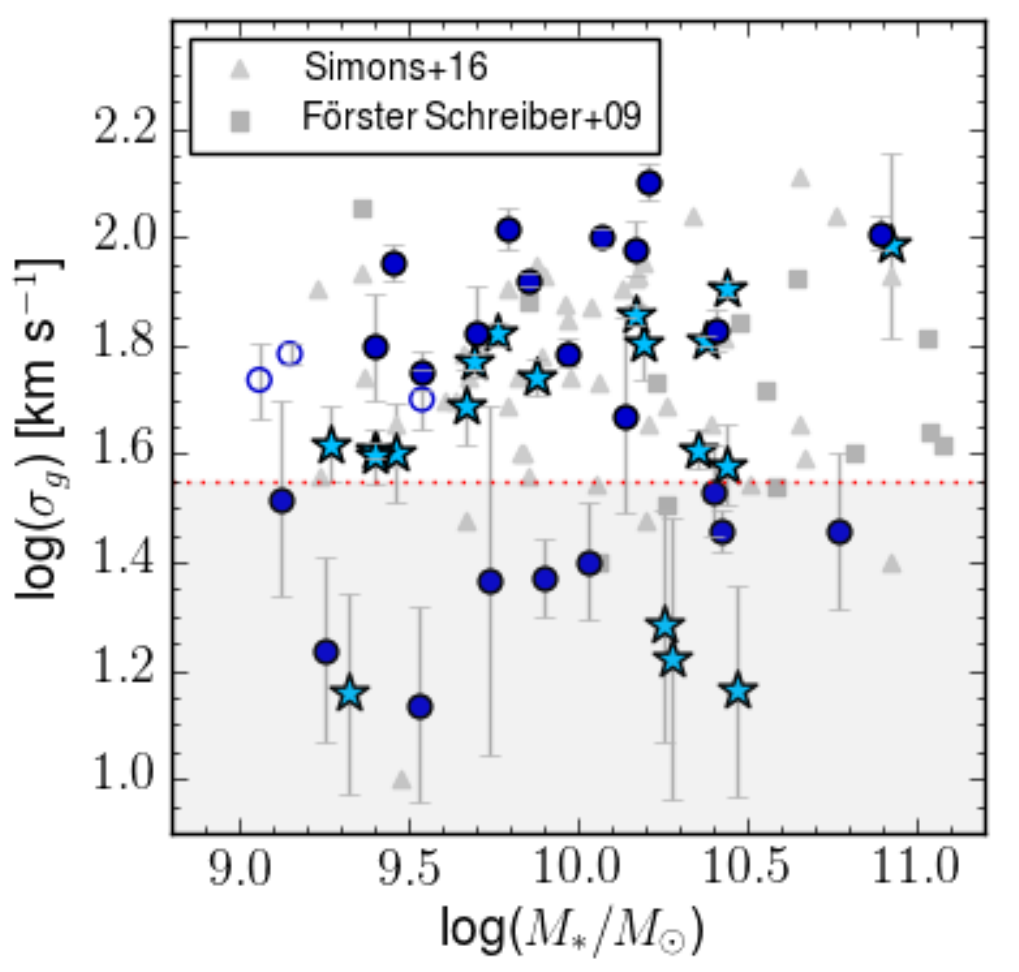}
	\caption{\sigmag\ plotted against \Mstar, values as determined by HELA models. Colors and markers are as described in Figure 5. Areas below MOSFIRE instrumental resolution are shown in the shaded region, marked by the red dotted line.}
\end{figure}

Using our environmentally-diverse sample, our findings are consistent with the results of \citet{Simons2016a}.
In all populations, at low stellar mass, we see evidence of less rotational support.
As stellar mass increases, SFGs have increasing amounts of rotational support, no matter their morphology.
Despite the large scatter in recovery of simulated \voversig, we can still observe a relation between rotational support and stellar mass.

\subsection{Comparison to Disk-Formation Models} \label{sec:modelcompare}

\begin{deluxetable*}{lllrrrrr} 
\tabletypesize{\footnotesize} 
\tablecolumns{7} 
\tablewidth{0pt} 
\tablecaption{Values for all weighted least-square linear fits\tablenotemark{a} to the stellar-mass Tully-Fisher Relation and \sof\ Relation and $j$-\Mstar\ Relation, of the form $\log(y)=A(\log(x)-10.)+B$} 
\tablehead{ 
\colhead{Population} & \colhead{$x$} & \colhead{$y$} & \colhead{$A$} & \colhead{$B$} & \colhead{$B$, fixed $A$\tablenotemark{b}} & \colhead{$\sigma_{RMS}$} & \colhead{$N$}} 
\startdata 
Total & $M_{\ast}$ & $V_{2.2}$ & 0.29$\pm$0.1 & 2.19$\pm$0.04 & 2.19$\pm$0.04 & 0.5$\pm$0.02 & 44 \\
Regulars & $M_{\ast}$ & $V_{2.2}$ & 0.28$\pm$0.07 & 2.24$\pm$0.03 & 2.23$\pm$0.02 & 0.6$\pm$0.02 & 25 \\
Irregulars & $M_{\ast}$ & $V_{2.2}$ & 0.3$\pm$0.15 & 2.16$\pm$0.06 & 2.16$\pm$0.06 & 0.39$\pm$0.03 & 19 \\
Total & $M_{\ast}$ & $S_{0.5}$ & 0.38$\pm$0.07 & 2.04$\pm$0.03 & 2.05$\pm$0.03 & 0.15$\pm$0.01 & 44 \\
Regulars & $M_{\ast}$ & $S_{0.5}$ & 0.31$\pm$0.05 & 2.08$\pm$0.02 & 2.08$\pm$0.02 & 0.16$\pm$0.01 & 25 \\
Irregulars & $M_{\ast}$ & $S_{0.5}$ & 0.43$\pm$0.1 & 2.01$\pm$0.04 & 2.03$\pm$0.04 & 0.16$\pm$0.01 & 19 \\
Total & $M_{\ast}$ & $j$ & 0.36$\pm$0.12 & 2.8$\pm$0.05 & 2.72$\pm$0.07 & 0.52$\pm$0.02 & 44 \\
Regulars & $M_{\ast}$ & $j$ & 0.39$\pm$0.11 & 2.8$\pm$0.05 & 2.73$\pm$0.06 & 0.56$\pm$0.03 & 25 \\
Irregulars & $M_{\ast}$ & $j$ & 0.33$\pm$0.20 & 2.81$\pm$0.07 & 2.71$\pm$0.11 & 0.48$\pm$0.05 & 19
\enddata 
\tablenotetext{a}{Objects more than 3$\sigma$ away from the fits are rejected from the fits to minimize the influence of outliers.}
\tablenotetext{b}{$A=0.29$ for the TFR, $A=0.34$ for \sof, and $A=0.67$ for $j$.}
\vspace{-0.8cm} 
\end{deluxetable*}

\begin{figure}[h]  \label{fig:krumholz}
	\centering
	\figurenum{6}
	\plotone{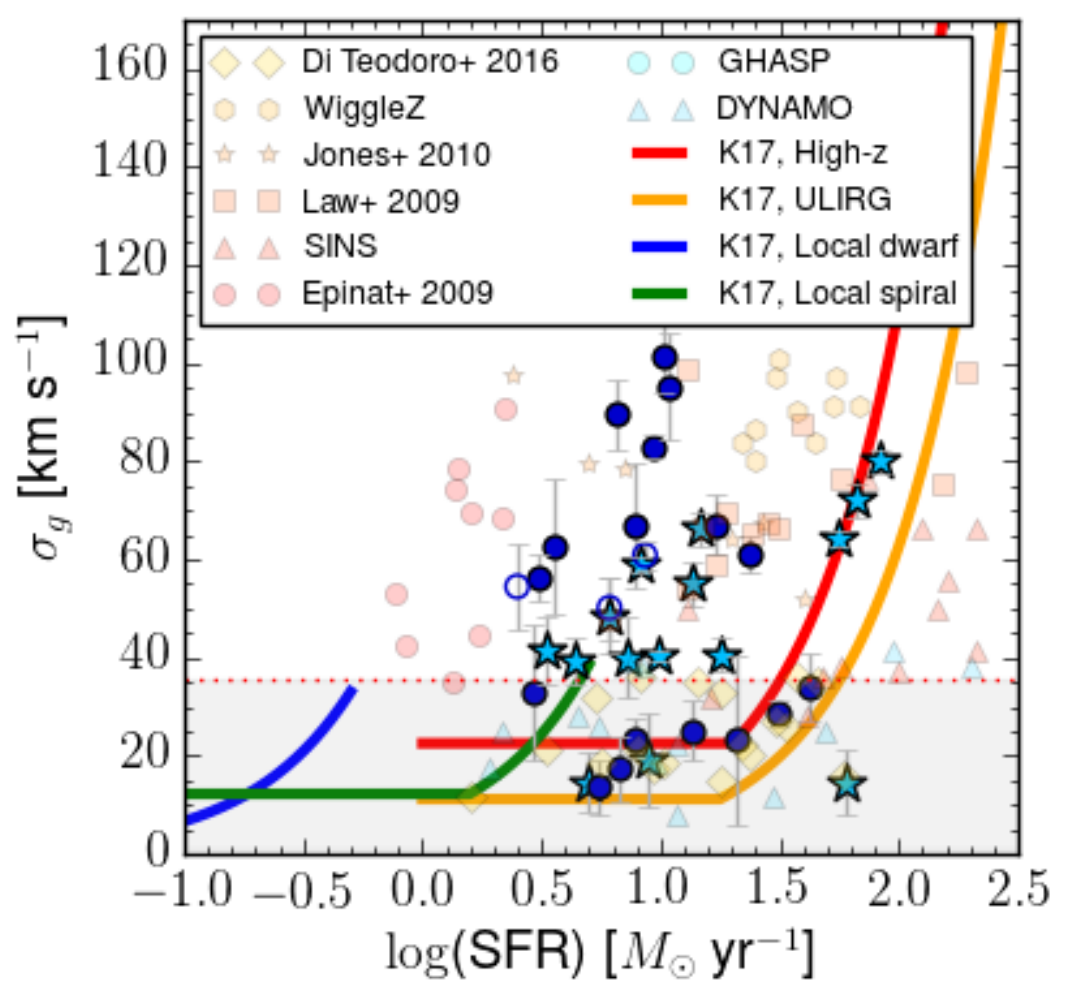}
	\caption{Relationship of our modeled \sigmag\ values against dust-corrected \halpha\ star-formation rate from \citet{Tran2016}. We compare our results to the models derived in \citet{Krumholz2017} for local disks and high-$z$ disks. Local and high-z samples with \halpha\ SFRs featured in \citet{Krumholz2017} are also shown here.}
\end{figure}

\citet{Krumholz2017} introduces a mathematical model for the evolution of gas in the disks of SFGs, which attempts to explain the nature of gas turbulence in these disks.
According to this model, gas turbulence can be fed through star formation feedback, radiative transport, or both.
The underlying prediction is that in gravitationally unstable galaxies, instability-driven mass transport will move mass inward toward the galaxy center until stability is restored.
In this model, disks are never more than marginally gravitationally unstable, and maintain a balance between turbulence driven by star-formation feedback and gravitational instability and the dissipation of turbulence.
It predicts that at high redshift, turbulence is mostly gravitationally-driven, whereas in local disks there is a minimum floor of \sigmag\ ($\sim 6-10$ \kms) where the disks settle that is driven by star-formation feedback.

Our values for \sigmag\ are determined through modeling with HELA, and our star-formation rates (SFR) are determined from dust-corrected \halpha\ flux, assuming a \citet{Calzetti1999} dust law \citep{Tran2016}.
In Figure 6, we compare these values to four theoretical models created assuming properties described in \citet{Krumholz2017}: a local dwarf (fraction of the ISM in the star-forming phase [$f_{sf}$] = 0.2, rotational velocity at 100\kms), a local spiral ($f_{sf}$ = 0.5, rotational velocity at 200\kms), a high-redshift galaxy ($f_{sf}$ = 1.0, rotational velocity of 200\kms), and an Ultra-Luminous InfraRed Galaxy (ULIRG, $f_{sf}$ = 1.0, rotational velocity of 300\kms).
Our sample maintains a similar shape to the high-$z$ and ULIRG models, but SFRs are lower, perhaps indicating that smaller SFRs can drive turbulence in high-$z$ objects.
However, this is consistent with the other high-$z$ observations seen in the text and plotted in Figure 6 \citep{Epinat2008,Epinat2009,Schreiber2009,Law2009,Jones2010, Green2013, Wisnioski2015,Stott2016,DiTeodoro2016}.

The model calculated for a local disk assumes that the dispersion is driven mostly by star-formation feedback, and the ULIRG and high-z models are driven primarily by mass transfer to the core of the galaxy. In this case, it could show that there is more turbulence driven by star formation feedback and mass transfer plays less of a role in high-z galaxies than predicted. \citet{Krumholz2017} assumes these objects are disks and are never more than marginally unstable. The offset of these galaxies from these predictions could mean these objects are unstable and are possibly not even disks. Instead turbulence may be driven at least partially by external factors such as a recent merger or disk instabilities caused by rapid gas accretion.

\subsection{Angular Momenta of SFGs at \zsimtwo} \label{sec:angularmomentum}

Using the maximum rotational velocity (assuming ideal disks, this is \vtwo), and scale radius, we can estimate specific angular momenta of our galaxies given the formula:
\begin{equation} \label{eq:jdisk}
	j_{disk}=K_n r_s V_{2.2},
\end{equation}
where $j_{disk}$ is the specific angular momentum (angular momentum per solar mass), and $K_n$ is defined as
\begin{equation} \label{eq:jdisk}
	K_n=1.15+0.029n+0.062n^2,
\end{equation}
where $n$ is the \sersic\ index of the galaxy \citep{Romanowsky2012}.
We recognize that in the case of galaxies with complex kinematics and morphological structure, that $r_s$ may not be the best representation of the disk radius, but to obtain a consistent sample we apply this to all galaxies.

Generally angular momentum measurements are taken using IFU spectroscopy. 
As such, our results may not be the same as what would be measured in an IFU survey. 
We hope to follow these results up with IFU observations of some of these objects, to determine if the 3D data-cube fitting method yields more accurate measurements of \jdisk\ than traditional velocity curve-fitting methods for slit spectroscopy.
Despite this disclaimer, our simulated slit observations (Appendix B) demonstrate that we can reliably recover our input \jdisk\ to within an offset of -5\% (Figure 13).
This small offset from our input is consistent over all simulated $\Delta\alpha$, inclination, and sizes, and only becomes unreliable at line masking $>50$\% and SNR$<$10.

Additionally, we assume that the angular momentum of the gas disk traces the angular momentum of the stellar disk and older stellar populations. Local kinematic studies usually make this assumption due to the difficulties of measuring the angular momentum of stellar populations \citep{Romanowsky2012, Obreschkow2014}, and these difficulties increase at high redshift. Simulations show that the stellar disk rotates slower than the gaseous disk in late-type galaxies  \citep{El-Badry2017}. In contrast, some observational studies of spatially resolved low-redshift clumpy star-forming disks show that the ionized gas and stellar kinematics are coupled \citep{Bassett2014}. The validity of our assumption is still under debate, but for consistency with local kinematic surveys we apply this assumption.

\begin{figure*}[t] \label{fig:jplots}
\centering
\figurenum{7}
\includegraphics[width=\textwidth]{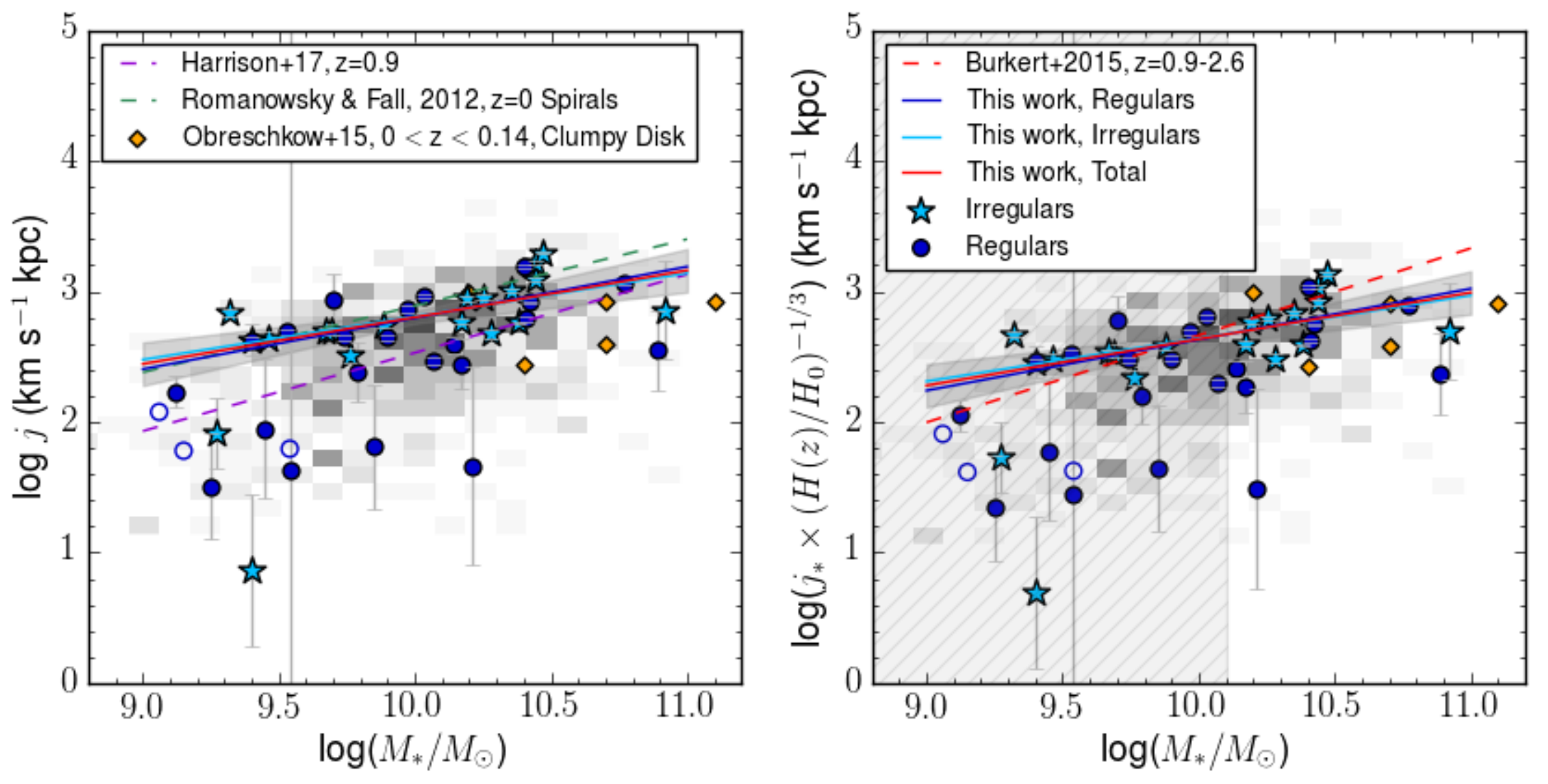}
\caption{Specific angular momenta of ZFIRE galaxies. Left: Specific angular momenta $j$ against \Mstar. We compare to the $z=0.9$ KROSS survey (purple dashed) \citep{Harrison2016}, the $z=0$ spiral galaxies from \citet{Romanowsky2012} (green dashed line), and the $z=0.1$ clumpy, turbulent disk sample of \citet{Obreschkow2015}. The shaded squares show the density of objects from the KROSS $z=0.9$ survey. Right: We correct our values of $j$ for redshift and compare to the results of \citet{Burkert2016} (red dashed). The shaded region shows the mass limit for the selection of galaxies used in the \citet{Burkert2016} sample.}
\end{figure*}

In Figure 7, left panel, we see our estimated \jdisk\ compared to lower-redshift observations. 
We note a shallower slope than \citet{Romanowsky2012} at $z=0$ and KROSS \citep{Harrison2016} ($z=0.9$).
For the total population, we find a slope of $0.36\pm0.12$ and intercept of $2.80\pm0.05$.

There are no significant differences between regulars and irregulars, although scatter in regulars ($0.56\pm0.03$ dex) is higher than irregulars ($0.48\pm0.03$).
The difference in scatter is due to the slow-rotating low-mass regulars.
We see similar slow rotators in the irregular population, but we have fewer in our sample.
In both cases, we find a similar, shallow slope of $0.39\pm0.12$ for regulars and $0.33\pm0.20$ for irregulars.
The shallow slope is from weighting of our linear fits, since low-rotation objects tend to have higher uncertainties in their measurements.
When we perform a linear fit without weighting, we find values much closer to the predicted ($A=0.63\pm0.14$, for the total sample, $0.56\pm0.15$ for regulars, and $0.66\pm0.27$ for irregulars).
When we fix $r_t=1/3 r_s$, we find the slope to move to $0.44\pm0.12$ with no significant differences between irregulars and regulars.
We find similar results when $r_t=0.4 r_s$.

When we hold the slope to be 2/3, we obtain a normalization of $2.72\pm0.07$, which is a normalization offset of $0.12\pm0.09$, or little to no redshift evolution from $z=0$.
This is in conflict with the \citet{Harrison2016} measurement of a 0.3 dex offset from $z=0$.
However if we perform the linear fit without weighting, we find a consistent offset with \citet{Harrison2016}.
In order to conclusively measure the slope and normalization of the line, we will need to explore the kinematics of low-rotation galaxies with greater precision, to bring these minimize our uncertainties.
It is expected that for $\Lambda$CDM disks, $\log{j}\propto \log({M_{\odot}}^{2/3})$ unless there is mass-dependent angular momentum buildup of the disk \citep{Romanowsky2012}.
If these results are confirmed, it is suggestive that stellar mass has a larger effect on angular momentum than morphology at \zsimtwo.

Angular momentum is expected to decrease with increasing redshift due to cosmic expansion as 
\begin{equation} \label{eq:redev}
j\propto (1+z)^{-1/2}, 
\end{equation}

\citep{Obreschkow2015}.
To determine if our sample shows any evolution apart from the theoretical $\Lambda$CDM evolution we scale our sample to local galaxies using Equation 6.
After correcting for any redshift evolution (Figure 7, right panel), we compare our findings to the work of \citet{Burkert2016}.
We again see a shallower slope than the $\log{j}\propto \log({M_{\odot}}^{2/3})$ trend, but when holding the slope to 2/3 we find an offset with the \citet{Burkert2016} results of $0.12\pm0.07$ dex.
If we set $r_t$ to fixed positions relative to $r_s$, find no significant difference from free $r_t$.
Given the scatter in this relation (0.52 dex), we do not find this to be a significant difference from the \citet{Burkert2016} result, which is not expected to evolve with redshift.

A two-population KS test confirms that to a 95\% confidence level, irregular galaxies have higher specific angular momenta than regular galaxies at equivalent stellar mass. Further observations are needed to confirm these results due to low numbers and possible unresolved irregular structure in regular galaxies. Most of this offset is on the low-mass (\Mstar$<$10) end of the j-\Mstar relation, on the high-mass end (\Mstar$>$10) these relationships tighten. When low-rotation resolved objects are removed, the irregular and regular populations are not significantly different.

Additionally, we compare our sample to the clumpy, turbulent galaxies of \citet{Obreschkow2015}, often considered high redshift analogs in the local universe. 
We can confirm that at least kinematically, \zsimtwo\ galaxies have similar properties to these local galaxies.

\section{Discussion} \label{sec:discussion}

\subsection{Morphology and Kinematics}

In some cases it appears that irregulars, including merger candidates, show ordered rotation fields, and as such cannot be identified by kinematics alone.
This is also observed in the IFU-based work of KMOS Deep Survey (KDS) \citep{Turner2017a}, who describe a similar phenomenon of merger candidates with ordered rotation fields.
In \citet{Hung2015} local merging galaxies are artificially redshifted and their rotation is examined.
All mergers with the exception of those with strong tidal features and two nuclei showed ordered rotation fields.
This could explain the similarity of the kinematic scaling relations for regular and irregular galaxies, which could include mergers, derived in our results.
We demonstrate that our irregular galaxies are often well-described by ordered rotation, as our models are derived from rotation-dominated isolated galaxies, and our kinematic extractions assume ordered rotation.

However as irregular galaxies are not well described by photometric modeling (Figure 1), these measurements could be incorrect from assuming that our morphological and kinematic PAs are consistent, and that our intrinsic axis ratio is 0.18.
Similarly, in our modeling, we assume that all galaxies are infinitely thin disks with \sersic\ indices of 1, which is not true for most of our measured galaxies, and for irregular galaxies, the \sersic\ profile is unreliable.

Given these caveats in our analysis, we expect different behaviors in our kinematic relationships if growth is dominated by major mergers or smooth gas accretion. Mergers, depending on the geometry of the system, could cause a system to abruptly gain or lose angular momentum, and would increase the scatter around kinematic scaling relations \citep{Vitvitska2002, Naab2014, Rodriguez-Gomez2017}. Assuming that these mergers are not happening in a preferred direction, we would expect a larger scatter in our velocity and angular momentum relations in merging galaxies (which we are assuming are represented by irregulars). We would also expect these galaxies to have higher values for \sigmag\ than galaxies that have not undergone a recent merger.

If growth is dominated by smooth accretion, the angular momentum of galaxies would again be subject to the direction of gas falling onto the disk. If gas is accreted along a filament, it would exert a torque causing an increase in angular momentum \citep{White1984a, Keres2004, Sales2012}.

Kinematic surveys are often biased toward galaxies with ordered rotation and a relatively small contribution of \sigmag\ toward overall kinematics at an observed redshift.
This is partially because these galaxies are usually intrinsically brighter, as they are more massive.
In addition to brightness, the size of a galaxy can have an effect on its kinematics.
\citet{Newman2012} demonstrated that spatially unresolved galaxies in kinematics surveys can have underestimated rotational velocities.
In our sample, we rejected four compact galaxies with unreliable measurements for \vtwo.
This could bias our sample and our results, underestimating the prevalence of low \voversig\ galaxies.
Additionally, we could be classifying galaxies with unresolved irregular structure as regular galaxies.

We find similar levels of scatter between regular and irregular populations in the TFR, \sof, (Figure 3) and \jdisk\ relations, but irregular galaxies have higher \jdisk\ values at given stellar mass (Figure 7), and do not have higher values of \sigmag\ (Figure 5). Due to our limited sample, more observations are needed to confirm these results. Given that these galaxies have clear irregularities and sometimes show obvious signs of merging close companions, these results are puzzling. We have yet to find simulations which show results like our observations.

In the case that irregulars have higher \jdisk\ than regular galaxies, a significant portion of our sample is in an over-dense proto-cluster region, and this may affect the direction of gas infall or orientation of mergers. Our assumption was that in the case of merger-dominated or accretion-dominated growth, orientation would be random, and would create a stochastic scatter. However it is possible that these interactions may have a preferred orientation, possibly due to the filamentary structure of the cosmic web \citep{Keres2004, Sales2012, Stewart2013, Danovich2015}. More observations are needed for a robust analysis of our conjecture, and knowledge of the cosmic web surrounding this structure would be beneficial.

\subsection{The Reliability of Kinematics From Slit Spectroscopy}

Some of the scatter in our kinematic scaling relations and angular momentum is possibly related to the scatter in our ability to recover our simulated \vrot\ and \sigmag, and the inherent issues with recovering velocities in unresolved galaxies.
This was likely because in unresolved emission lines, the position of the turnover radius is unclear, so we tend to overestimate the positions of simulated $r_t$ and $V_t$.
In other surveys, it is assumed $r_t=0.4r_s$, as observed in \citet{Miller2011}.
However, this is an empirical observation at $z\sim1.7$, when disks are settling.
Whether this assumption holds at $z>2$ is unclear, but the position of $r_t$ in an arctangent velocity curve will affect the derived rotational velocities of a galaxy.

Our simulations (Appendix B) demonstrate that we tend to consistently overestimate \vtwo\ by around 10\% at high data quality (Appendix B.1) and inclination $>25^o$ (Figure 11).
When we can fix $r_t$ to a known value, our recovery is more accurate, to 5\%.
Similarly we underestimate \sigmag\ by 10\%.
Small deviations from our inputs in either of these values lead to overestimated values for \voversig\ with a high scatter in recovered values of our simulations, meaning recovered \voversig\ values may be unreliable (Figure 12).
However, these offsets lead to only slightly overestimated values for \sof, which are reliably offset at high data quality and inclination $>25^o$.
Similarly, our recovery of \jdisk\ is reliable within 5\% of the input with small scatter in our results (Figure 13).
These results show that given the degeneracies seen in modeling emission lines from slit spectroscopy, we can reliably recover values for \sof\ and \jdisk\ if these offsets are accounted for.

We suggest that current slit observations and data analysis can reliably measure \sof\ and specific angular momentum of spatially resolved galaxies at \zsimtwo.
Unresolved galaxies can give unreliable velocity measurements, so increased spatial resolution in multi-object spectrographs are necessary to progress in our understanding of high-redshift kinematics.
The James Webb Space Telescope (JWST) will benefit kinematics due to the NIRSPEC instrument for this reason. NIRSPEC shutter resolution will be at 0.1$\arcsec$, but more importantly these data will not be seeing-limited.
Multi-object slit spectroscopy and JWST provide the opportunity for larger sample sizes, and increased sensitivity to low-mass and faint objects.
As we enter the era of large astronomical surveys, slit spectroscopy will prove an invaluable tool for building large samples of galaxies.

\section{Summary} \label{sec:summary}

We examine an environmentally diverse sample of \zsimtwo\ star-forming galaxies in the COSMOS field observed by the ZFIRE survey.
Complementary NIR imaging in the F160W bandpass from HST/WFC3 as part of the CANDELS project allow for morphological analysis of this sample.
This sample is made up of \totnum\ galaxies: 14 are associated with an over-dense region at $z=2.095$ and 30 are in the field from $2.0<z<2.5$. 
These galaxies are split into two morphological sub-samples, termed ``regulars'' (\disknum) and ``irregulars'' (\irrnum) (Figure 1). 
This classification is based on the presence of excess residual emission from a single-S{\'e}rsic fit where a galaxy is classified as an irregular if residual levels are above twice the nearby sky levels, and greater than 25\% of the original flux levels.

The \halpha\ emission lines are used to extract kinematic components using HELA \citep{Straatman2018}.
HELA simulates a 3D data cube, collapses it into a 0.7$\arcsec$ slit, and runs an MCMC simulation to determine the best-fit model to the emission line, assuming an arctangent rotation curve and a constant gas velocity dispersion.
HELA recovers the velocity of simulated galaxies (Appendix B) at $2.2r_s$ (\vtwo) to within 10\% of our input and \sigmag\ to within -10\% of its input (Figure 11).
Using recovered kinematics, HELA can reliably recover \sof\ to a minor offset of within -10\% of the input, and \jdisk\ (specific angular momentum) to within -5\% of the input (Figures 12 and 13).
\voversig\ tends to be overestimated by 30\% with a high scatter in recovery.
When we constrain the location of the kinematic turnover radius $r_t$ to a known position relative to the scale radius $r_s$, our offsets decrease by 5\% from inputs.

Using the values for \vtwo\ derived from our fitting method, we determine a stellar-mass TFR of $\log(V_{2.2}) = (0.29\pm0.1)\log(M/M_{\odot}-10) + (2.19\pm0.04)$ (Figure 3).
There are no significant differences between regulars and irregulars.
When we include the contribution of \sigmag, in the case of \sof, we find $\log(S_{0.5}) = (0.38\pm0.07)\log(M/M_{\odot}-10) + (2.04\pm0.03)$.
The scatter of the overall sample is consistent with other measurements of \sof\ at $z > 1.5$ \citep{Price2015, Straatman2017}.

To measure pressure against rotational support, we determine \voversig\ (Figure 4), and measure a trend of increasing rotational support with increasing stellar mass, similar to the results of \citet{Simons2016a}.
However there is high scatter in our recovery of simulated \voversig values, leading us to believe that the significant scatter in our results (0.6 dex) may be driven by measurement uncertainties.

We compare our results to the mathematical modeling of \citet{Krumholz2017}, which are based on a balance between turbulence driven by star-formation feedback and gravitational instability, and the dissipation of turbulence by mass transport (Figure 6).
Our sample shows a similar shape in the dust-corrected SFR and \sigmag turbulence but the models overpredict the SFR necessary to produce high gas turbulence in high redshift galaxies.

We also estimate specific angular momentum values (Figure 7), and determine that galaxies have a shallower relationship (slope $A=0.36\pm0.12$) between \jdisk\ and \Mstar\ than predicted ($A=0.67$), either due to undersampling low-rotation low-mass galaxies, or due to a mass-dependent angular momentum buildup in the disk \citep{Romanowsky2012}. 
Additionally, we do not find any evidence of angular momentum offsets with redshift at consistent stellar mass.
More observations of these galaxies will clarify our results, as well as more precise measurements of the kinematics of pressure-dominated SFGs.
Our irregular and regular populations were consistent.
Our simulated observations demonstrate reliable recovery of input kinematics, and we achieve similar \jdisk\ measurements to $z\sim0.1$ high-$z$ analogs \citep{Obreschkow2015}.

Our work demonstrates that slit spectroscopy can reliably recover kinematics measurements such as \vtwo, \sof, or \jdisk\ to either a consistent offset that can be corrected, or to a small offset from simulated inputs.
Low spatial resolution can limit our ability to recover kinematics, but with an increase in resolution, MOS spectroscopy can provide robust kinematic measurements.
In the coming age of large astronomical datasets, the reliability of slit spectroscopy will be instrumental in building large spectroscopic samples at high redshift and using the Near Infrared Spectrograph, NIRSPEC on the James Webb Space Telescope.

\acknowledgments

We would like to thank the anonymous referee who provided helpful comments and suggestions, greatly improving our analysis of this work.
We acknowledge S. Kassin, J. Walsh, and R. Quadri for helpful conversations, and the Mitchell family, particularly the late George P. Mitchell, for their continuing support of astronomy. 
L. Alcorn thanks the LSSTC Data Science Fellowship Program, her time as a Fellow has benefited this work.
This work was supported by a NASA Keck PI Data Award administered by the NASA Exoplanet Science Institute. 
Data presented herein were obtained at the W. M. Keck Observatory from telescope time allocated to NASA through the agency's scientific partnership with the California Institute of Technology and the University of California. 
This work is supported by the National Science Foundation under Grant \#1410728.  
GGK acknowledges the support of the Australian Research Council through the award of a Future Fellowship (FT140100933). 
The authors acknowledge the Texas A\&M University Brazos HPC cluster that contributed to the research reported here.

The authors wish to recognize and acknowledge the very significant cultural role and reverence that the summit of Mauna Kea has always had within the indigenous Hawaiian community. 
We are most fortunate to have the opportunity to conduct observations from this mountain.

\bibliography{library}

\appendix

\section{HELA Modeling}
Here we describe our method of fitting our emission lines, using HELA (Heidelberg Emission Line Algorithm), provided by its developer, C.M. Straatman \citep{Straatman2018}, which uses the prescription of \citet{Price2015}.

The emission line fit is generated from a 3D data-cube. 
This data cube is generated given an input inclination, slit offset, redshift, emission line wavelength, and an estimated scale radius, turnover radius, asymptotic velocity, and \sigmag. 
Given bounds in spatial and wavelength space ($x_{init}$ and $y_{init}$), we create an x-y grid of velocity space, face-on with a galaxy, or at $i=0^o$.
\begin{figure*}[t] \label{fig:display}
\centering
\figurenum{8}
\includegraphics[width=\textwidth]{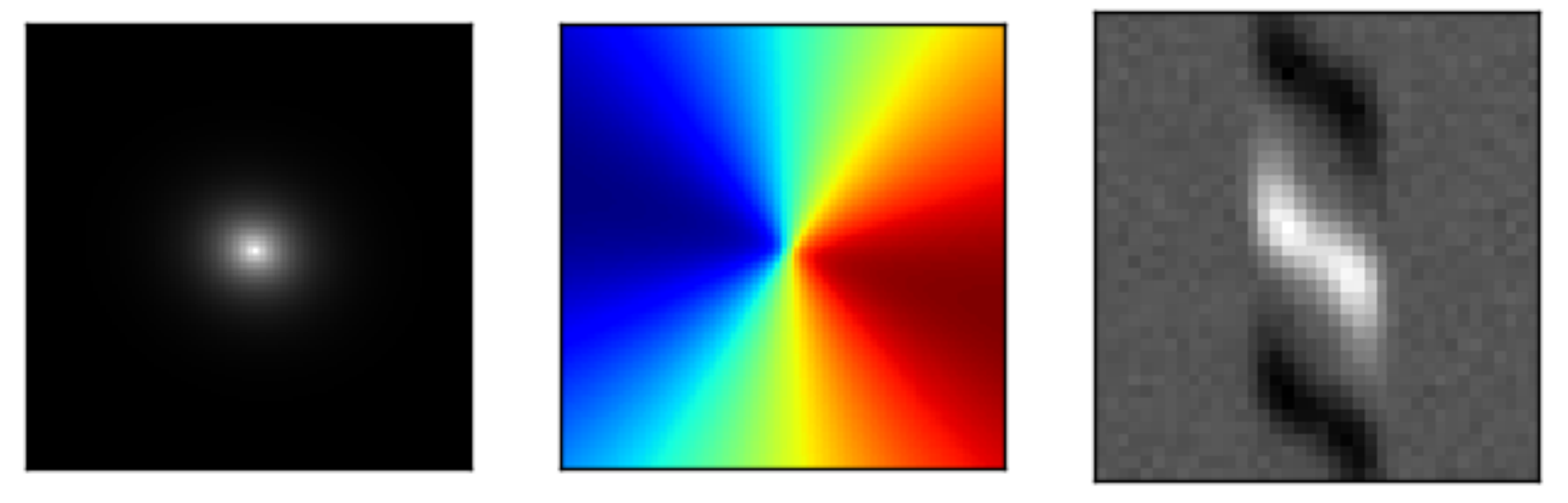}
\caption{An example of our models created in HELA. Left: Spatial intensity profile of an infinitely thin disk galaxy, with $V_t=300$ \kms, $r_s=0.5 \arcsec$, $r_t=0.15 \arcsec$, \sigmag$=25$ \kms, $i=30^o$, and $\Delta \alpha= 15^o$. Center: The line of sight velocity field of the galaxy to the left. Right: Emission line of the galaxy described, convolved with a 2D Moffat profile at 0.7$\arcsec$ seeing.}
\end{figure*}
With our input $\Delta \alpha$, we transform our model using
\begin{eqnarray} \label{eq:patransform}
	x_0=x_{init}\cos{\Delta \alpha}-y_{init} \sin{\Delta\alpha} \\
	y_0=x_{init}\sin{\Delta \alpha}+y_{init}\cos{\Delta\alpha},
\end{eqnarray}
to account for our offset between the galaxy major axis and our slit $PA$.
We transform our values using our input inclination with
\begin{eqnarray} \label{eq:incltransform}
	x_i=x_p/\cos{i},
\end{eqnarray}
rotating our galaxy into its correct inclination.
We define a variable $r$, the distance from the center of the galaxy, as,
\begin{eqnarray} \label{eq:distance}
	r^2=\sqrt{x^2_{i}+y^2_{0}},
\end{eqnarray}
and the angle $\psi$ as,
\begin{eqnarray} \label{eq:angle}
	\cos{\psi}=y_{p}/r.
\end{eqnarray}

A velocity profile is created assuming an infinitely thin disk, an arctangent rotation curve 
\begin{equation} \label{eq:vcurve}
	V_{rot}(r)=\frac{2}{\pi}V_t\ arctan \frac{r}{r_t},
\end{equation}
where $V_t$ is the asymptotic velocity and $r_t$ is the turnover radius.
This equation is then used to determine the line-of-sight velocity ($V_{LOS}$)
\begin{equation} \label{eq:vlos}
	V_{LOS}=V_{rot}(r)\sin{i}\cos{\psi}.
\end{equation}

To map our kinematic components into a 2D emission-line observation, as would be seen from slit spectroscopy, we create a spatial exponential intensity profile,

\begin{equation} \label{eq:iprofile}
	I(r)=I_0 \exp{\frac{-(r)}{r_s}},
\end{equation}
where $r_s$ is the intensity scale radius.
The intensity profile is then mapped onto $V_{LOS}$ using
\begin{equation} \label{eq:emdisp}
	I(r,\lambda)=\frac{I(r)}{\sqrt{2\pi}\sigma_g}\exp(-\frac{(\lambda-\lambda_{LOS})^2}{2\sigma_g^2}),
\end{equation}
where \sigmag\ is the intrinsic gas velocity dispersion.

We convolve this intensity profile with a Moffat 2D PSF if Moffat parameters $\alpha$ and $\beta$ are provided, as in the Moffat PSF profile:

\begin{equation} \label{eq:moffmodel}
	PSF(r)=\frac{\beta-1}{\pi \alpha^2}\left[1+\left(\frac{r}{\alpha}\right)^2\right]^{-\beta}.
\end{equation}

If Moffat parameters are not provided, then a Gaussian profile of given seeing can be used in place of a Moffat profile.
Then we collapse the model over a slit width of $0.7\arcsec$, and scale to our preferred intensity signal.
During fitting to MOSFIRE data or simulated observations, this scaling is determined from a weighted least-squares fit of the model to the data or simulation, weighted by the measurement errors from the weight images.

Our best-fit models for our sample can be seen in Figure 9.

\begin{figure}
   \centering
 \begin{tabular}{l}
 \includegraphics[width=\textwidth]{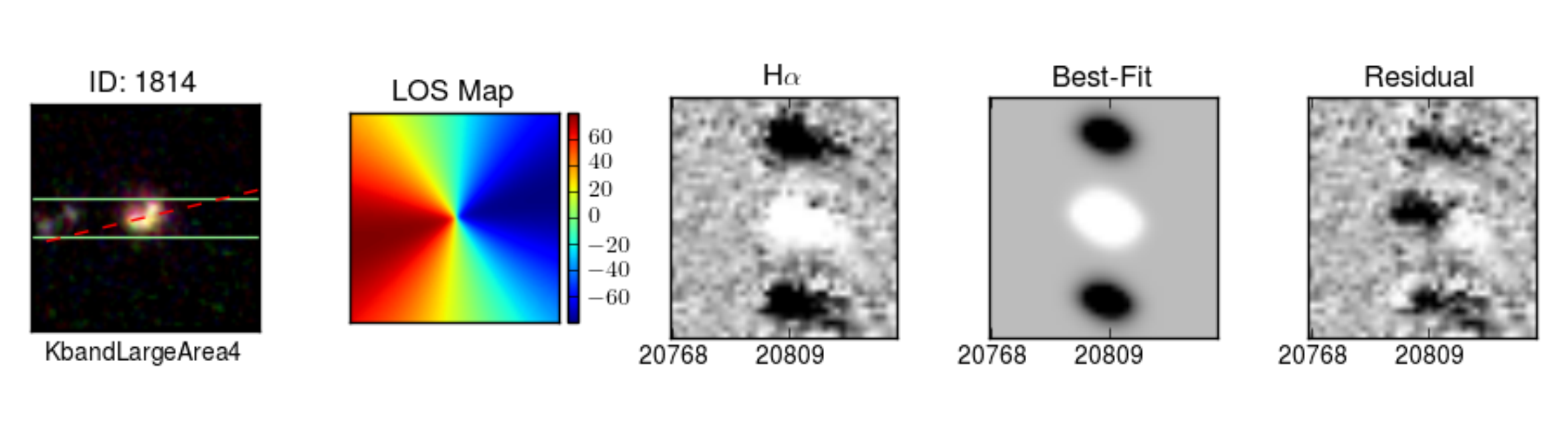}\\
 \includegraphics[width=\textwidth]{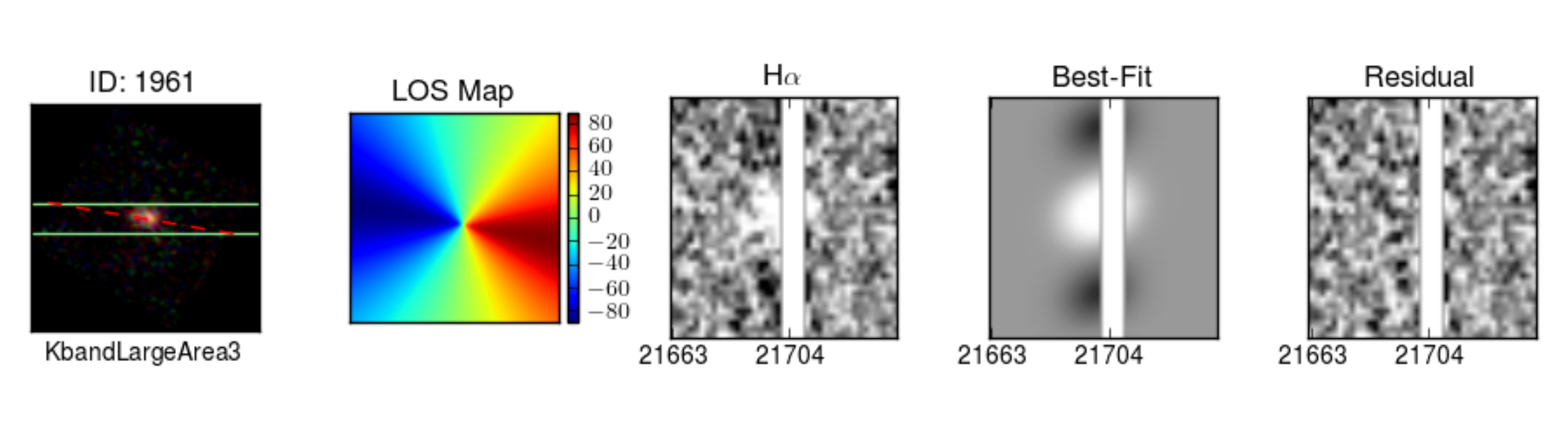}\\
 \includegraphics[width=\textwidth]{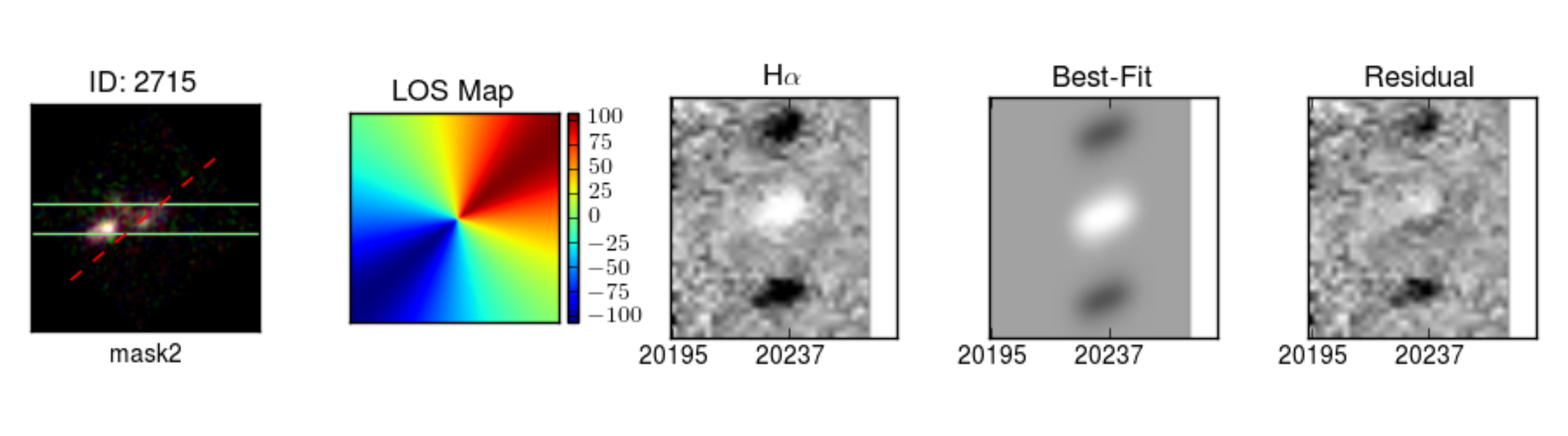}\\
  \includegraphics[width=\textwidth]{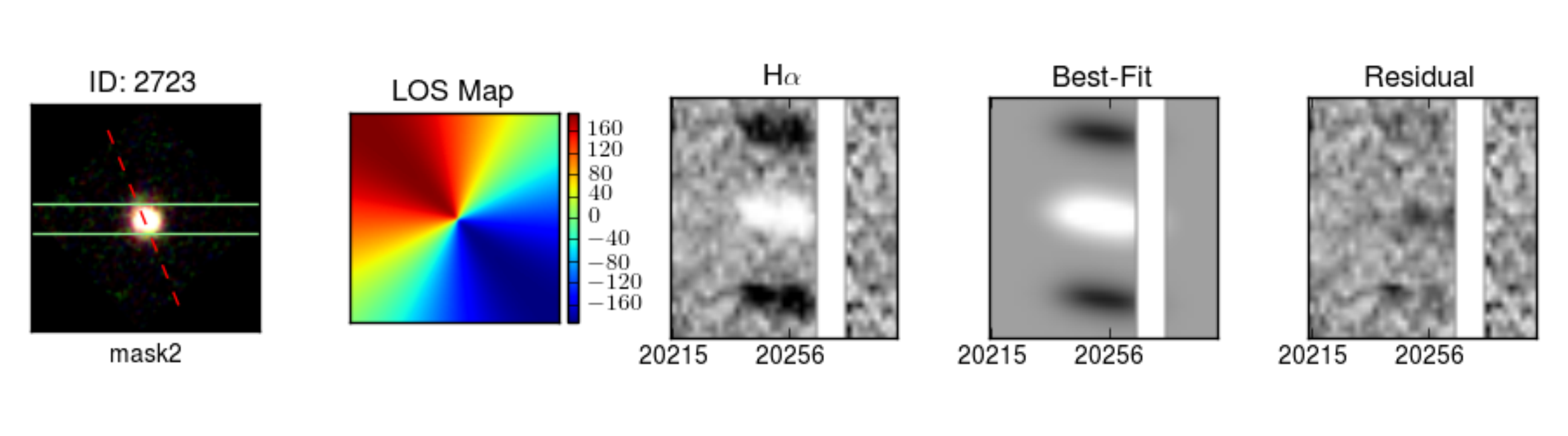}\\
       \end{tabular}
     \caption{Imaging and best fits of galaxies in our sample. From Left: RGB images are from F160W (red), F140W (green), and F125W (blue). The slit overlay is shown in green and the major axis of the galaxy is shown in red. Second from left: The LOS map is aligned with the RGB image. Center: The \halpha\ emission line with sky emission masked in white and continuum removed, if present. Second from right: Best-fit emission line from HELA modeling, characterized by the LOS map. Right: Residual from the best-fit line.}
\end{figure}
\begin{figure}
   \centering
  \addtocounter{figure}{-1}
 \begin{tabular}{l}
 \includegraphics[width=\textwidth]{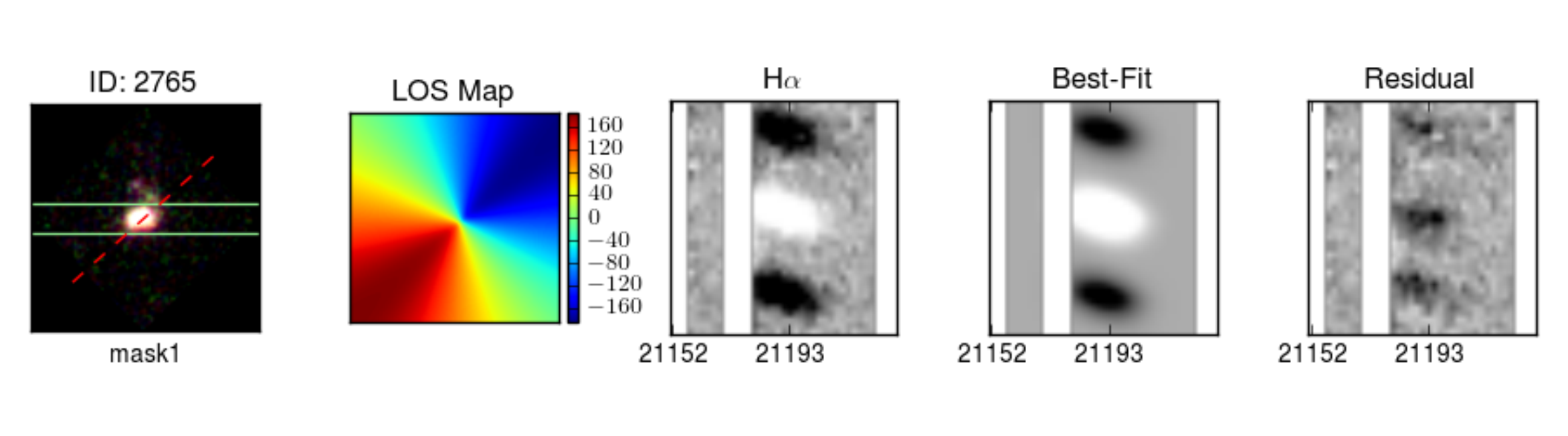}\\
\includegraphics[width=\textwidth]{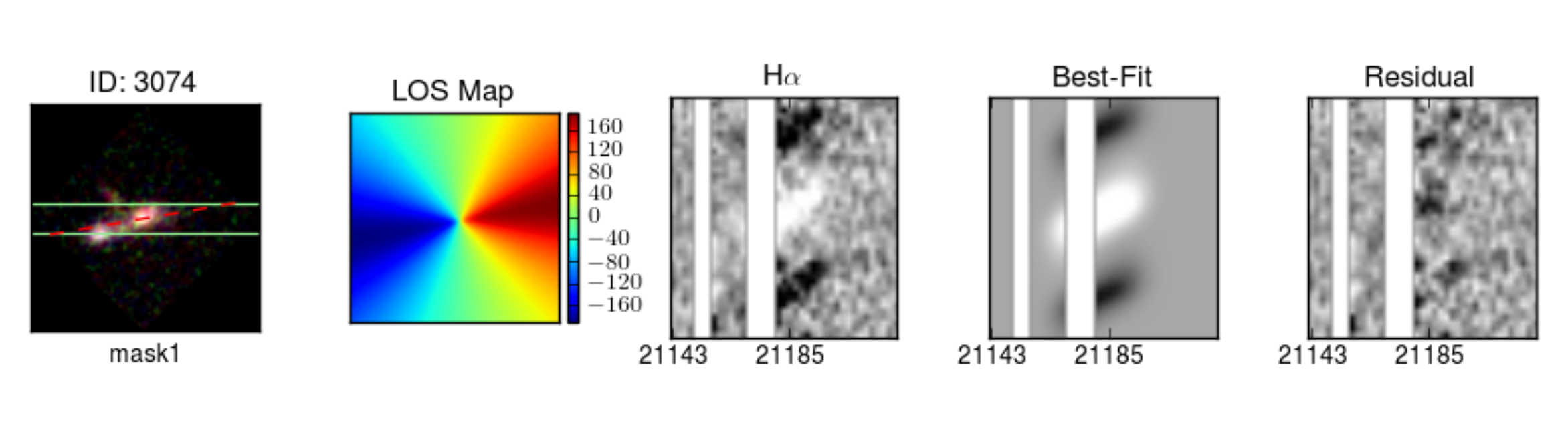}\\
 \includegraphics[width=\textwidth]{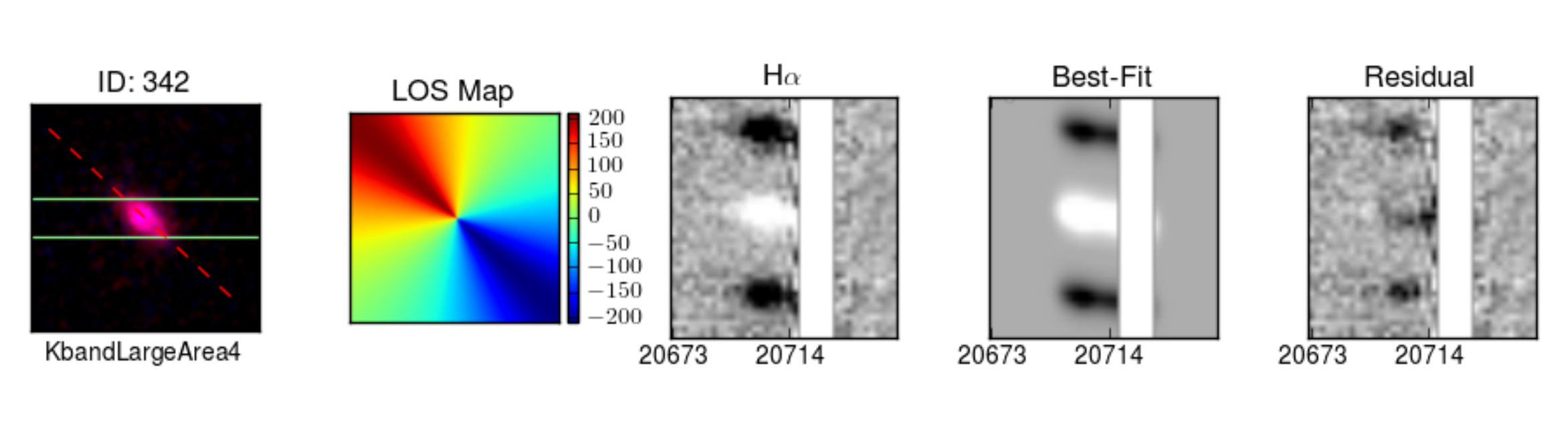}\\
 \includegraphics[width=\textwidth]{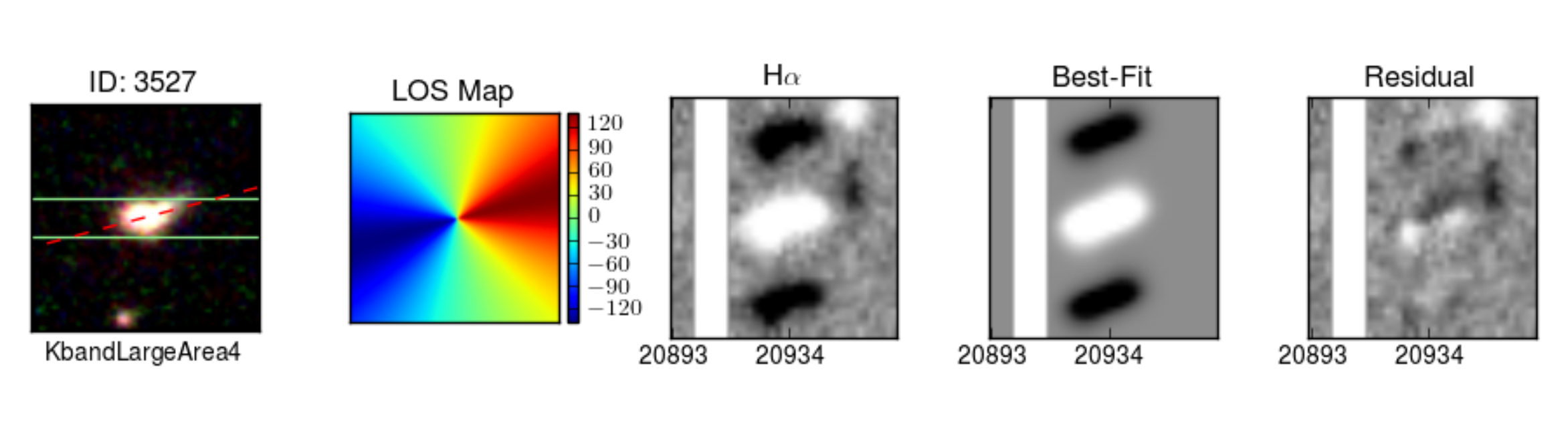}\\
       \end{tabular}
     \caption{Continued}
\end{figure}
\begin{figure}
   \centering
  \addtocounter{figure}{-1}
 \begin{tabular}{l}
\includegraphics[width=\textwidth]{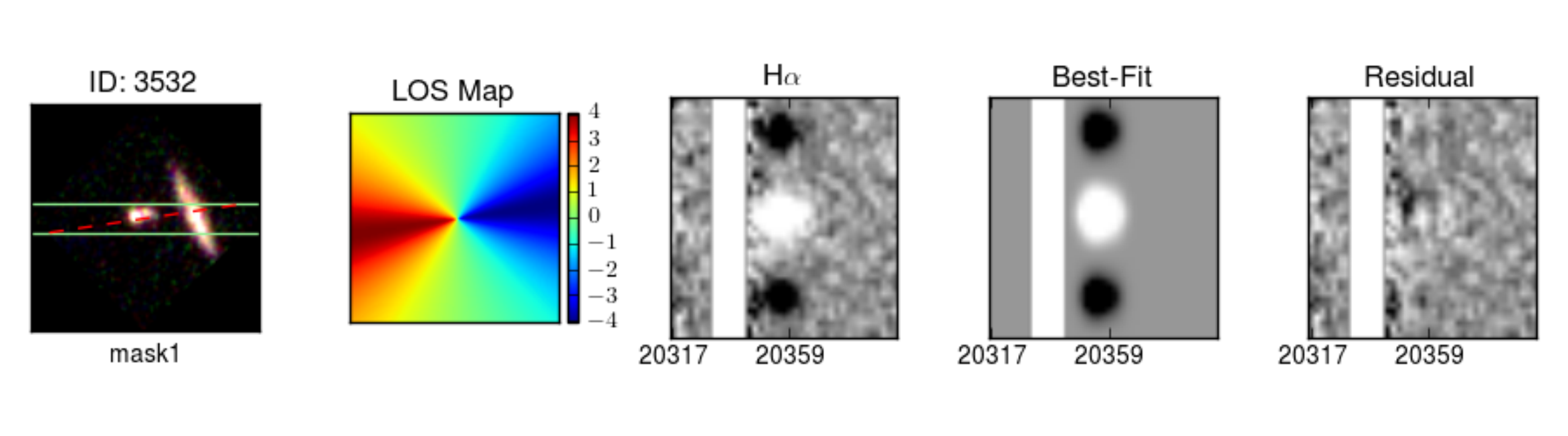}\\
 \includegraphics[width=\textwidth]{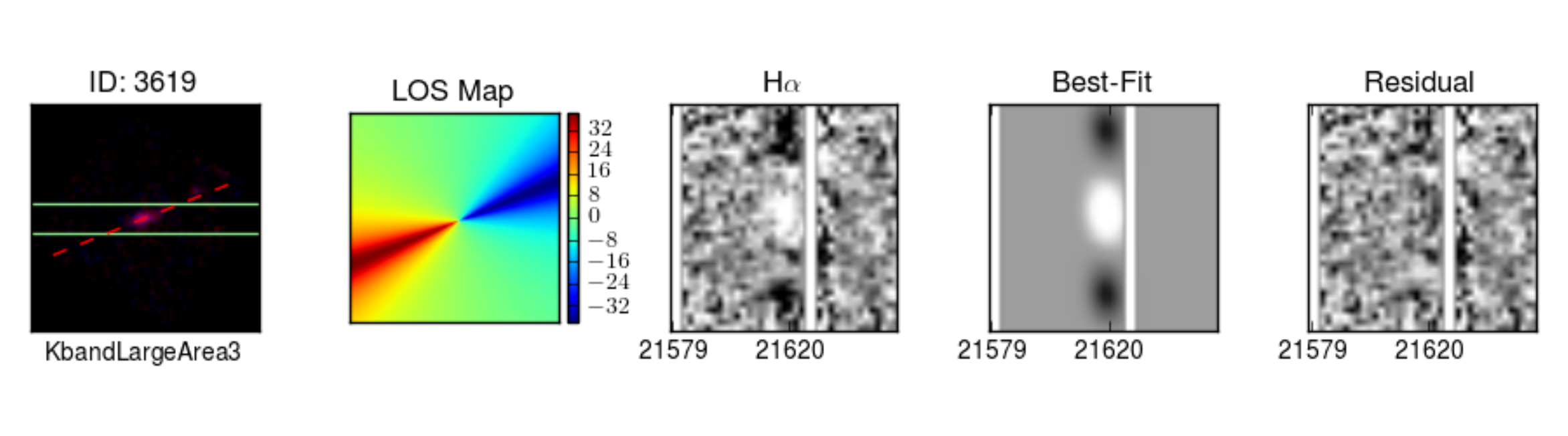}\\
  \includegraphics[width=\textwidth]{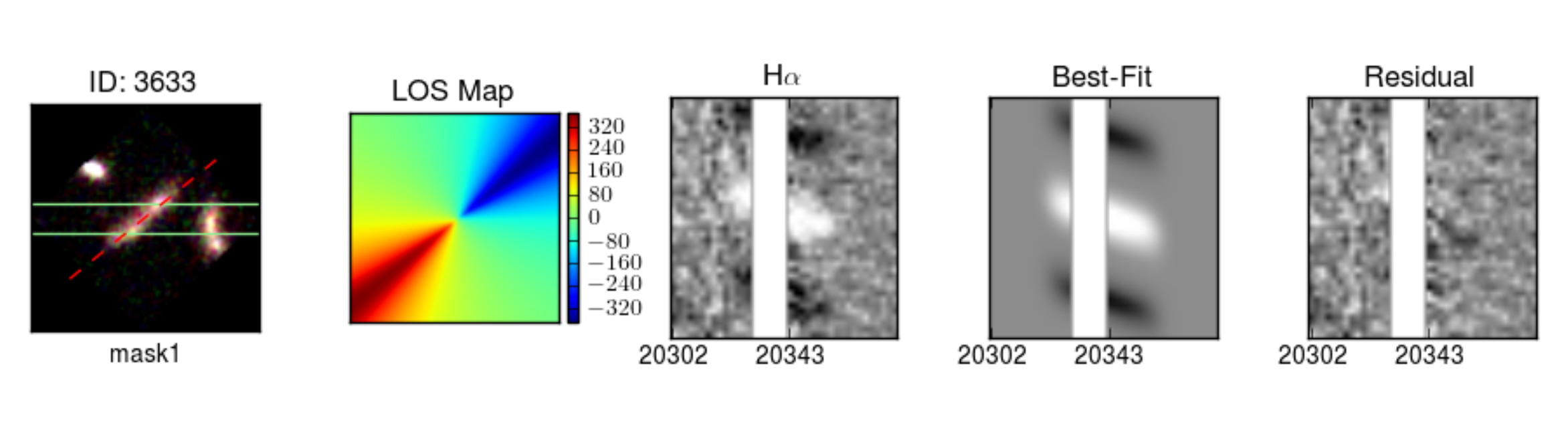}\\
  \includegraphics[width=\textwidth]{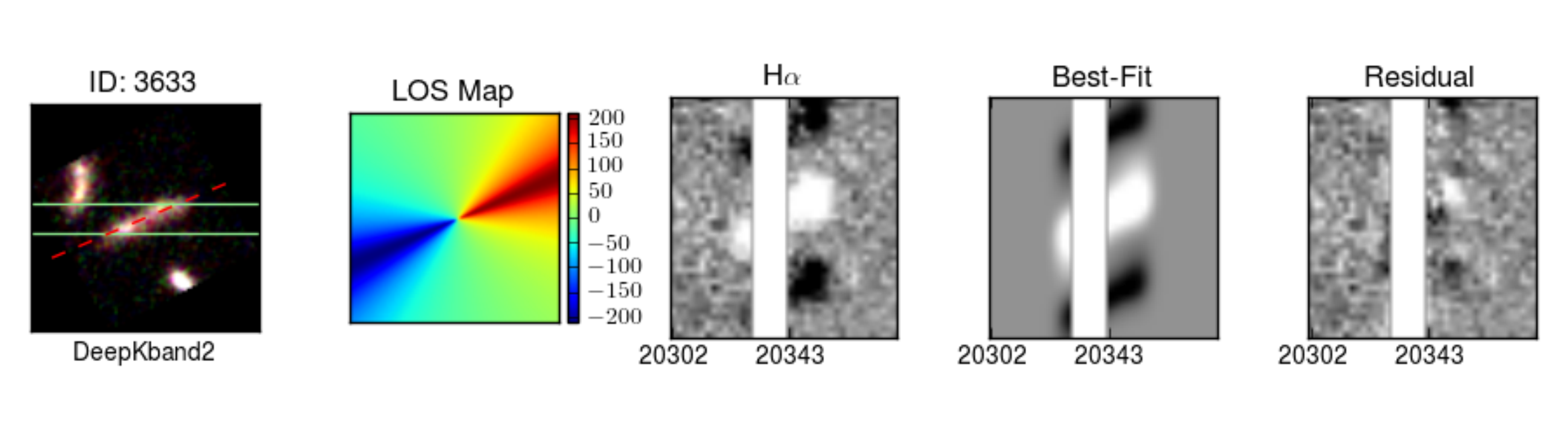}\\
       \end{tabular}
     \caption{Continued}
\end{figure}
\begin{figure}
   \centering
  \addtocounter{figure}{-1}
 \begin{tabular}{l}
 \includegraphics[width=\textwidth]{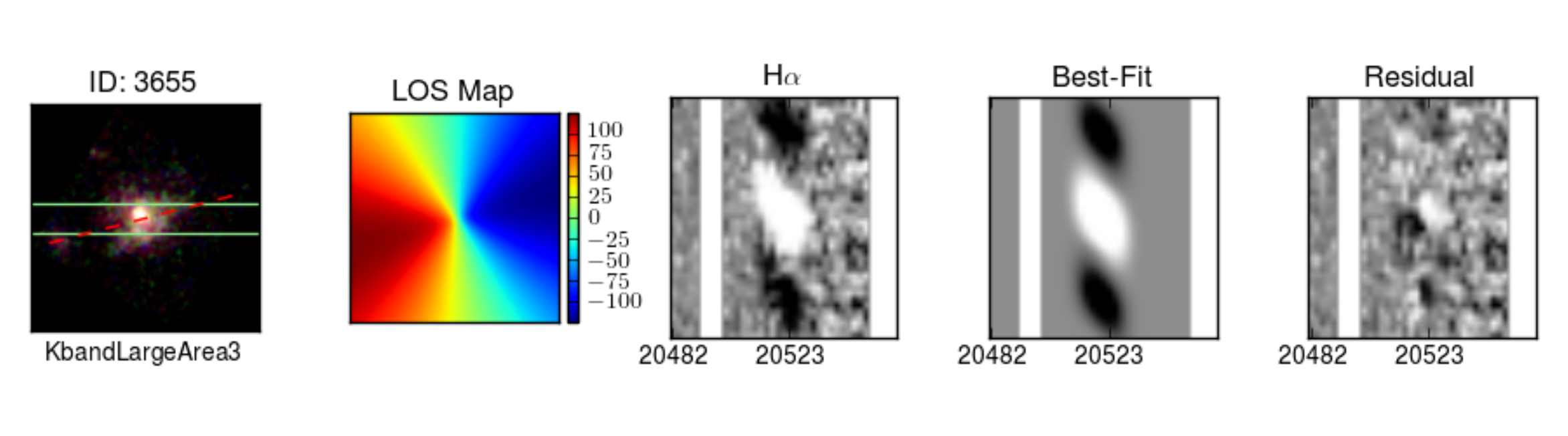}\\
 \includegraphics[width=\textwidth]{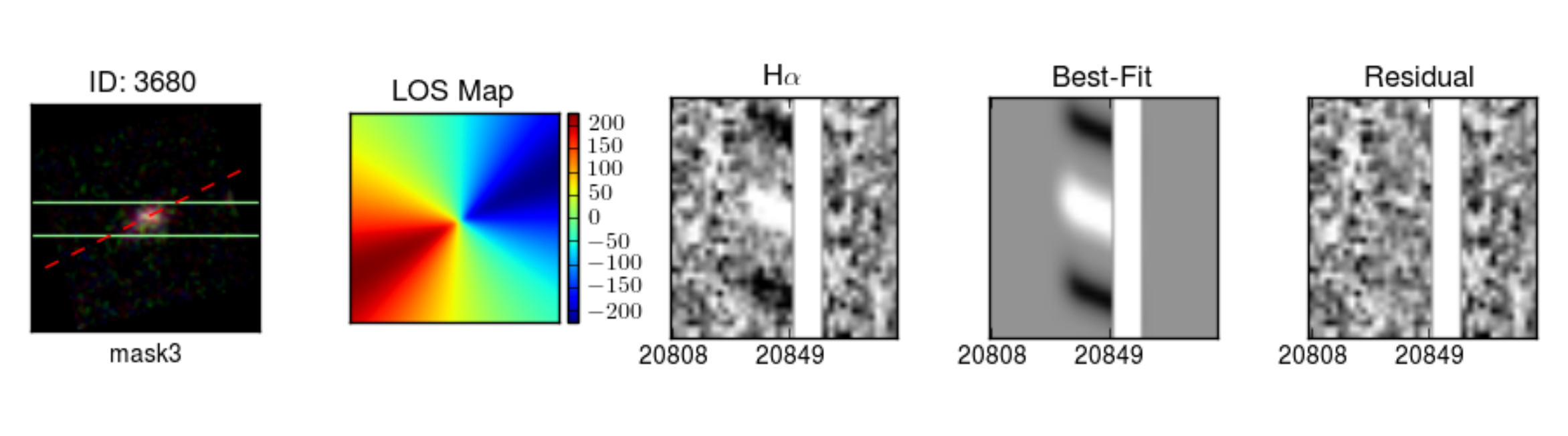}\\
 \includegraphics[width=\textwidth]{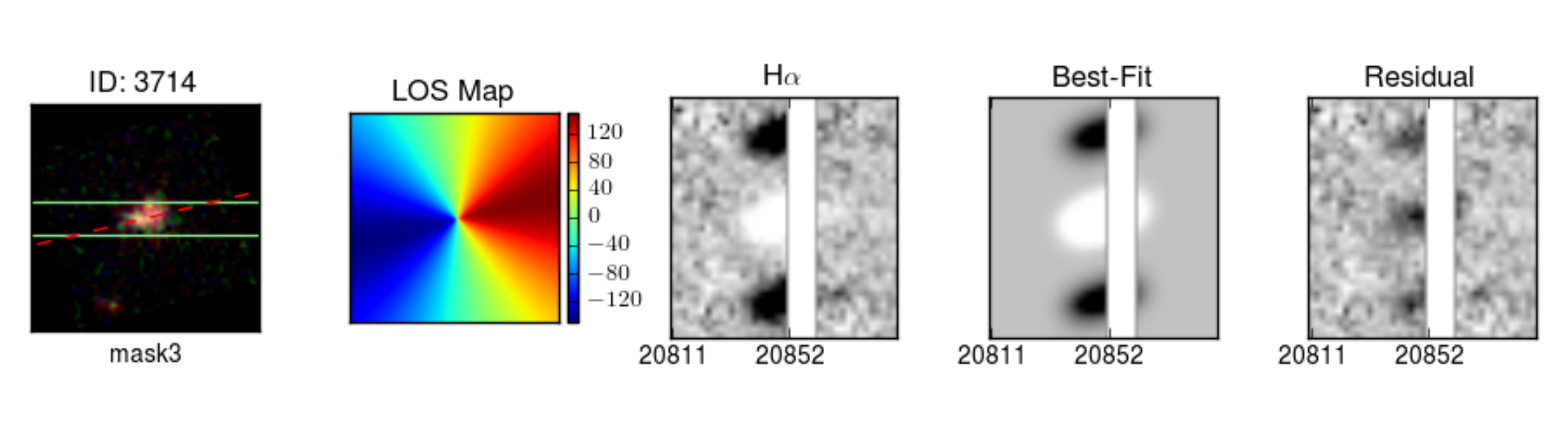}\\
 \includegraphics[width=\textwidth]{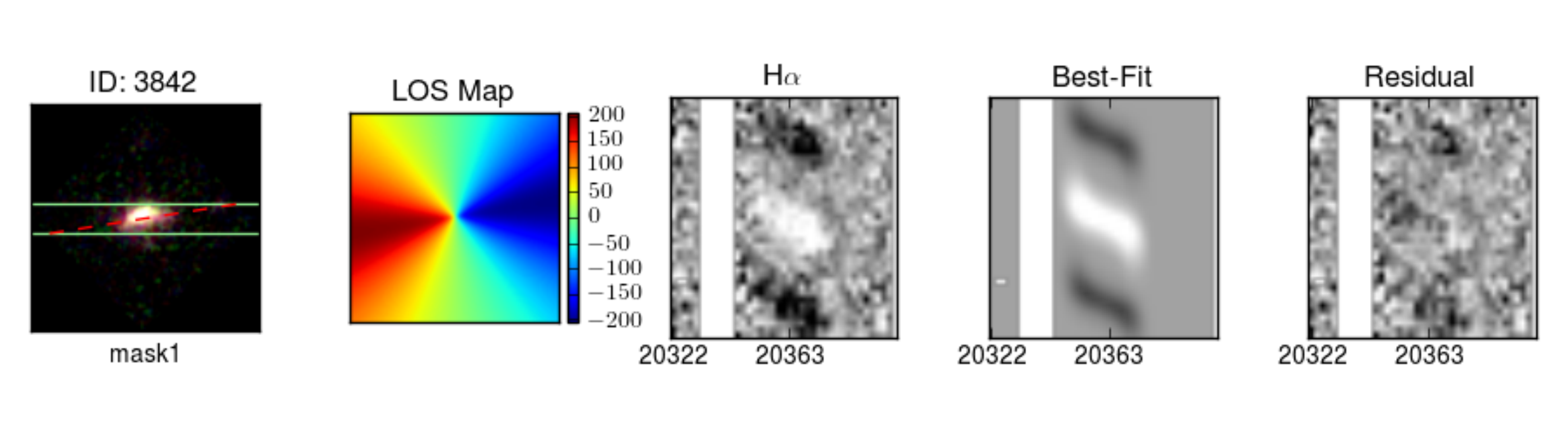}\\
       \end{tabular}
     \caption{Continued}
\end{figure}
\begin{figure}
   \centering
  \addtocounter{figure}{-1}
 \begin{tabular}{l}
\includegraphics[width=\textwidth]{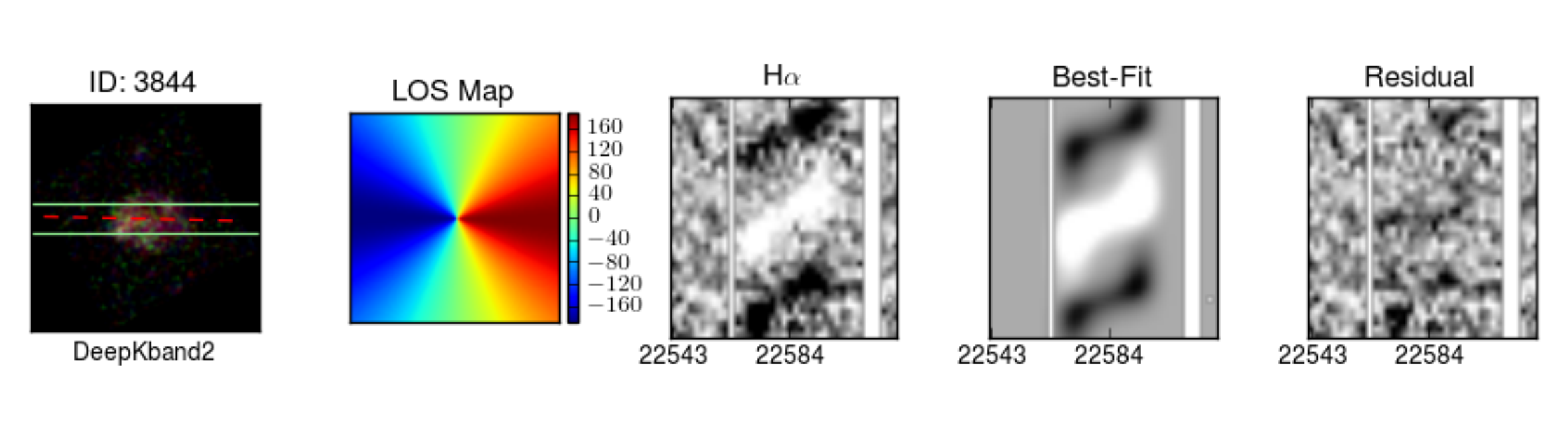}\\
\includegraphics[width=\textwidth]{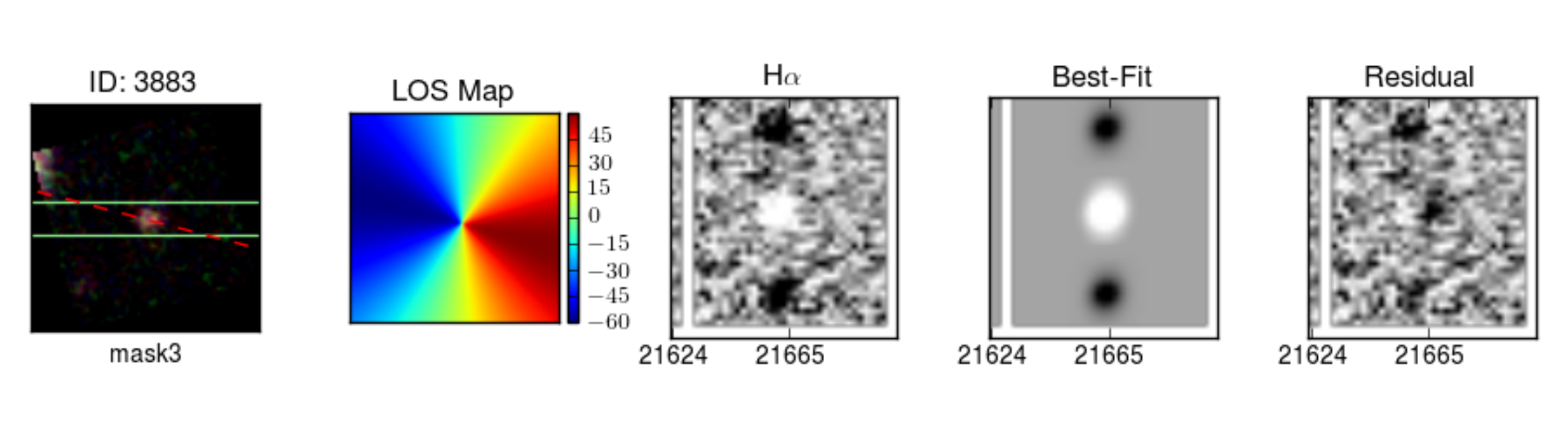}\\
\includegraphics[width=\textwidth]{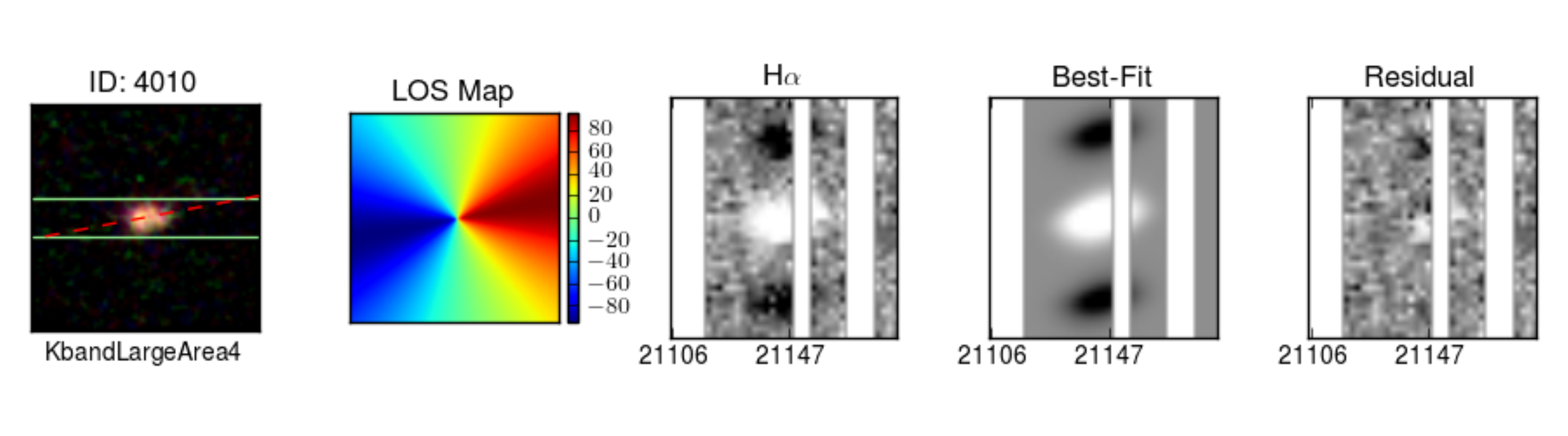}\\
\includegraphics[width=\textwidth]{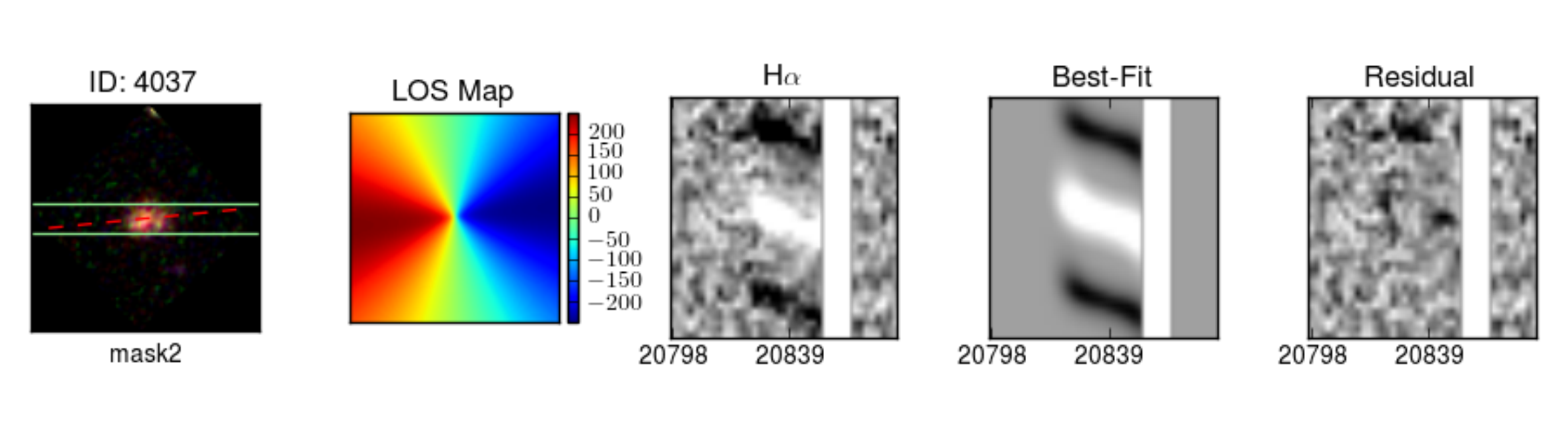}\\
       \end{tabular}
     \caption{Continued}
\end{figure}
\begin{figure}
   \centering
  \addtocounter{figure}{-1}
 \begin{tabular}{l}
\includegraphics[width=\textwidth]{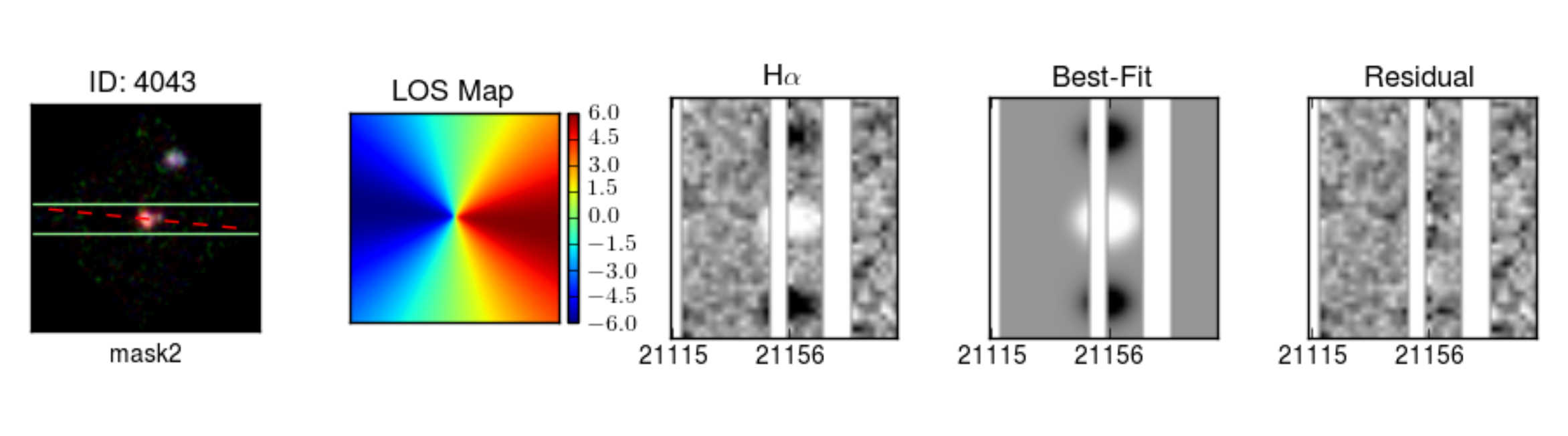}\\
 \includegraphics[width=\textwidth]{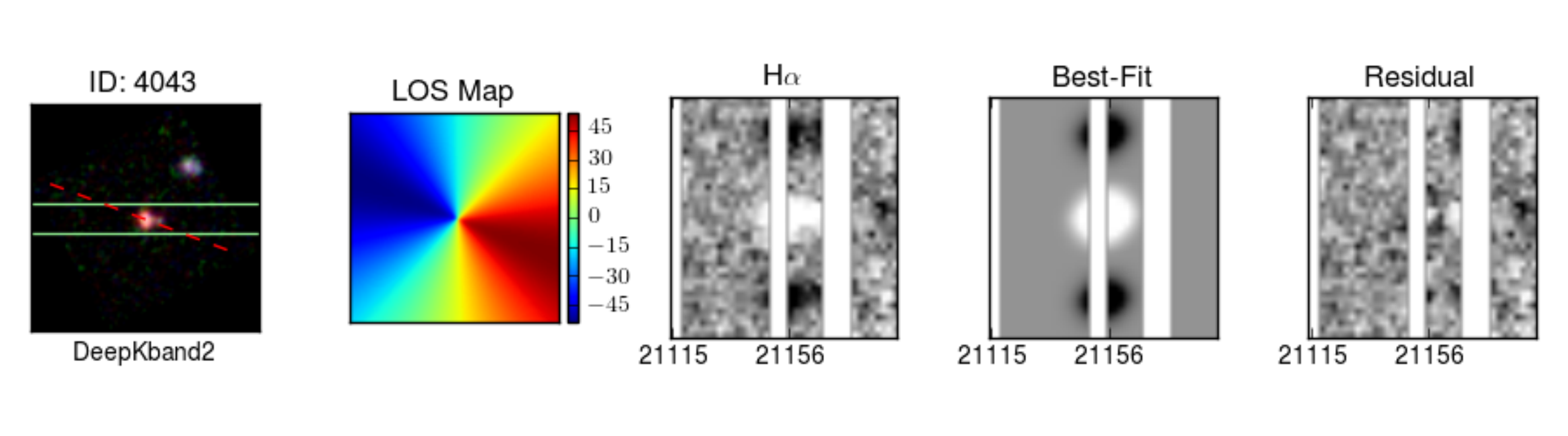}\\
  \includegraphics[width=\textwidth]{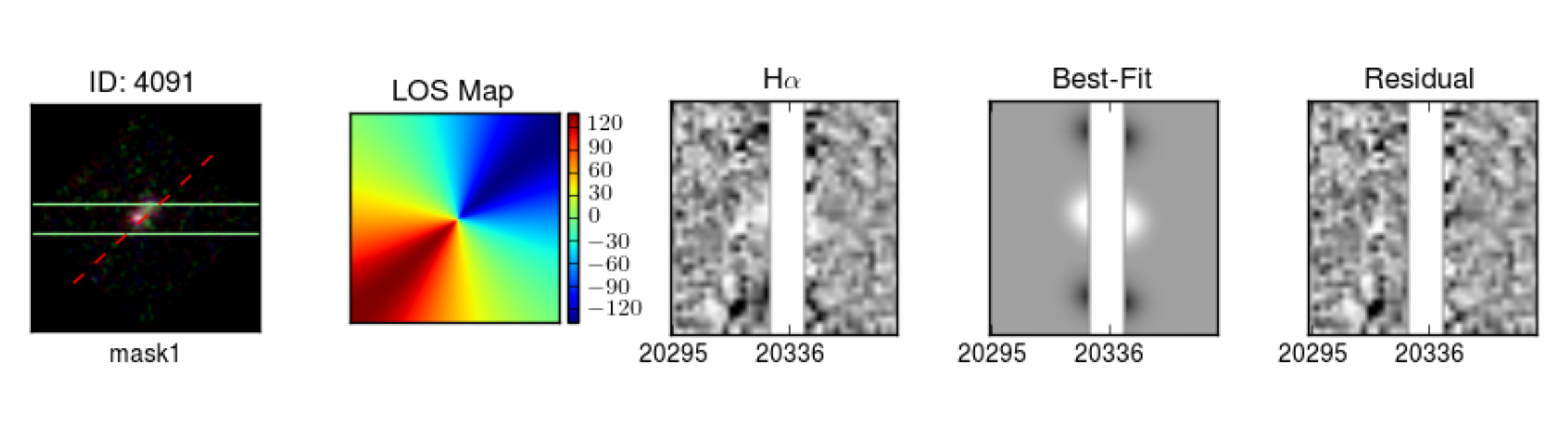}\\
 \includegraphics[width=\textwidth]{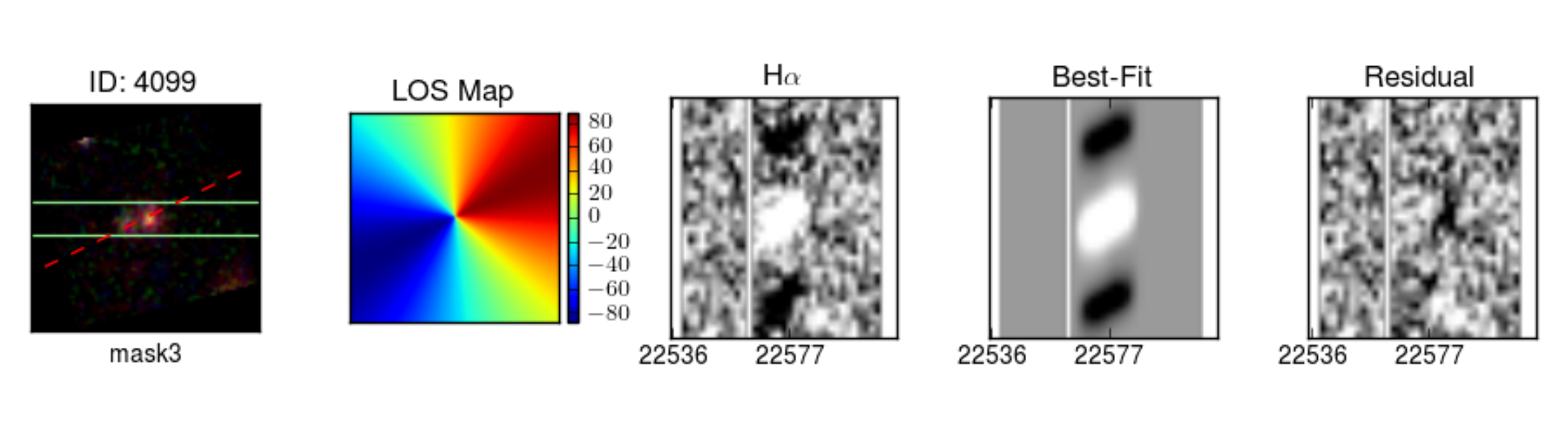}\\
       \end{tabular}
     \caption{Continued}
\end{figure}
\begin{figure}
   \centering
  \addtocounter{figure}{-1}
 \begin{tabular}{l}
\includegraphics[width=\textwidth]{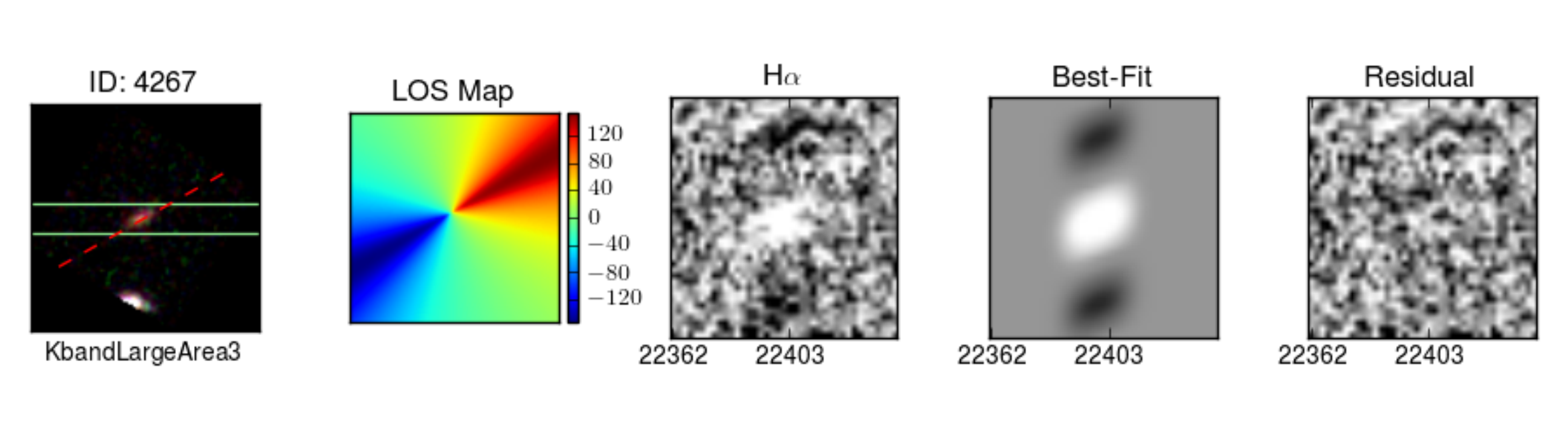}\\
 \includegraphics[width=\textwidth]{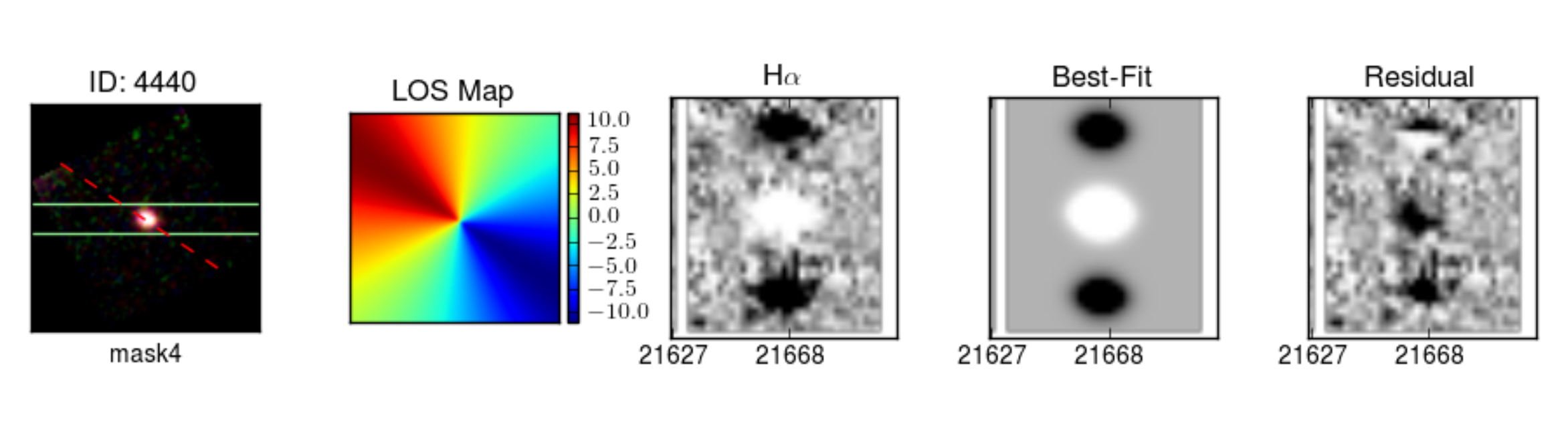}\\
  \includegraphics[width=\textwidth]{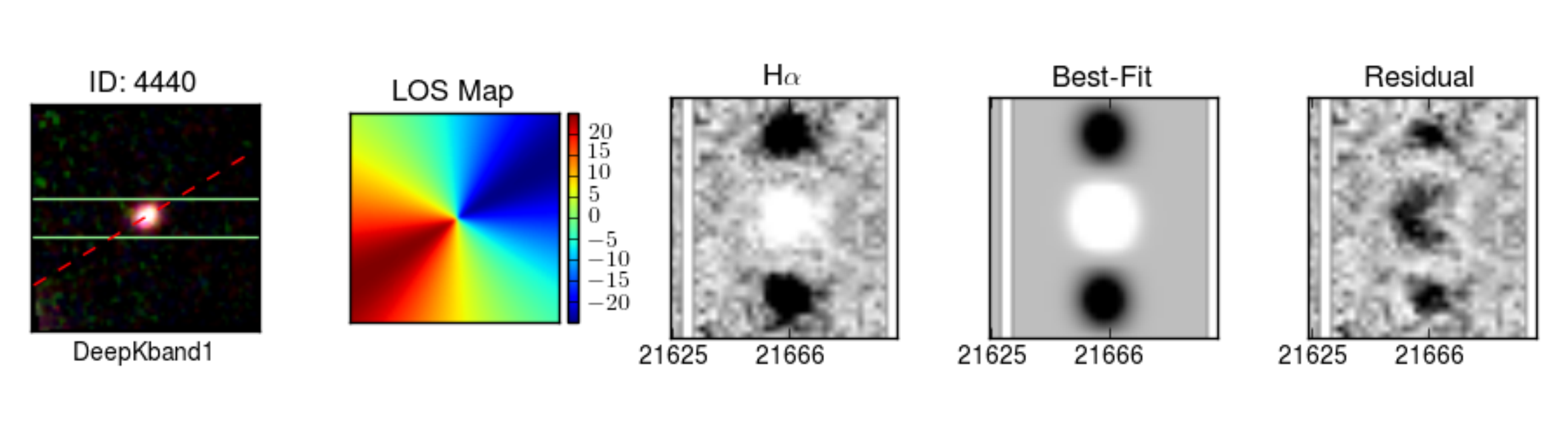}\\
 \includegraphics[width=\textwidth]{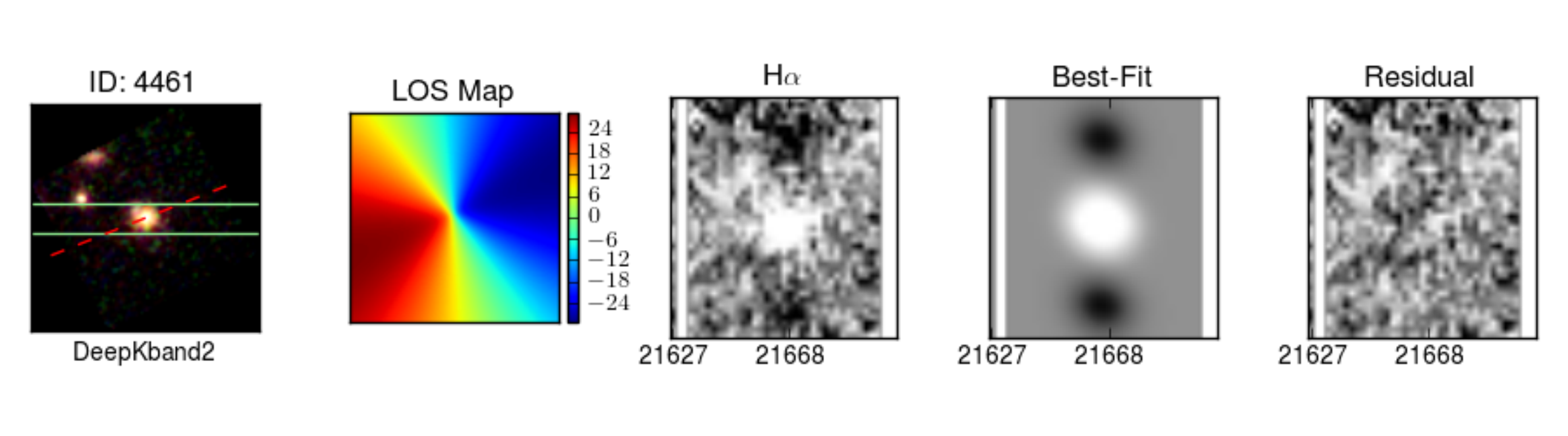}\\
       \end{tabular}
     \caption{Continued}
\end{figure}
\begin{figure}
   \centering
  \addtocounter{figure}{-1}
 \begin{tabular}{l}
\includegraphics[width=\textwidth]{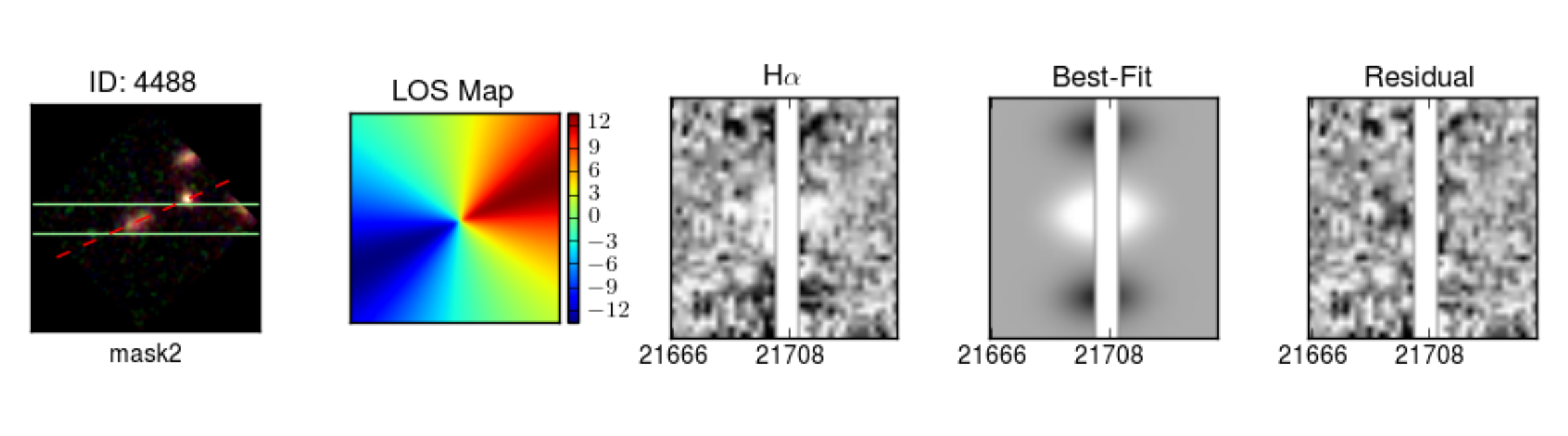}\\
 \includegraphics[width=\textwidth]{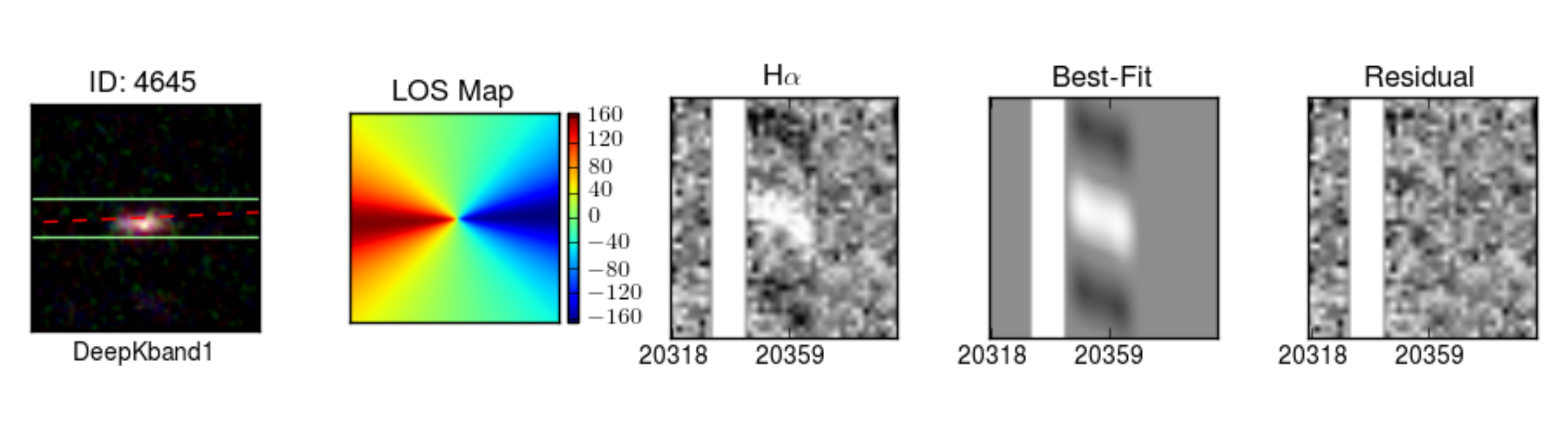}\\
 \includegraphics[width=\textwidth]{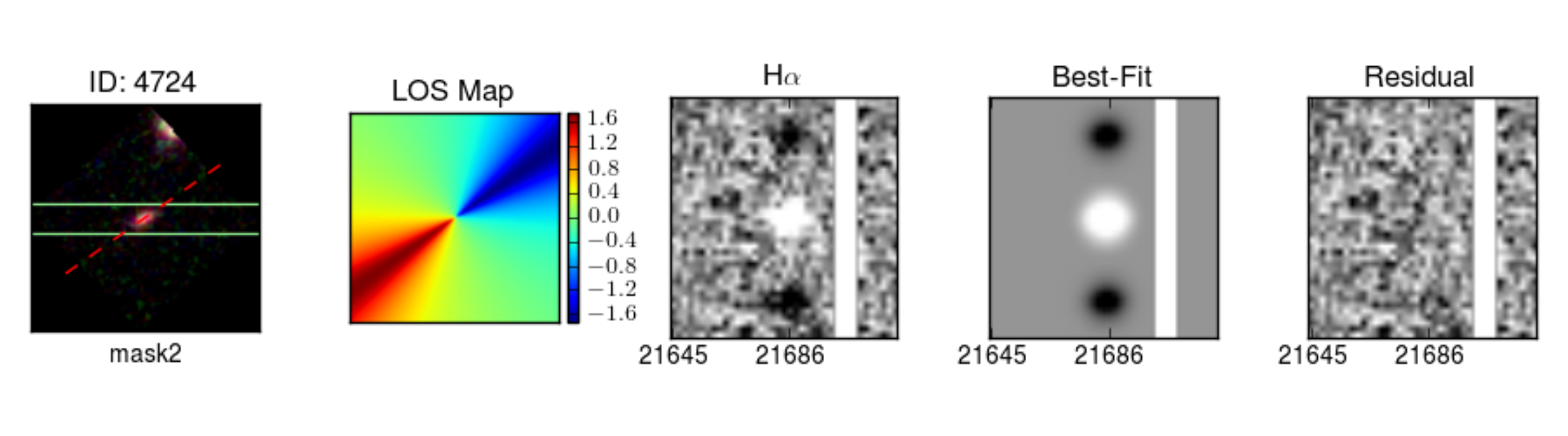}\\
 \includegraphics[width=\textwidth]{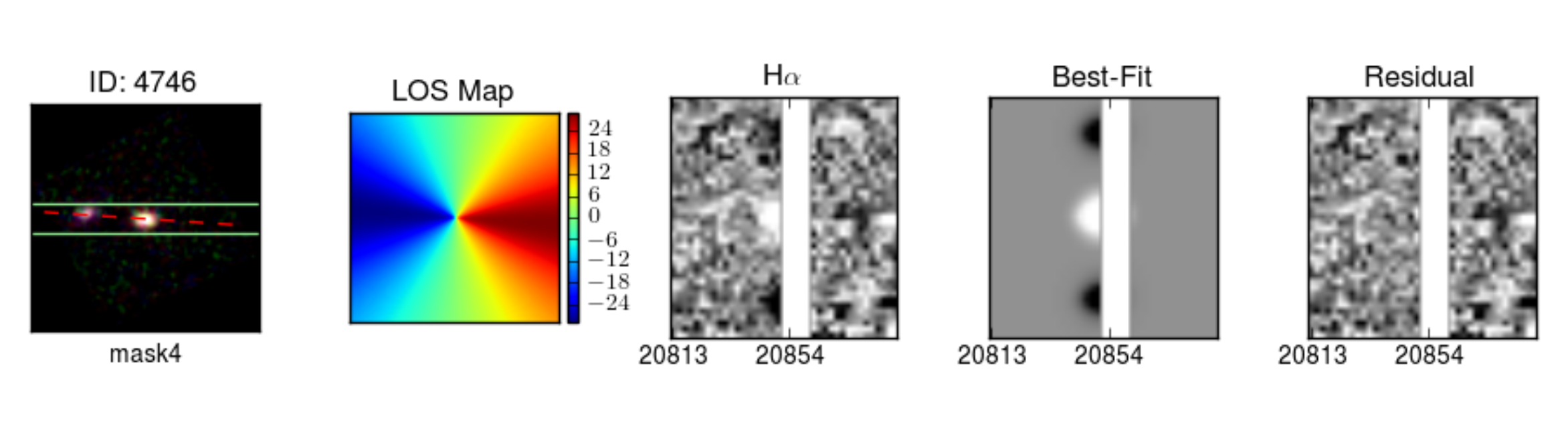}\\
       \end{tabular}
     \caption{Continued}
\end{figure}
\begin{figure}
   \centering
  \addtocounter{figure}{-1}
 \begin{tabular}{l}
 \includegraphics[width=\textwidth]{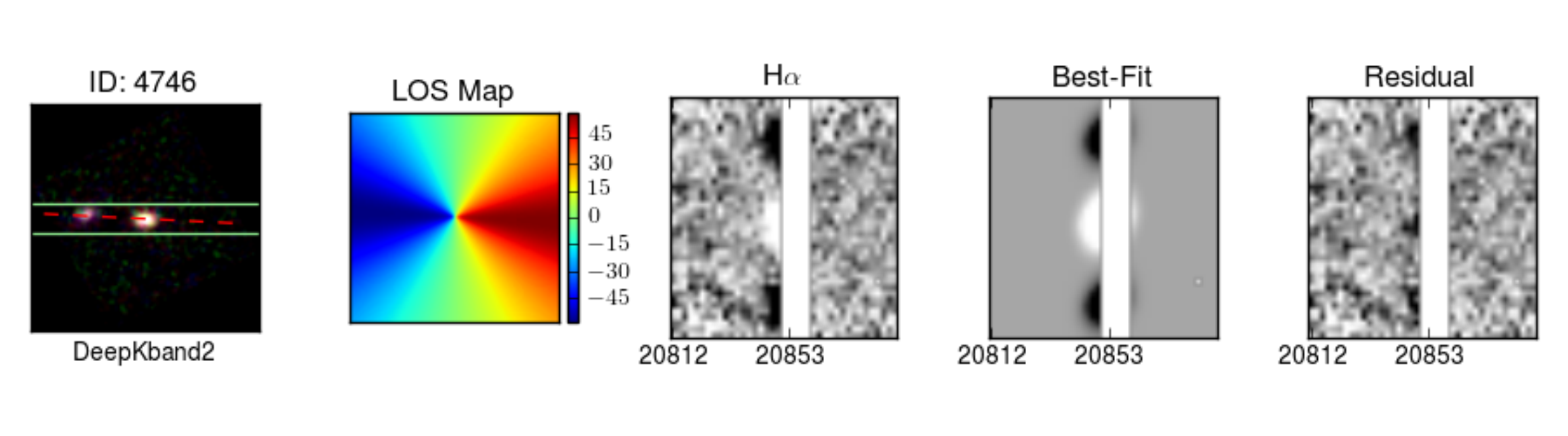}\\
 \includegraphics[width=\textwidth]{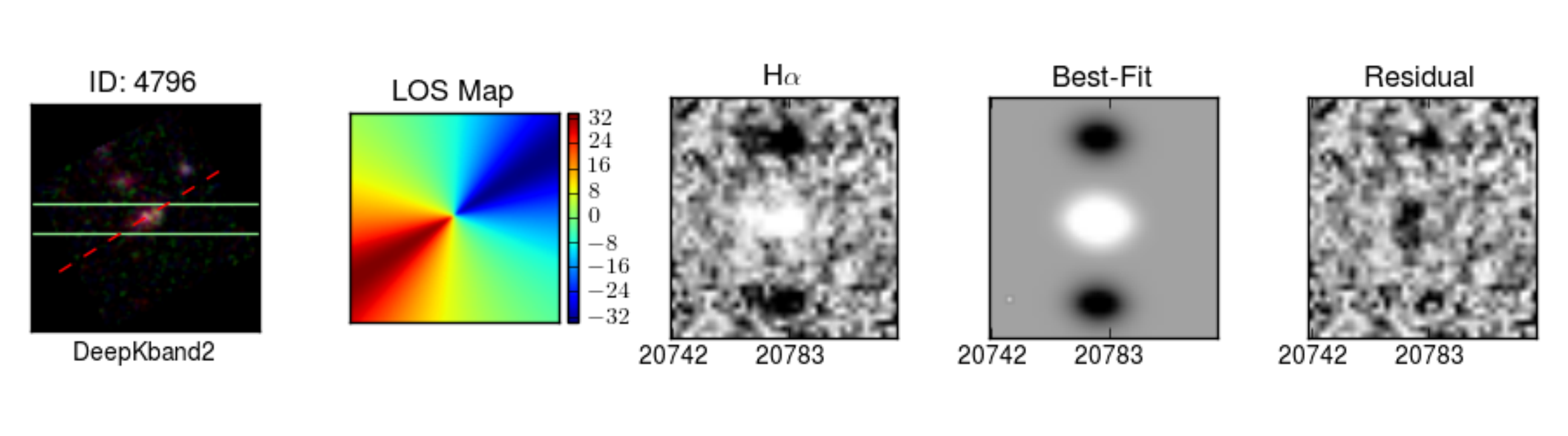}\\
 \includegraphics[width=\textwidth]{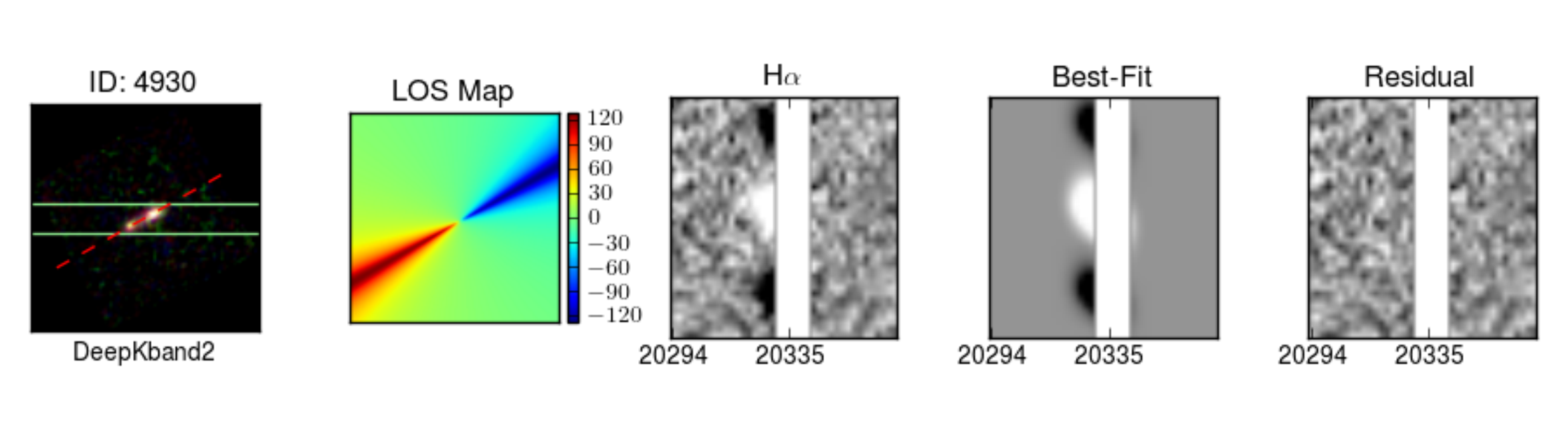}\\
 \includegraphics[width=\textwidth]{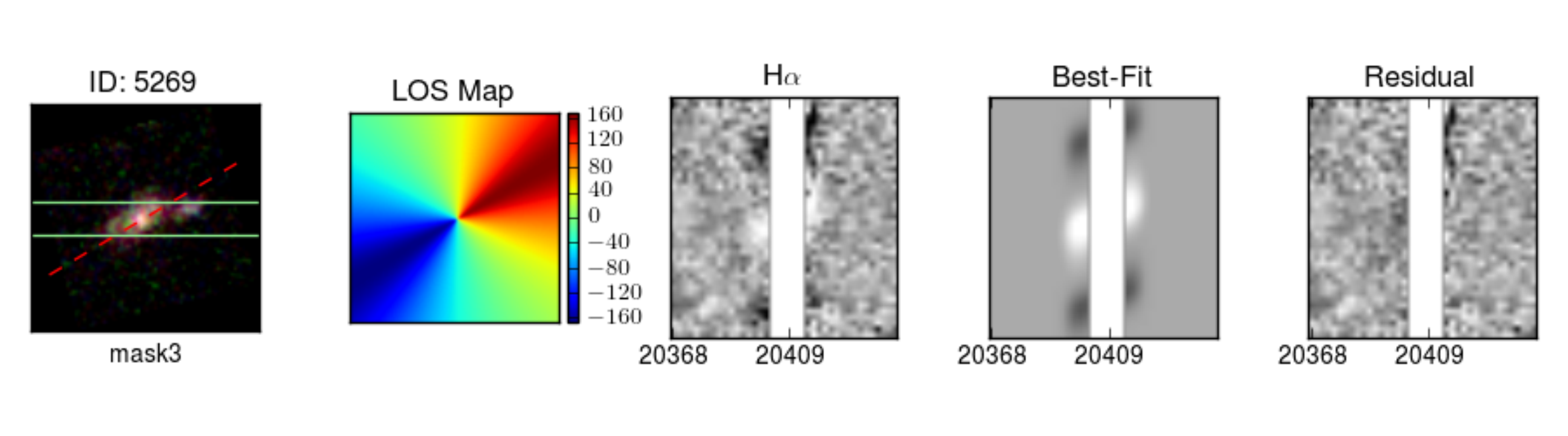}\\
       \end{tabular}
     \caption{Continued}
\end{figure}
\begin{figure}
   \centering
  \addtocounter{figure}{-1}
 \begin{tabular}{l}
\includegraphics[width=\textwidth]{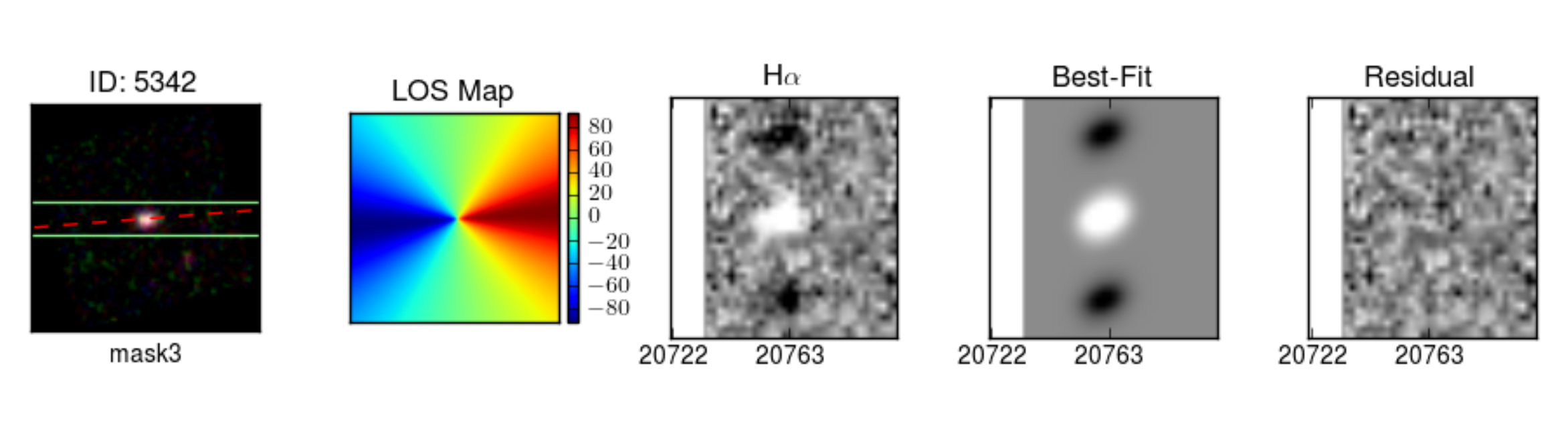}\\
\includegraphics[width=\textwidth]{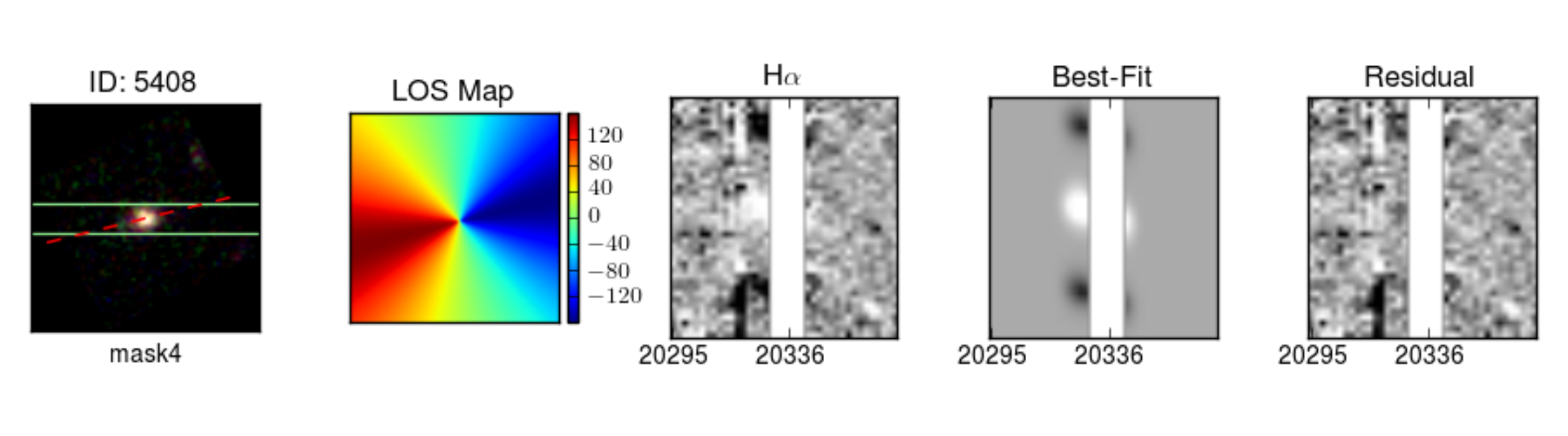}\\
\includegraphics[width=\textwidth]{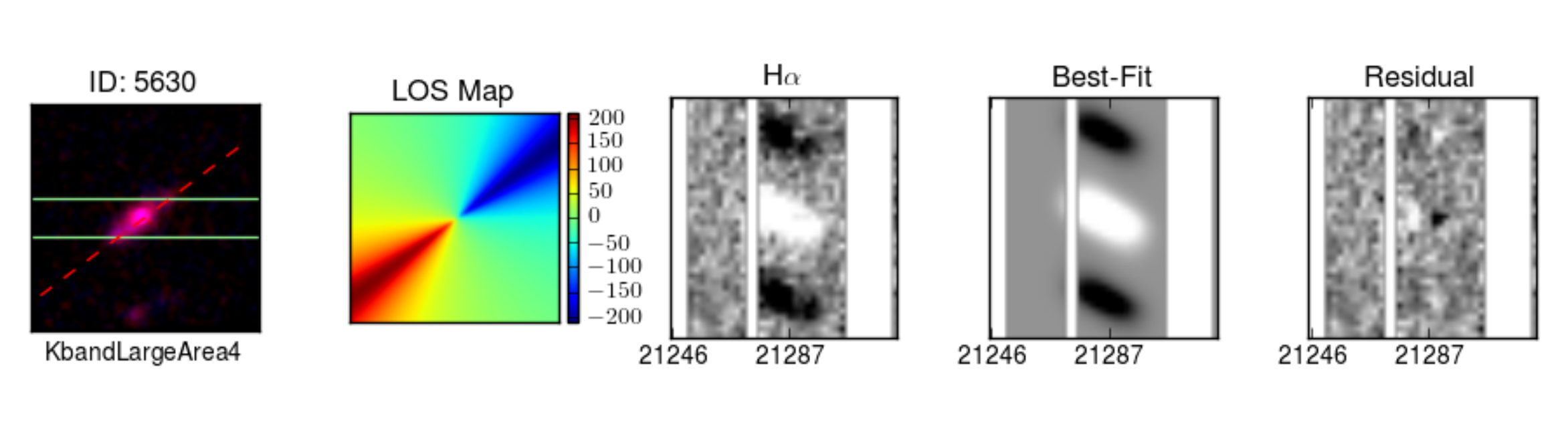}\\
\includegraphics[width=\textwidth]{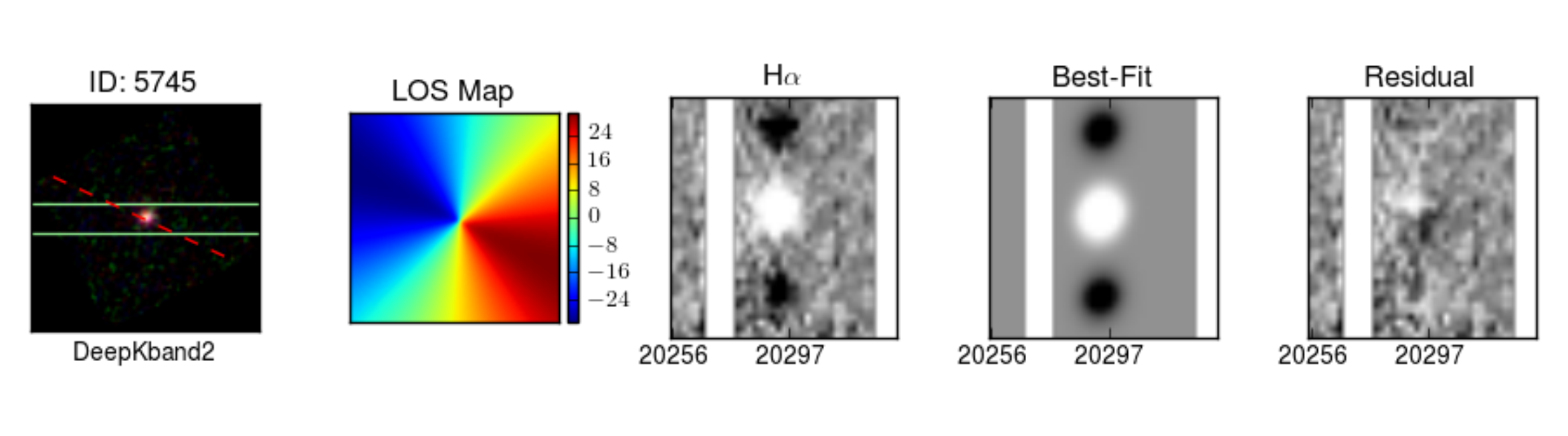}\\
       \end{tabular}
     \caption{Continued}
\end{figure}
\begin{figure}
   \centering
  \addtocounter{figure}{-1}
 \begin{tabular}{l}
\includegraphics[width=\textwidth]{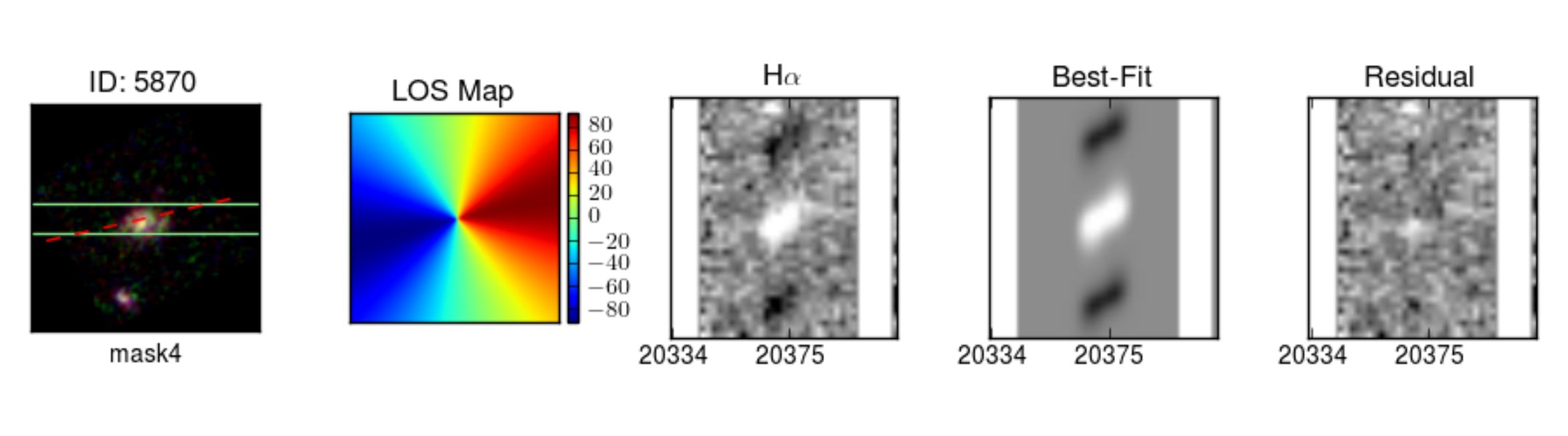}\\
\includegraphics[width=\textwidth]{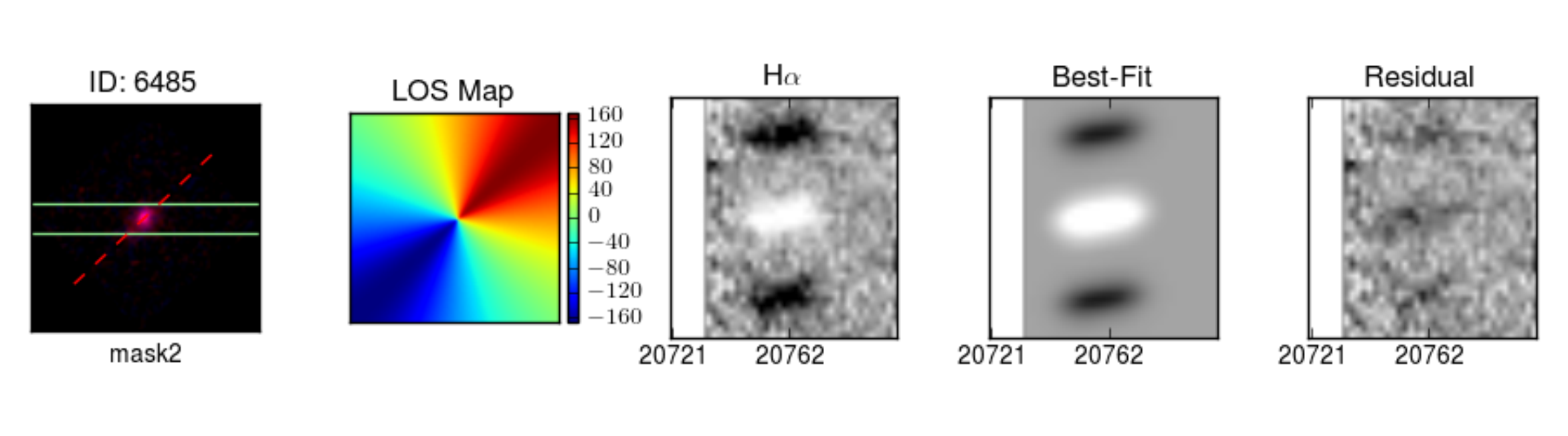}\\
\includegraphics[width=\textwidth]{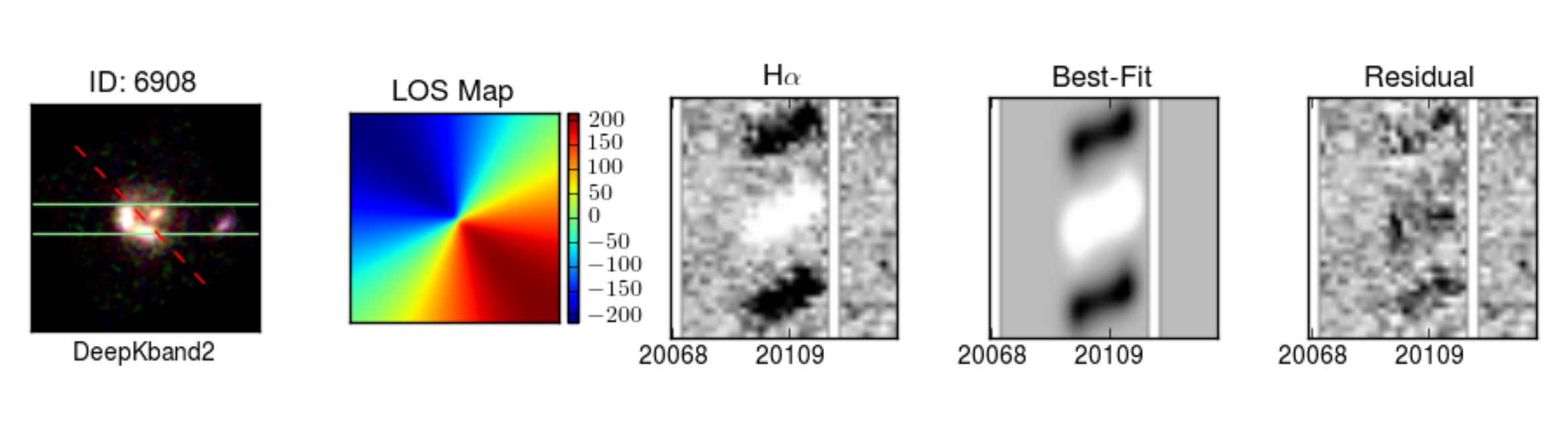}\\
\includegraphics[width=\textwidth]{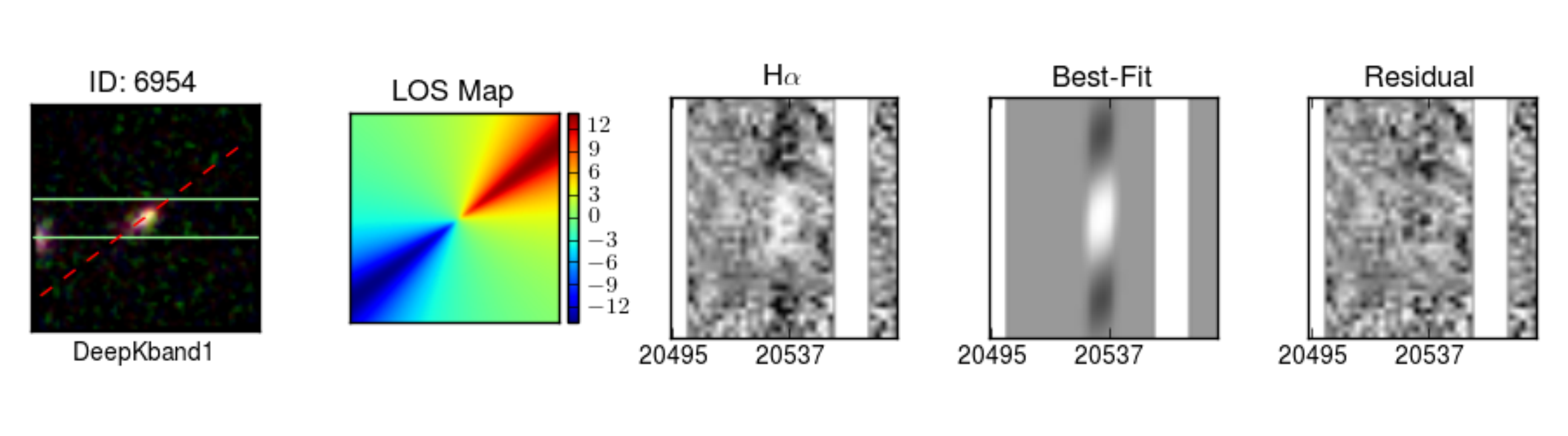}\\
       \end{tabular}
     \caption{Continued}
\end{figure}
\begin{figure}
   \centering
  \addtocounter{figure}{-1}
 \begin{tabular}{l}
\includegraphics[width=\textwidth]{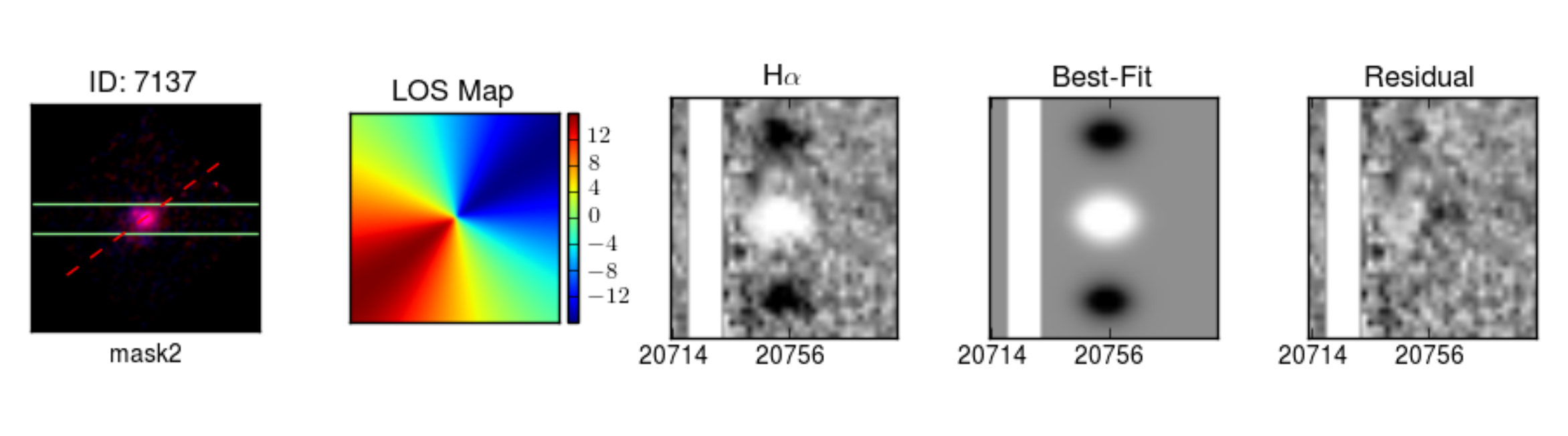}\\
\includegraphics[width=\textwidth]{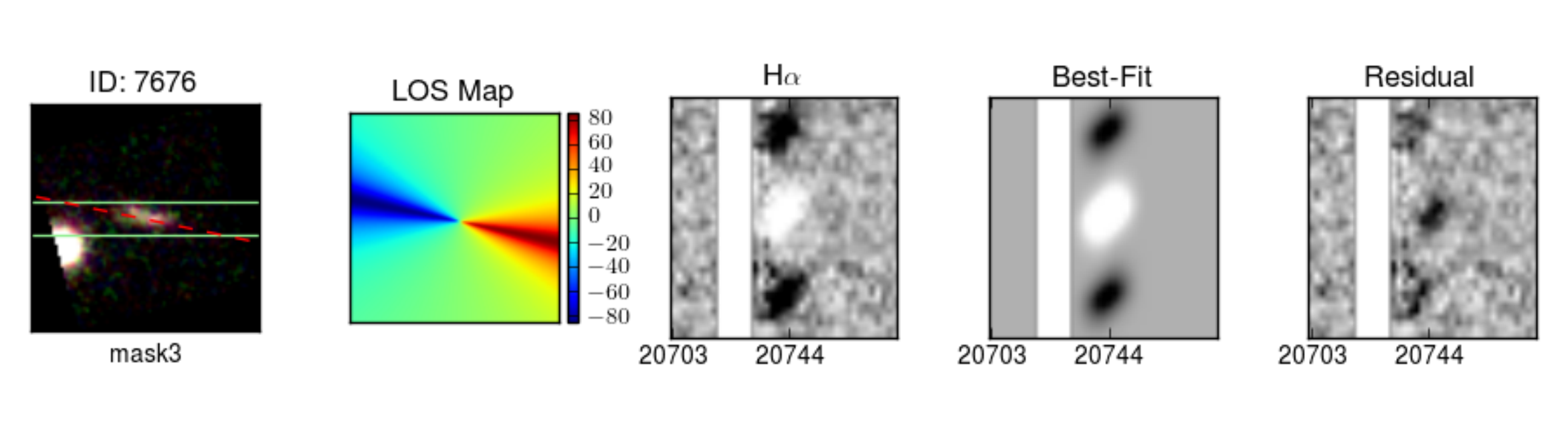}\\
\includegraphics[width=\textwidth]{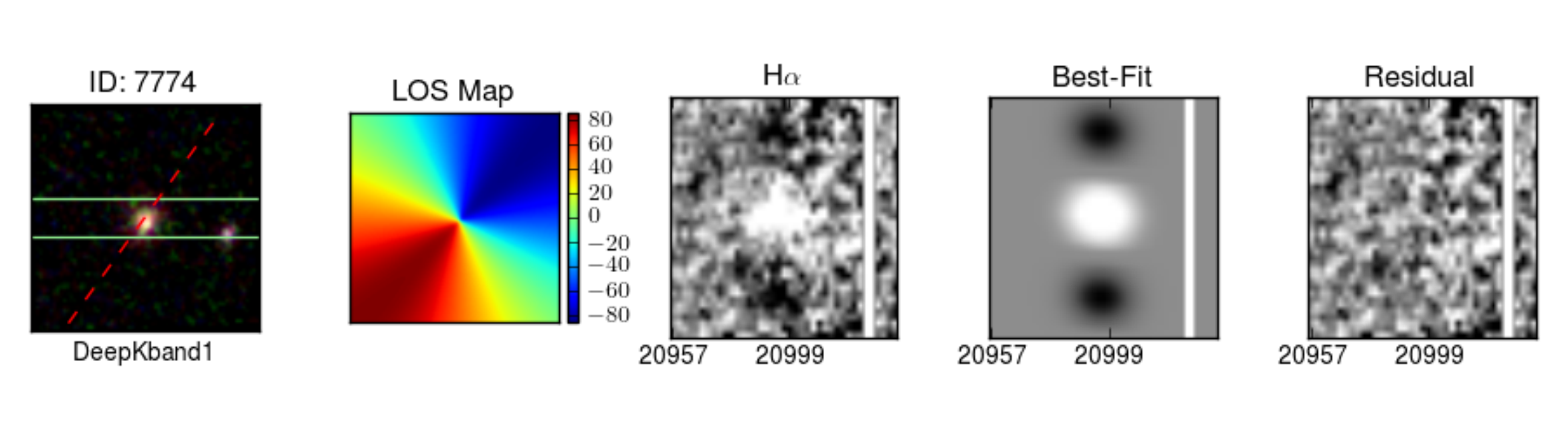}\\
\includegraphics[width=\textwidth]{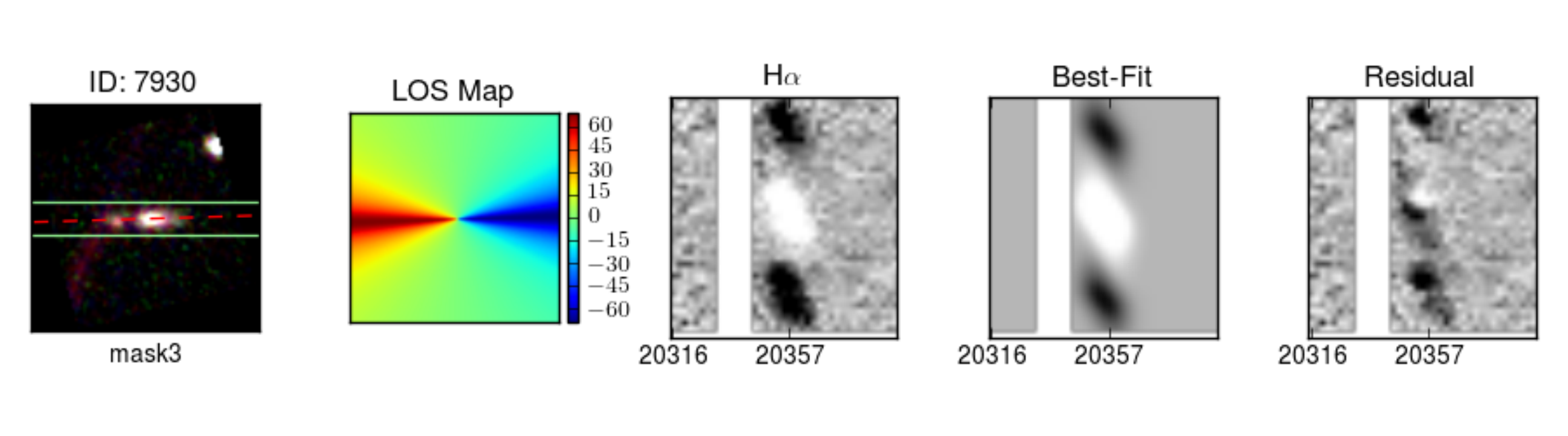}\\
       \end{tabular}
     \caption{Continued}
\end{figure}
\begin{figure}
   \centering
  \addtocounter{figure}{-1}
 \begin{tabular}{l}
\includegraphics[width=\textwidth]{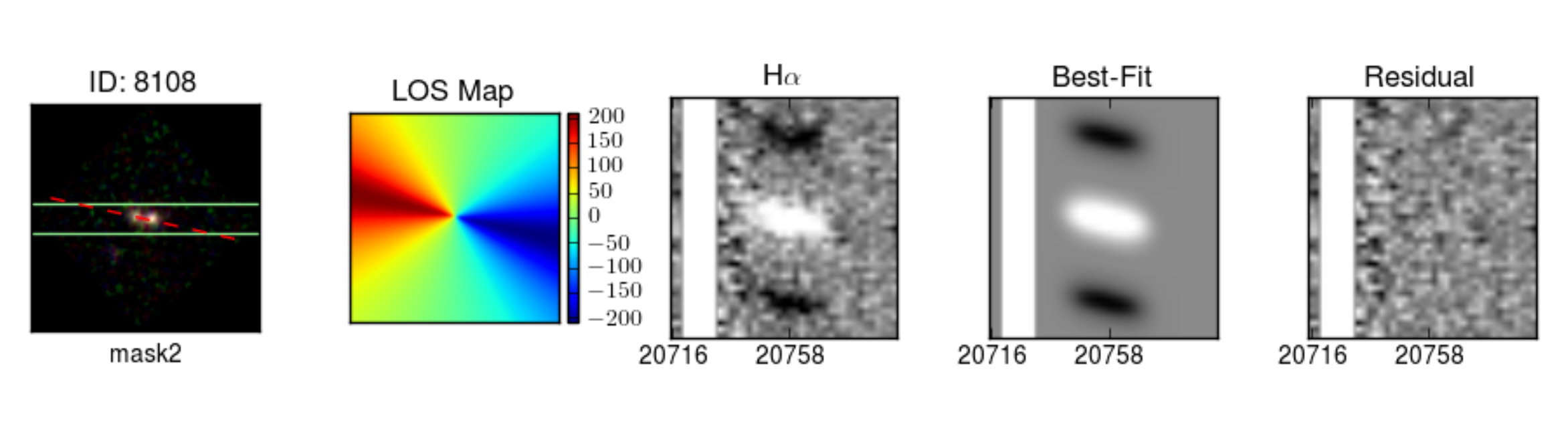}\\
\includegraphics[width=\textwidth]{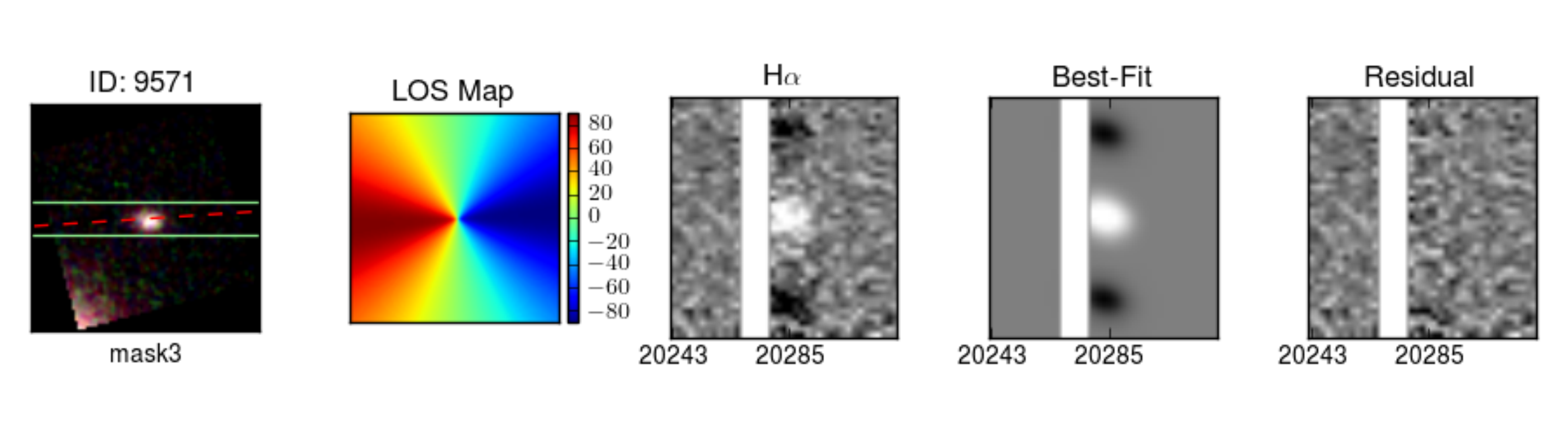}\\
       \end{tabular}
     \caption{Continued}
\end{figure}

\section{Fitting Simulated Emission Lines} \label{sec:simfitting}

We test our fitting procedure on a sample set of simulated MOSFIRE observations.
We use 1000 simulated emission lines of galaxies created from the GBKFIT program \citep{Bekiaris2015}. 
Three examples of these simulated emission lines are in Figure 4.

\begin{figure}[h] \label{fig:specexamples}
\centering
\figurenum{10}
\includegraphics[width=0.4\textwidth]{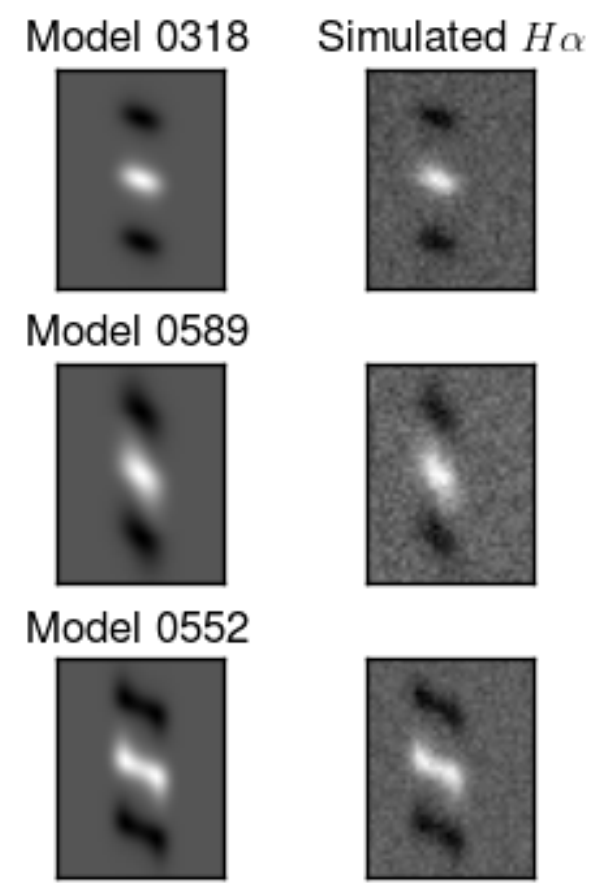}
	\caption{Examples of models used in our model library. Left column: Models from GBKFIT with 1.25$\arcsec$ dither patter. Right column: Models with low MOSFIRE-level sky noise added, with no sky emission. These are examples of our simulated observations, used to test the effectiveness of our method.}
\end{figure}
GBKFIT creates simulated 3D data cubes of galaxies given initial properties such as galaxy redshift ($z$), scale length $r_s$ (1-5 kpc), turnover radius $r_t$ ($\frac{r_s}{3}$), turnover velocity $V_t$ (100-400 \kms), gas sigma \sigmag\ (20-100 \kms), inclination $i$  (0-90$^{o}$), and offset from the PA of the slit $\Delta\alpha$ (-45-45$^{o}$). 
Galaxies are all infinitely thin exponential disks with arctangent rotation curves, 
\begin{equation} \label{eq:vcurve}
	V_{rot}(r)=\frac{2}{\pi}V_t\ arctan \frac{r}{r_t}.
\end{equation}
All objects have a constant intrinsic gas velocity dispersion. 
These models are convolved with the desired seeing and projected through a  0.7$\arcsec$ wide slit. 
In this case, we used 2D Moffat at 0.7$\arcsec$ seeing and $\beta=2.5$. 
The values of these properties in our sample span the range of possible values in all cases, providing a diverse sample of disk galaxies, with $\frac{v}{\sigma}=1-20$.

We measure pixel-to-pixel RMS from 2D MOSFIRE K-band observations and add simulated sky noise to each model (Figure 3). We do not simulate a continuum. Scale models to the sky noise to create mock observations at varying signal-to-noise (SNR) values (from SNR=5-60). 
If part of the line is masked from simulated sky emission, the SNR drops depending on the amount of line coverage. The SNR was calculated by summing all pixels of the spectrum within defined limits and dividing by the summed squares of the equivalent pixels in the corresponding noise spectrum. This region was defined as within $5r_s$ and 1.26$\arcsec$ of the center of the object, and within $3FWHM$ of the emission line.

\subsection{The Effects of SNR and Masking Sky Emission}

When masking sky emission, we do not perform any operations on masked pixels. 
The fraction of pixels masked does affect recovery rates of our input models, and through our simulations we have found that if more than half of the emission line is masked at any SNR, we underestimate our input \vtwo\ by 12\% at half masked to 83\% at 80 - 100\% masked (Figure 4, Row 2, far right).
Similar results are found in \sigmag\ recovery: at 50\% masked, we tend to overestimate \sigmag\ by 20\%, increasing to up to 70\% overestimated at 80 - 100\% masked (Figure 4, Row 4, far right).
SNR correlates with recovery as well, although less significantly.
At SNR $>$ 10, we overestimate \vtwo\ by $\sim10$\% at a 20\% scatter, and at lower SNR we find the scatter to increase to $\sim$ 70\% (Figure 4, Row 1, far right).
For \sigmag\ recovery, we find at SNR $>$ 10, we tend to underestimate \sigmag\ by 10\% at a scatter of 15\%, and at lower SNR the scatter can increase to $\sim$70\% (Figure 4, Row 3, far right).

\begin{figure*}[h] \label{fig:simrecover}
\centering
\figurenum{11}
\includegraphics[width=0.8\textwidth]{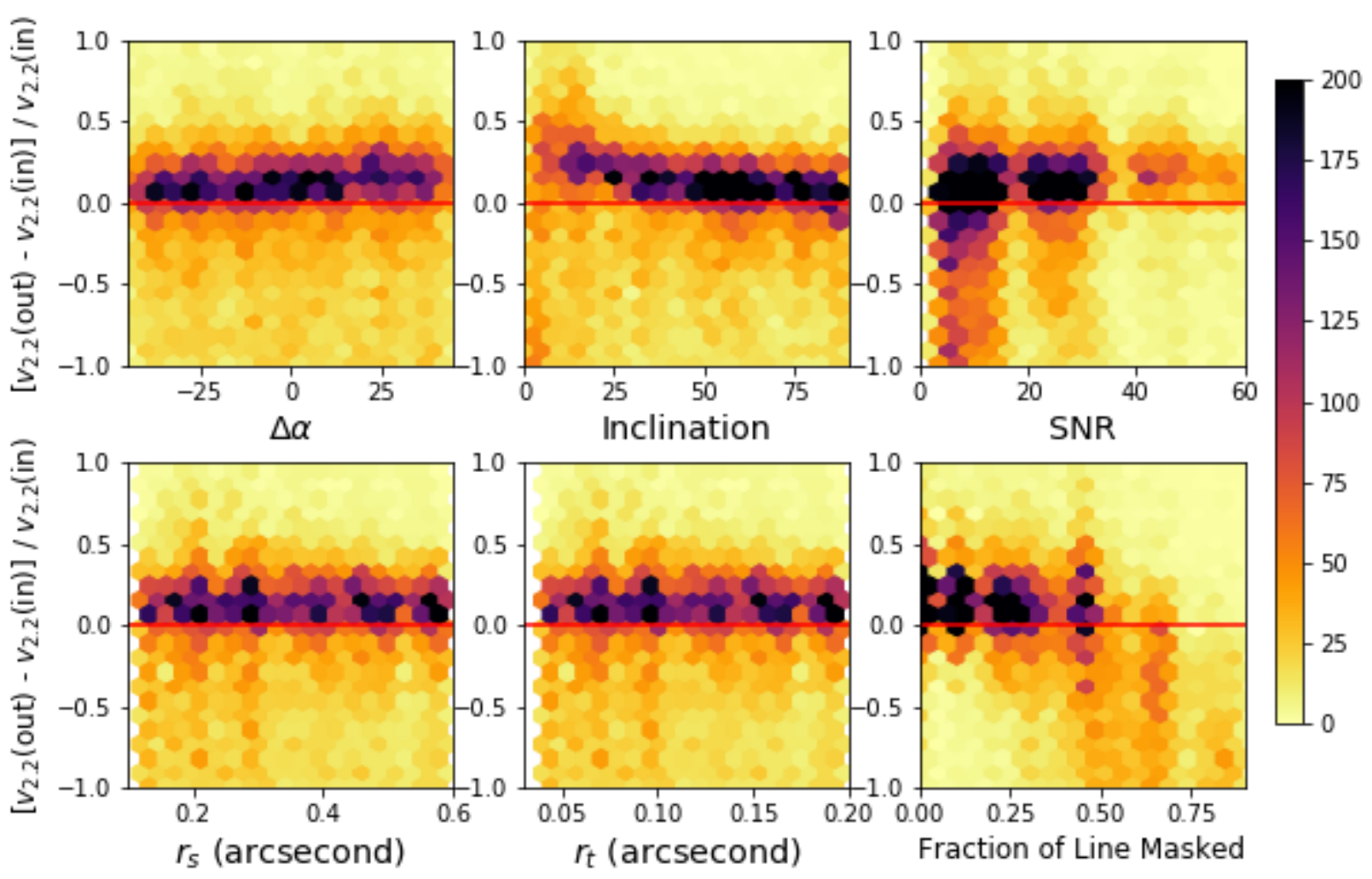}

\includegraphics[width=0.8\textwidth]{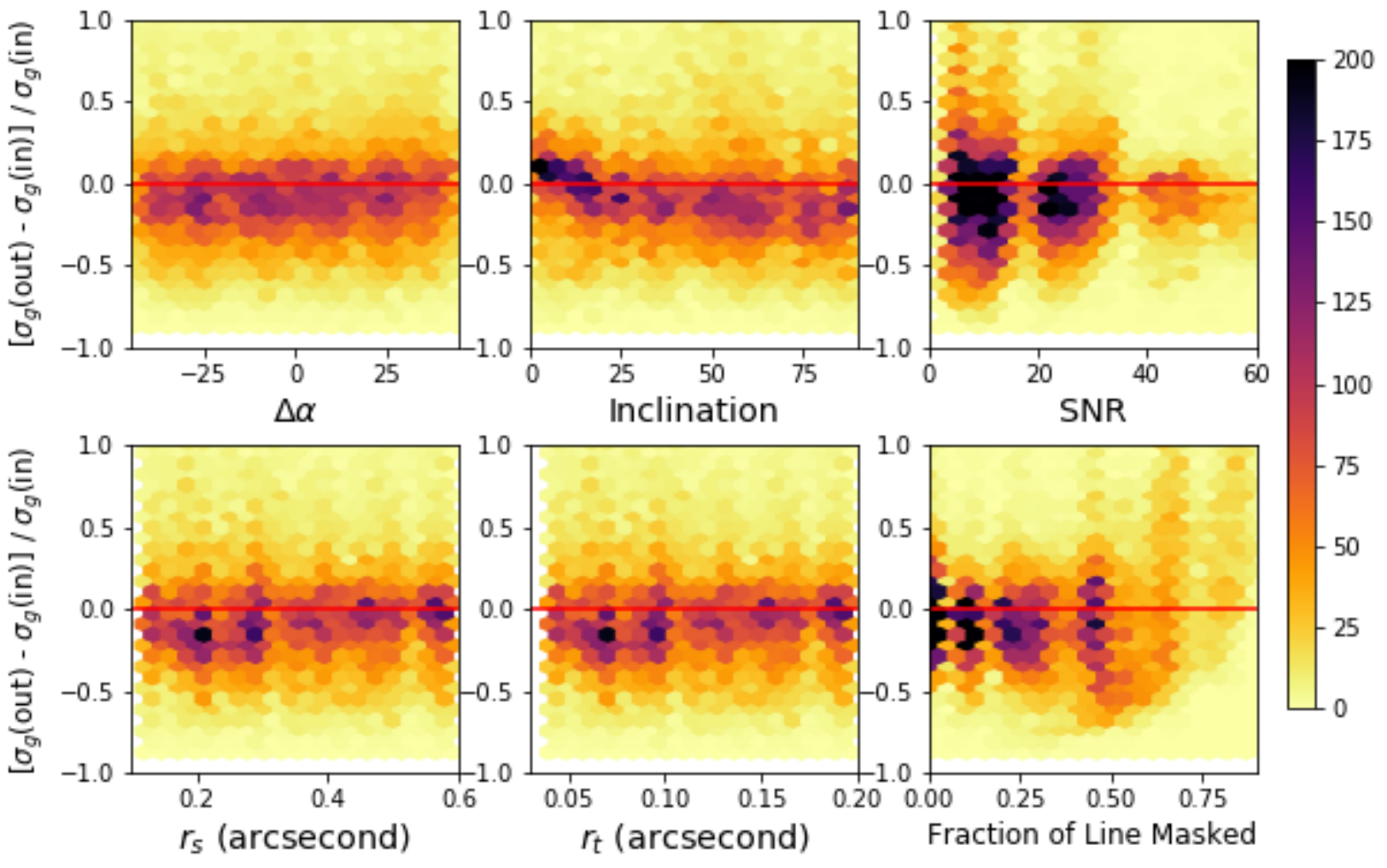}

\caption{Recovery rates of \vtwo\ and \sigmag\ for simulated MOSFIRE observations at varying SNR and portion of the emission line masked (due to sky emission). Simulations are emission line models generated by GBKFIT, and embedded in MOSFIRE-level sky noise. Using HELA modeling, we test our recovery rate against (from left, top row) $\Delta\alpha$ (slit and morphological PA offset), inclination, SNR, (from left, bottom row) $r_s$ (disk scale radius), $r_t$ (turnover radius), and emission line masked fraction. All 2D histograms are plotted on the same color scale. We tend to overestimate \vtwo\ by $\sim$10\%, and underestimate \sigmag\ by 10\%. Inclination tends to have an effect at an inclination of 30$^o$, where we begin overestimating our \vtwo\ by up to 30\%. At more than half the emission line masked, our recovery is unreliable.}
\end{figure*}

\subsection{Fixed and Free Turnover Radius}

The recovery of $r_t$ is significant in the recovery of rotational velocity, as $V_t$ is correlated with $r_t$.
However, \vtwo\ is a more reliable measurement due to a smaller offset from predicted.
Similar surveys fix $r_t$ in comparison to $r_s$ e.g. $r_t=0.4r_s$ \citep{Price2015}. 
We have decided our final sample will not hold $r_t$ fixed, and instead will allow $r_t$ to free values where $r_t< r_s$.
However we include results if we fix $r_t=0.33r_s$ and $r_t=0.4r_s$ in our analysis.

In the case where we allow $r_t$ to vary freely at any length below $r_s$, we find we overestimate $r_t$ by around 30\% of the input with a large scatter, while recovering our input $r_s$ to a median offset of -20\% of the input, and within a 1$\sigma$ scatter of 15\% of the input value.
However we tend to overestimate our velocity at $r=2.2 r_s$, to within $\sim$10\%.
We recover \sigmag\ to a small bias ($\sim$10\% underestimated from the input), at a 1$\sigma$ scatter of 15\%, increasing to 70\% scatter at high line coverage and low SNR.
Therefore, if we have bias in our results, we are overestimating the velocities in the \Mstar-TFR and in \voversig.
We also determine our ability to recover specific angular momentum, \jdisk\ (underestimated by only $\sim$5\% at low line coverage) and \voversig\ (overestimated by 25\% at low line coverage).
Interestingly, the rotational velocity and the velocity dispersion are both recovered well below $r_s<0.2\arcsec$. The size (both $r_s$ and $r_t$) of the modeled galaxy seems to be uncorrelated with the recovery rate, possibly because all our modeled galaxies are smaller than the seeing they are convolved to.

In our simulated observations from GBKFIT, $r_t$ is constantly held to be $r_s=3r_t$.
To determine our ability to recover the velocity, we try holding $r_t$ to be at this fixed distance relative to $r_s$.
When we recover our kinematic parameters while holding $r_t=1/3r_s$, we find that we underestimate both $r_t$ and $r_s$, but \vtwo\ is recovered with only minor offsets (overestimated by $\sim$5\% with a scatter of $\sim$20\% at low line coverage).
\sigmag\ is still recovered at minor offsets (underestimated by 5\%).
\jdisk\ is underestimated by 10\% and \voversig\ is overestimated by 25\%.
We find that we can reliably recover \vtwo\ and \sigmag\ at small offsets, as well as \sof\ and \jdisk.
However, due to the small scatter in the recovered values for \vtwo\ and \sigmag, our \voversig\ values have high scatter and are overestimated, and are thus likely unreliable.

\begin{figure*}[h] \label{fig:relationrecover}
\centering
\figurenum{12}
\includegraphics[width=0.8\textwidth]{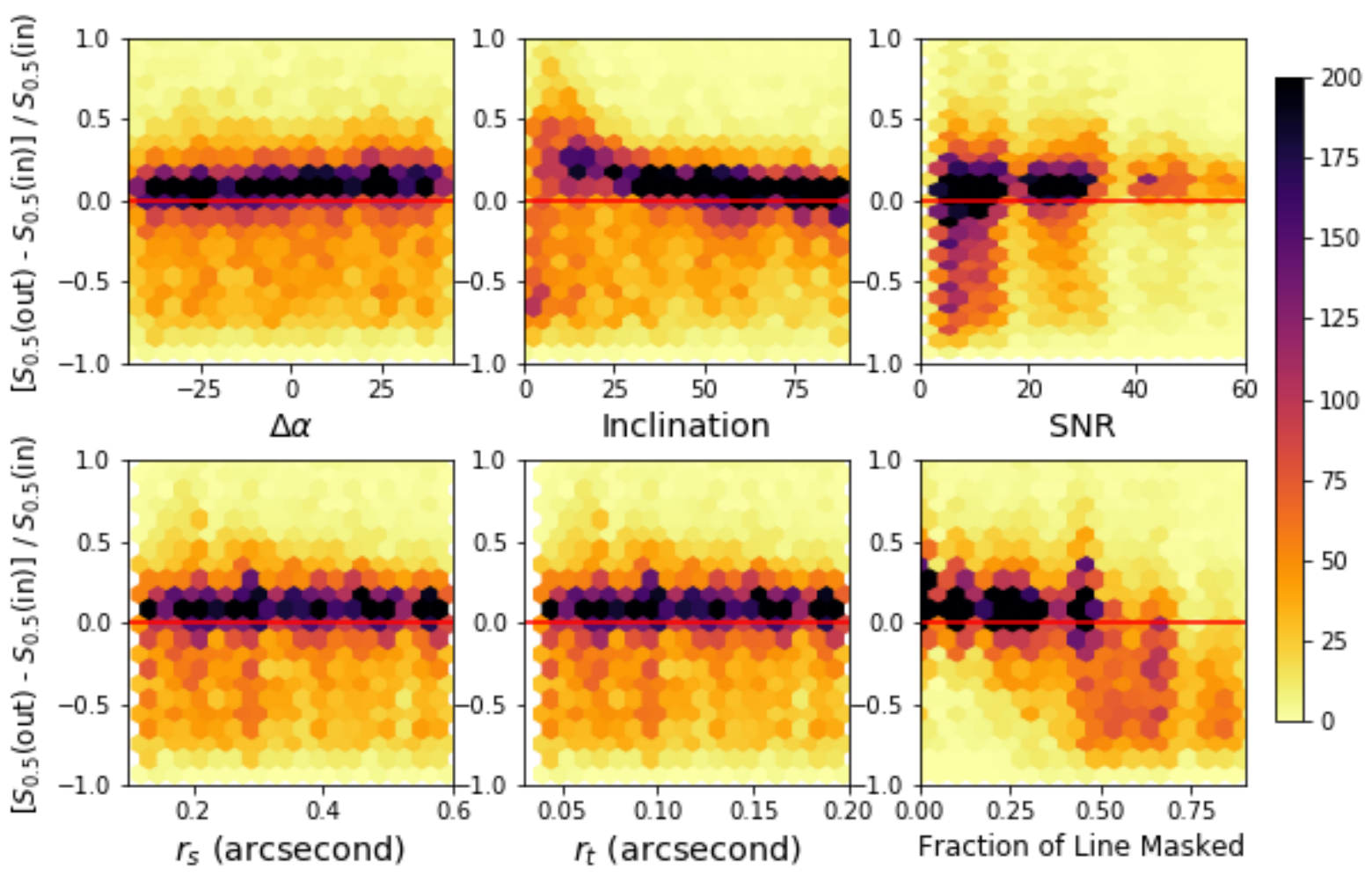}

\includegraphics[width=0.8\textwidth]{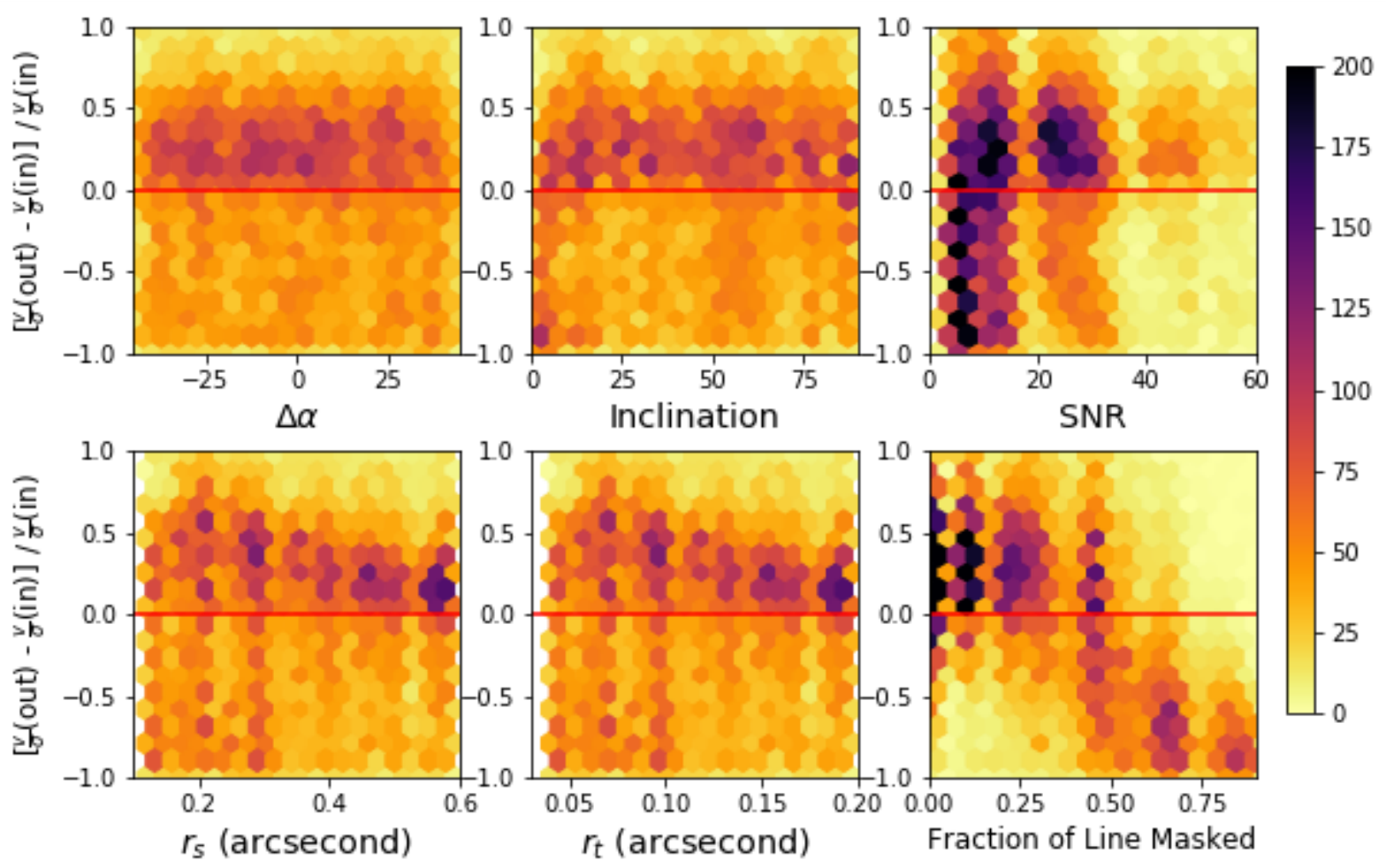}

\caption{Recovery rates of \sof\ and \voversig\ for simulated MOSFIRE observations. Top: We overestimate \sof\ by within 10\% of the input values. Inclination affects recovery starting at around 30$^o$, where we begin overestimating \sof\ by 20\%. Bottom: \voversig\ recovery is less reliable, where we tend to overestimate our values at around 25\% of our input value with significant scatter. These results indicate that the \sof\ parameter is by far the more reliable method of measuring kinematics, and \voversig\ values are possibly biased too high and at high scatter.}
\end{figure*}

\begin{figure*}[t] \label{fig:jrecover}
\centering
\figurenum{13}
\includegraphics[width=0.8\textwidth]{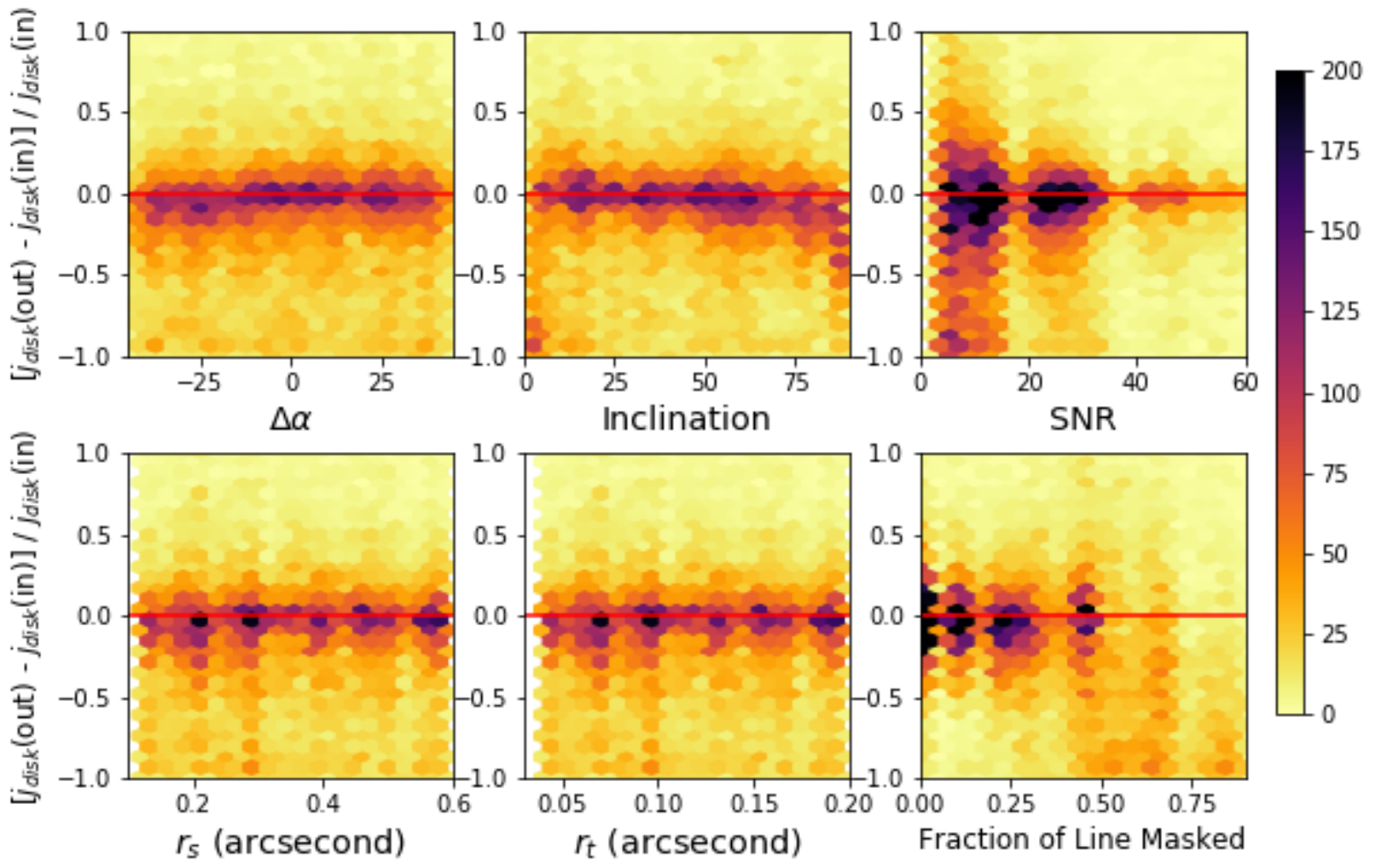}

\caption{Recovery rate of \jdisk\ for simulated MOSFIRE observations. We can reliably recover input \jdisk\ for objects with less than 50\% of the line masked, or with SNR$>$10.}
\end{figure*}

\end{document}